\documentclass[a4paper, 12pt]{article}
\usepackage{jheppub}
\usepackage[numbers,sort&compress]{natbib}
\usepackage{graphicx, subfigure}  
\usepackage{bm}        
\usepackage{amssymb}   
\usepackage{amsmath,mathtools,mathrsfs} 
\usepackage{mathalfa}
\usepackage{braket} 
\usepackage[british]{babel}
\usepackage[applemac]{inputenc}
\usepackage{bbold}
\usepackage{marginnote}
\usepackage{verbatim}

\usepackage{empheq}
\newlength\dlf  

\usepackage{tikz}
\usepackage{pdf14}
\usepackage{color}
\usepackage{pgfplots}
\usetikzlibrary{patterns}
\usepackage{tkz-euclide}
\usetkzobj{all}
\usepackage{subfigure}
\newcommand{\be}{\begin{equation}}
\newcommand{\ee}{\end{equation}}
\newcommand{\bea}{\begin{eqnarray}}
\newcommand{\eea}{\end{eqnarray}}

\newcommand{\del}{\partial}

\newcommand{\m}{{\ensuremath {\cal M}}}
\newcommand{\h}{{\ensuremath {\cal H}}}

\newcommand{\bW}{\ensuremath{\overline W}}

\newcommand{\bF}{\ensuremath{\bar F}}
\newcommand{\bZ}{\ensuremath{\overline Z}}
\newcommand{\bU}{\ensuremath{\overline U}}

\newcommand{\bb}{\ensuremath{\bar b}}
\newcommand{\ba}{\ensuremath{\bar a}}
\newcommand{\bc}{\ensuremath{\bar c}}
\newcommand{\bu}{\ensuremath{\bar u}}

\newcommand{\N}{\ensuremath{\vec{N}}}
\newcommand{\Nt}{\ensuremath{\vec{\tilde N}}}
\newcommand{\Pv}{\ensuremath{\vec{\Pi}}}
\newcommand{\G}{\ensuremath{{\cal G}}}
\newcommand{\D}{\ensuremath{{ D}}}

\newcommand{\Cm}{\ensuremath{{ \cal C}}}

\newcommand{\w}{\ensuremath{{\wedge}}}

\newcommand{\ks}{\varkappa }

\newcommand{\Sp}{\ensuremath{\cal{S}}}

\newcommand{\mabb}{\ensuremath{\left[ {\bf \cal{S} }\right]}}

\newcommand{\bi}{\ensuremath{\bar \imath}}
\newcommand{\bj}{\ensuremath{\bar \jmath}}

\title{The Spectra of Type IIB Flux Compactifications at Large Complex Structure}

\author[1]{Callum Brodie,}
\author[2]{M.C.~David Marsh}
\affiliation[1]{Rudolf Peierls Centre for Theoretical Physics, University of Oxford,\\ 1 Keble Road, Oxford OX1 3NP, United Kingdom}
\affiliation[2]{Department of Applied Mathematics and Theoretical Physics,
University of Cambridge, Cambridge, CB3 0WA, United Kingdom}

\emailAdd{callum.brodie@bnc.ox.ac.uk}
\emailAdd{m.c.d.marsh@damtp.cam.ac.uk}

\abstract{
We compute the spectra of the Hessian matrix, ${\cal H}$,
 and the matrix ${\cal M}$ that governs the critical point equation of the low-energy effective supergravity, as a function of the complex structure and axio-dilaton moduli space in type IIB flux compactifications at large complex structure. 
We find both spectra analytically in  an $h^{1,2}_-+3$ real-dimensional subspace 
 of the moduli space, and show that they exhibit a universal  structure with highly degenerate eigenvalues, independently of the choice of flux, the details of the compactification geometry, 
 and the number of complex structure moduli.
 In this subspace, the 
 spectrum of the Hessian matrix 
 contains no tachyons, but there are also no critical points.  We show numerically  that  the spectra of ${\cal H}$ and ${\cal M}$ 
 remain highly peaked
 over a large fraction of the sampled moduli space of explicit Calabi-Yau compactifications with 2 to 5 complex structure moduli.  In these models,
 the scale of the supersymmetric contribution to the scalar masses is strongly linearly correlated with the value of the superpotential over almost the entire moduli space, with particularly strong correlations arising for $g_s < 1$. We contrast these results with the expectations from the much-used continuous flux approximation, and comment on the applicability of Random Matrix Theory to the statistical modelling of the string theory landscape. 
}

\begin{document}
\maketitle

\newlength{\tempwidth}
\newlength{\tempheight}
\newlength{\densityplotsize}
\setlength{\tempwidth}{.32\linewidth}
\setlength{\tempheight}{.25\linewidth}
\setlength{\densityplotsize}{.38\linewidth}
\definecolor{mycolor}{rgb}{0.6, 0.63, 0.9}
\pgfplotsset{mz plot style main/.style={ 
ybar, bar width=1.4, height=.3\textwidth,width=.55\textwidth, hide y axis, axis x line*=left, xtick={-6,-4,...,6}, minor xtick={-5,-3,...,5}, xmin=-7, xmax=7,ymin=0,ymax=1,
    } 
}
\pgfplotsset{htot plot style main/.style={ 
ybar, bar width=1, height=.3\textwidth,width=.55\textwidth, hide y axis, axis x line*=left, xtick={-8,-4,...,8}, minor xtick={-6,-4,...,6}, xmin=-12, xmax=12,ymin=0,ymax=1,
    } 
}
\pgfplotsset{mz plot style appendix/.style={ 
ybar, bar width=1, height=\tempheight, width=\tempwidth, hide y axis, ticklabel style = {font=\tiny}, axis x line*=left, xtick={-6,-4,...,6}, minor xtick={-5,-3,...,5}, xmin=-7, xmax=7,ymin=0,ymax=1,
    } 
}
\pgfplotsset{htot plot style appendix/.style={ 
ybar, bar width=1, height=\tempheight, width=\tempwidth, hide y axis, ticklabel style = {font=\tiny}, axis x line*=left,  xtick={-8,-4,...,8}, minor xtick={-6,-4,...,6}, xmin=-12, xmax=12,ymin=0,ymax=1,
    } 
}
\pgfplotsset{list plot style/.style={ 
only marks, height=.4\textwidth, width=.45\textwidth, xlabel=\scriptsize{$m_s$}, ylabel=\scriptsize{$m_{3/2}$}, ticklabel style = {font=\scriptsize}, ylabel style={yshift=-.2cm},
    } 
}
\pgfplotsset{corrcoff plot front style/.style={ 
height=\densityplotsize, width=\densityplotsize, xlabel=Re$(u^i)$, ylabel=Im$(u^i)$, ylabel style={yshift=-.4cm},xmin=\xMin,xmax=\xMax,ymin=\yMin,ymax=\yMax,xtick={-2,-1,...,2},ytick={1,...,4},minor xtick={-2,-1.8,...,2},minor ytick={.4,.6,...,5},
    } 
}
\pgfplotsset{corrcoff plot back style/.style={ 
height=\densityplotsize, width=\densityplotsize,xmin=\xMin,xmax=\xMax,ymin=\yMin,ymax=\yMax, xtick=\empty,ytick=\empty,
    } 
}
\pgfplotsset{densityplot legend style/.style={ 
height=.65\densityplotsize, width=.02\textwidth, xmin=-1, xmax=0, tick style={opacity=0.6},minor tick style={opacity=0.3},scale only axis,yticklabel pos=right,ytick pos=right,ytick align=outside,x axis line style=-,y axis line style=-,ticklabel style = {font=\scriptsize},
    } 
}
\newcommand{\rowname}[1]
{\rotatebox{90}{\makebox[\tempheight][c]{\scriptsize{#1}}}}
\newcommand{\columnname}[1]
{\makebox[\tempwidth][c]{\scriptsize{#1}}}

\section{Introduction}
Flux compactifications
provide a
promising framework for
connecting string theory with 
 the models and phenomena of particle physics and cosmology  \cite{Grana:2000jj, Gubser:2000vg, Giddings:2001yu, deWolfe:2002nn, Kachru:2003aw, Kachru:2003sx}.  Fluxes give rise to an energy density that depends on the shape, and hence the moduli,  of the compactification manifold, and the minimisation of this energy allows for a controlled tree-level stabilisation of several moduli, 
which 
is helpful in
fixing moduli-dependent 
coupling constants 
in the low-energy theory
and in
bringing it  
into agreement with observational constraints on massless scalars. 
 Furthermore,  the fluxes backreact on the  geometry and may create regions 
 of the extra dimensions with significant warping.   
As there is 
a very large number of possible 
  compactification manifolds
  and 
  flux choices, 
the number of 
low-energy, four-dimensional
effective theories arising from flux compactifications is enormous 
and prompts 
 the notion of a `flux landscape' of effective theories (see e.g.~\cite{Douglas:2006es, Denef:2007pq, Denef:2008wq, Maharana:2012tu, Schellekens:2013bpa} for reviews). 

The existence of such a landscape 
raises
several challenges. First, deriving the  
explicit predictions from \emph{generic} flux compactifications  
with tens or hundreds of moduli fields
 is   computationally 
 prohibitively complicated:
 the explicit form of the (classical) flux-induced effective 
 potential has so far only been found for examples with a handful of moduli  (for which the period vector has been explicitly computed), and a systematic classification of the vacua for a given compactification manifold has only been possible for comparatively simple  examples
 with  few moduli (see e.g.~\cite{Tripathy:2002qw, Markus}). With generic vacua out of  
 reach, the generally applicable lessons from  
 these special,
 explicit constructions may at best be inferred  by extrapolation, which may be tenuous. 
   
   Second, 
   it is a priori possible that the landscape of flux vacua contains a large number of solutions that are compatible with the outcomes of any experiments and observations 
 that humankind may ever conduct.   
If so, it is unlikely that explicitly detailing the properties of
any given flux vacuum
will lead to profound insights,
even if the computational obstacles for constructing explicit, generic flux vacua were overcome. 
 
Rather, both these challenges motivate a statistical approach: 
by approximating the quantised fluxes as continuous variables,  
a great deal has been learned about the distribution of flux vacua without requiring the direct construction and enumeration of the corresponding solutions \cite{Ashok:2003gk, Denef:2004ze, Denef:2004cf, Eguchi:2005eh} (see also~\cite{Douglas:2006es, Denef:2007pq, 
Schellekens:2013bpa} and references therein).
The employment of statistical tools opens up the possibility of finding 
  limits of the theory where the relevant distributions  
 take relatively simple forms due to  
 some form of 
 central-limit-theorem type of behaviour.  The spectra of matrix ensembles provide a particularly compelling target in this respect  since for
 large matrices with randomly distributed entries,
  the spectra  
  quickly approach `universal' limits that are largely independent of the statistical input. 
 For example, the spectrum of 
large
 Hermitian matrices with statistically independent, normally distributed entries is famously given by the Wigner semi-circle law \cite{Wigner, Mehta}, but so is the spectrum of any Hermitian matrix ensemble with independent and identically distributed \emph{non-Gaussian} entries as long as the moments of the distribution are sufficiently bounded \cite{Deift, Kuijlaars}, and so is the  spectrum of random Hermitian matrices with a large number of \emph{statistically correlated} matrix entries \cite{Wigner2, 2005SchenkerShulzBaldes,  2007HofmannCredner}. The existence of  strong universality theorems has motivated the employment of random matrix theory (RMT) techniques in the study of the flux landscape \cite{Denef:2007pq, Marsh:2011aa, Chen:2011ac,Bachlechner:2012at, Sousa:2014qza} (see also \cite{MH:2004gm,Aazami:2005jf, Easther:2005zr, Pedro:2013nda, Long:2014fba, Marsh:2013qca, Bachlechner:2014gfa, Bachlechner:2014hsa, Battefeld:2014qoa} for additional string theory motivated  applications of RMT to cosmology), thus potentially providing a significant extension of the computational reach of the statistical methods. 
  While  the relevant matrices arising in 
  flux compactifications
are not  generic, random matrices, but rather carry a lot of structure inherited from the geometry and topology of the compactification, one may hope that for sufficiently large and complex systems, RMT universality will dominate the string theory correlations, and  comparatively simple spectra may emerge as a result \cite{Denef:2007pq, Marsh:2011aa}.

Two matrix ensembles play particularly prominent roles in the study of flux vacua: the first matrix, which we denote by \m, is formed from the second covariant derivatives of the superpotential $W$ as,
\be
\m= \left(
\begin{array}{c c }
0 & Z_{ab} e^{-i \vartheta} \\
\bZ_{\ba \bb} e^{i \vartheta} & 0
\end{array}
\right) \, , \label{eq:m}
\ee
where 
$Z_{ab} =  \D_a D_b W = \partial_a F_b + K_a F_b - \Gamma_{ab}^c F_c$ for $F_a = D_a W = \del_a W +K_a W$ and $K_a = \del_a K$. 
Here  
$\vartheta$ denotes the argument of the superpotential, and we have set $M_{\rm Pl} =1$. 
The importance of \m~is two-fold: the elements of $Z_{ab}$ 
set the scale of the supersymmetric contribution to the masses of the chiral fields,   
 and 
 \m~appears in
 the critical point equation, $\partial_a V= 0$, of the F-term scalar potential, $V= e^K \left(
K^{a \bar b} F_a \bar F_{\bar b} - 3 |W|^2
\right)$, as
 \cite{Denef:2004cf},
\be
\m \hat{F}_{\pm} = \pm 2|W| \hat{F}_{\pm} \, ,
\label{eq:cp}
\ee 
where,
\be
 \hat F_{\pm}  =
\left(
\begin{array}{c}
F^{\bar a} e^{-i\vartheta} \\
\pm \bF^a e^{i\vartheta} 
\end{array}
\right) \, .
\label{eq:v1}
\ee 
One of the main results of  \cite{Denef:2004cf} was to note that the symmetries of \m~are just those of the symmetry class CI in the classification of mesoscopic Hamiltonians by Altland and Zirnbauer \cite{Altland:1997zz}.
It was furthermore argued in  \cite{Denef:2004cf} that
 for sufficiently 
 generic compactifications with many moduli, 
 the rough features of the spectrum should be well-described by the 
 corresponding random matrix theory ensemble.

The second matrix of considerable interest for the counting of flux vacua is the Hessian matrix defined by,
\be
{\cal H} = 
\left(
\begin{array}{c c}
\nabla_{a}\nabla_{ \bb} V & \nabla_{a} \nabla_{ b} V \\
\nabla_{\bar a} \nabla_{ \bb} V & \nabla_{\bar a} \nabla_{ b} V
\end{array}
\right) \, .
\ee
For a critical point to be a metastable vacuum, the spectrum of \h~must be positive definite. Reference \cite{Marsh:2011aa} (by one of the present authors, and collaborators) showed that for a `random supergravity' in which the superpotential and K\"ahler potential are random functions, \h~is well-described by a random matrix model consisting of the sum of a Wigner matrix with two Wishart-type matrices. The spectrum of this `WWW-model' was obtained analytically by 
freely convolving the spectra of the independently contributing matrices (cf.~equation (4.8) of \cite{Marsh:2011aa}). In this model, typical critical points are unstable saddle-points and \h~has a significant fraction of negative eigenvalues. Since the assumptions of the random supergravity closely matched those proposed for the flux landscape in \cite{Denef:2004cf}, one may expect these random matrix results to be applicable for the complex structure and axio-dilaton sector of the flux landscape.

Clearly, it is important to verify the applicability of RMT techniques to the flux landscape, but doing so is hard for obvious reasons: the universal limits are expected to be applicable precisely for the large and generic  systems that are the most challenging to construct explicitly. 
The purpose of this paper is to compute the spectra of \m~and \h~in the flux landscape, focussing on
the axio-dilaton and complex structure moduli sector of
 type IIB flux compactifications in the large complex structure limit. Our main results are as follows:
\begin{itemize}
\item We numerically show that the spectra of \m~and \h~in explicit flux compactifications with two to five complex structure moduli differ significantly from the `universal' RMT spectra (cf.~section \ref{sec:num}). The string theory spectra exhibit strong peaks that are absent in the random matrix theory models.
\item For $h^{1,2}_->0$ complex structure moduli, 
we analytically compute the spectra of \m~and \h~in an $h^{1,2}_-+3$ real-dimensional subspace, \Sp, of the full moduli space (cf.~section \ref{sec:analytic}). The eigenvalues of the matrix ${\cal M}$ come in opposite sign pairs and, in this subspace, the positive branch of the spectrum is given by $h^{1,2}_-$ degenerate eigenvalues equal to $|W|$ and a single eigenvalue equal to $3|W|$. 
The 
spectrum of \h~is given by $h^{1,2}_-$ zero modes, $h^{1,2}_-+1$ degenerate eigenvalues equal to $2m_{3/2}^2$ and a single eigenvalue equal to $8 m_{3/2}^2$. Here, as in the rest of this paper, $m_{3/2}$ denotes the gravitino mass restricted to the complex structure and axio-dilaton sector. These results hold for all flux compactifications in the large complex structure limit, i.e.~independently of the number of complex structure moduli, the `Yukawa couplings' ($\ks_{ijk}$ not all vanishing),  and the  (not all vanishing) flux configuration. In this  sense these string theory spectra are universal, albeit very different from the random matrix theory expectations. See Figure \ref{fig1} for a comparison of the RMT spectra with the  analytic string theory spectra in \Sp. 

\begin{figure}[t!]
    \begin{center}
     \subfigure[Spectrum of \m~in units of $|W|$]{\label{fig:AZ-CI}
        \includegraphics[width=0.43\textwidth]{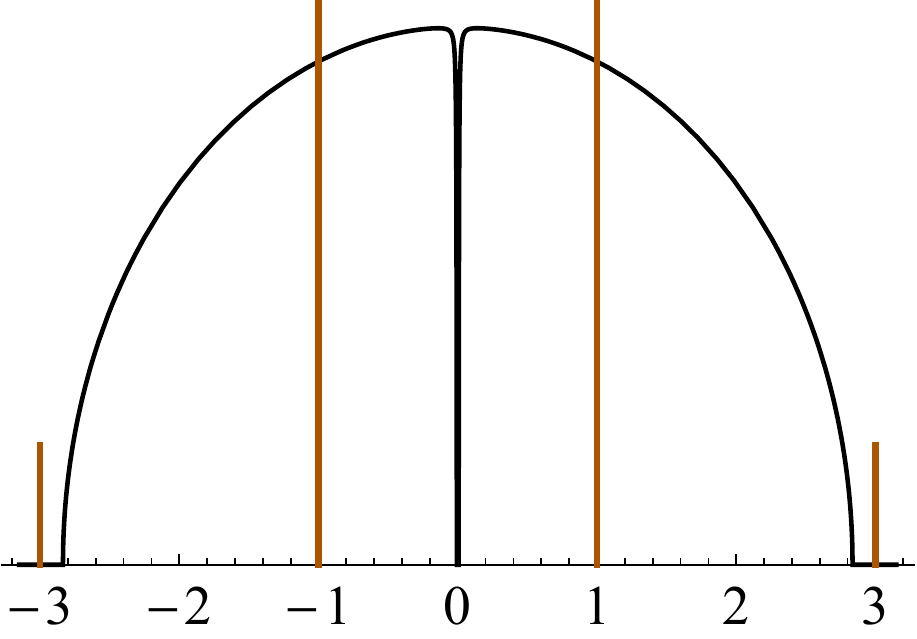}}
     \subfigure[Spectrum of \h~in units of $m_{3/2}^2$] {\label{fig:WWW}
        \includegraphics[width=0.43\textwidth]{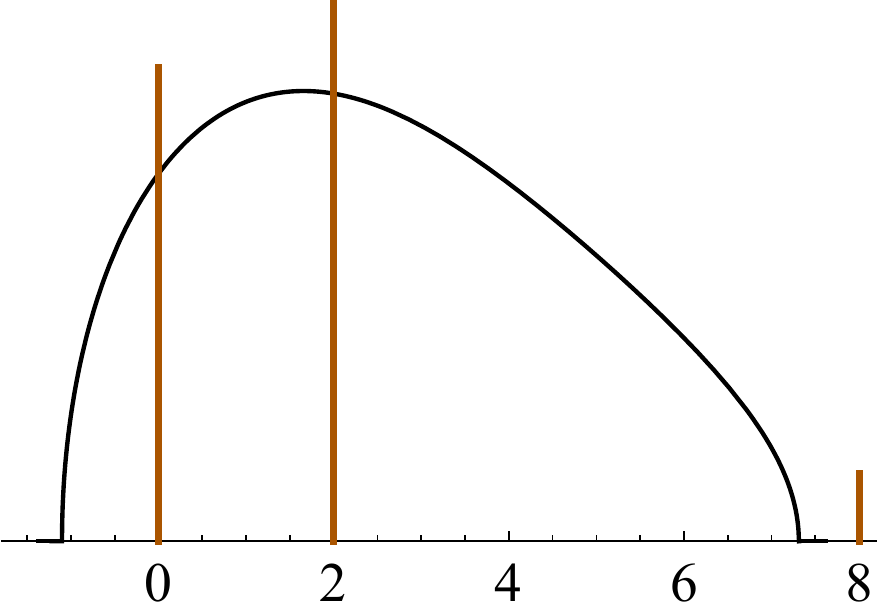}}
            \end{center}
    \caption{Random Matrix Theory spectra in black, string theory spectra in  \Sp~in orange.  Normalisation is arbitrary, and the relative heights of the delta-function peaks are taken to schematically indicate the degeneracy of each eigenvalue.
}
   \label{fig1}
\end{figure}

\item We show that there are no critical points -- neither supersymmetric nor non-supersymmetric -- in \Sp. We furthermore show that the slow-roll  parameters are universally given by $\epsilon = 4$, $\eta_{\parallel} =8$ 
(see section \ref{sec:nogo}).

\item We show numerically that the scale of the supersymmetric masses exhibits a strong positive linear correlation with the value of the superpotential in the explicit compactifications we consider (cf.~section \ref{sec:corr}). The continuous flux approximation that underpins many statistical results on the flux landscape predicts a vanishing or negative correlation between these quantities within a broad set of assumptions, and we discuss the break-down of this approximation (cf.~section \ref{sec:contFlux}). The strong correlation significantly reduces the frequency of compactifications with large flux induced hierarchies in this region of the moduli space. 


\item We consider flux compactifications beyond the large complex structure limit and show that
%
existing
universality theorems of random matrix theory do not by default apply to these compactifications. 
We suggest that
RMT techniques may nevertheless be applicable to more general compactifications if
 the geometric correlations that arise in string  theory compactifications 
 are taken into account
 (cf.~section \ref{sec:rmt}). 
\end{itemize}
These findings lead to several  directions of possibly very interesting future research, and we briefly discuss these together with our conclusions in section \ref{sec:concl}.

\section{Type IIB flux compactifications}
\label{sec:bkg}
 In this section, we briefly review the structure of the four-dimensional supergravities that arise as the low-energy limit of 
 type IIB string theory on Calabi-Yau orientifolds with non-trivial R-R and NS-NS flux.

The relevant low-energy degrees of freedom
for a compactification on the orientifold $\tilde M_3$ of the Calabi-Yau three-fold $M_3$
 include the axio-dilaton, $\tau = C_0 + i e^{-\phi}$,  the complex structure moduli, $u^i$, where $i=1,\ldots,h^{1,2}_-(\tilde M_3)$, and the K\"ahler moduli, $T^r$, where $r=1,\ldots, h^{1,1}_+( \tilde M_3)$.\footnote{ In addition, the spectrum may contain axion multiplets $G^{\alpha}$, with $\alpha = 1, \ldots, h^{1,1}_-(\tilde M_3)$, which we will not consider in this paper.} 
To leading order in the $g_s$ and $\alpha'$ expansions, the K\"ahler potential is given by,
 \be
 K = - \ln \left(i \int_{M_3} \Omega \wedge \bar \Omega \right)
 - \ln \left(-i(\tau - \bar \tau)\right)
 - 2 \ln {\cal V}
 \label{eq:K}
 \, ,
 \ee
where ${\cal V}$ denotes the compactification volume and $\Omega$ the holomorphic three-form. We will throughout this paper consider $\tau$ in the fundamental region of the torus, $\{ \tau \in \mathbb{C}, |{\rm Re}(\tau)| < \frac{1}{2}, |\tau|\geq1, {\rm Im}(\tau)>0\}$.  

The complex structure moduli arise from the periods of $\Omega$ as follows: for $I = 0, \ldots , h^{1,2}$, take $(A^I, B_I)$ to be a canonical homology basis of $H_3(M_3)$ and $(\alpha_I, \beta^I)$  the dual cohomology basis satisfying,
\bea
\int_{M_3} \alpha_I \wedge \beta^J= - \int_{M_3} \beta^J \wedge \alpha_I = \delta_I^J \, ,~~~~
\int_{M_3} \alpha_I \wedge \alpha_J = \int_{M_3} \beta^I \wedge \beta^J = 0\, .
\eea 
With respect to this basis, the periods of $\Omega$ are given by,
\be
\vec{\Pi} =
 \left(
 \begin{array}{c}
 \int_{A^I} \Omega \\
  \int_{B_I} \Omega
  \end{array}
  \right)
   \equiv 
\left(
\begin{array}{c}
z^I \\
 {\cal G}_I
 \end{array}
\right) \, .
\label{eq:Period0}
\ee
Here $z^I$ serve as projective coordinates on the complex structure moduli space. The periods  ${\cal G}_I$ satisfy the equation,
$
2 {\cal G}_I = \del_I (z^J {\cal G}_J)
$, so that 
 ${\cal G}_I$
 is the gradient of a homogeneous function of degree two: ${\cal G}_I = \partial_I {\cal G}$.

The complex structure moduli 
 can  be expressed as the
  inhomogeneous coordinates 
on the space of complex structure deformations, $u^i = z^i/z^0$ for $i = 1, \ldots, h^{1,2}$.  Upon setting $z^0 =1$, the period vector is given by,  
\be
\Pv = 
\left(
\begin{array}{c}
1 \\
u^i \\
2 F- u^j F_j \\
F_i
 \end{array}
\right) \, ,
\label{eq:Period}
\ee
with the prepotential $F = (z^0)^{-2} \G$. 
We will be particularly interested in the
 `large complex structure expansion', in which $F$ is given by,
\be
F = - \frac{1}{6} \ks_{ijk} u^i u^j u^k - \frac{1}{2} \ks_{ij} u^i u^j + \ks_i u^i + \frac{1}{2} \ks_0 + I \, .
\label{eq:F}
\ee
where $I$ denotes quantum instanton contributions (that we will be more specific about when considering explicit examples). In this limit, the expansion coefficients are given by the classical intersections of the mirror-dual Calabi-Yau, and the coefficients $\ks_{ijk} = \int_{M_3} \Omega \wedge \partial^3_{ijk} \Omega$  are traditionally referred to as the `Yukawa couplings'. The $d=4$, ${\cal N}=1$ low-energy supergravity is obtained by orientifolding $M_3$ to $\tilde M_3$ and  leaves the involution-odd complex structure moduli in the chiral spectrum \cite{Grimm:2004uq}. The details of this involution are not important to our general discussion, and
we will henceforth take $i$ to run from $1$ to $h^{1,2}_-$. 

The complex structure dependent part of the K\"ahler potential may now be written as,
\bea
K_{\rm c.s.} &=& - \ln \left(i \int_{M_3} \Omega \wedge \bar \Omega \right) = 
-\ln \left( i \vec{\Pi}^{\dagger}\, \Sigma\, \vec{\Pi} \right) \nonumber \\
&=&- \ln\left(
\frac{i}{6} \ks_{ijk}(u^i - \bar u^i)(u^j - \bar u^j)(u^k - \bar u^k)  
 -2 {\rm Im}(\ks_0)  
\right) \, ,
\eea
where $\Sigma$ denotes the symplectic matrix,
\be
\Sigma = 
\left(
\begin{array}{c c}
0 & \mathbb{1} \\
- \mathbb{1} & 0
\end{array}
\right) \, .
\ee
We will denote the joint moduli space of the axio-dilaton and complex structure moduli by \Cm.

We are interested in compactifications in which 
integrally quantised RR ($F_3$) and NS-NS ($H_3$) fluxes wrap some non-trivial three-cycles of $M_3$, 
\be
\frac{1}{(2\pi)^2 \alpha'} \int_{A^I, B_I} F_3 = \vec{N}_{\rm RR} \in \mathbb{Z}^{2(h^{1,2}_-+1)}\, ,~~~\frac{1}{(2\pi)^2 \alpha'} \int_{A^I, B_I} H_3 = \vec{N}_{\rm NS-NS} \in \mathbb{Z}^{2(h^{1,2}_-+1)}\, .
\ee
It is convenient to introduce the complex three-form flux $G_3 = F_3 - \tau H_3$, and define the complexified flux vector (without subscript) as,
\be
\vec{N} =  \Sigma \, 
\left(
\begin{array}{c}
 \int_{A^I} G_3 \\
 \int_{B_I} G_3
\end{array}
\right) \, .
\ee
The fluxes contribute to the D3-charge tadpole by,
\be
Q_{\rm flux} = \frac{1}{(2\pi)^4 (\alpha')^2} \int_{M_3} H_3 \wedge F_3 
= - \frac{1}{(2\pi)^4 (\alpha')^2} \frac{1}{\tau - \bar \tau} \vec{N}^{\dagger}\, \Sigma\, \vec{N} \, .
\label{eq:Qflux}
\ee 
Requiring that the total sum of D3 charge vanish in the internal space leads to a joint condition on the D3-brane content, the fluxes, and the D7-brane and O-plane configuration,
\be
Q_{\rm flux} + N_{\rm D3} = \frac{\chi}{24} - \frac{1}{4} N_{\rm O3} \, ,
\ee
where $N_{\rm D3}$, $N_{\rm O3}$  denote the net number of D3-branes and O3-planes, and $\chi$ is the Euler characteristic of the F-theory four-fold that corresponds to the given D7-brane and O7-plane content.  Fluxes that preserve some  supersymmetry contribute positively to the tadpole,  and when considering specific examples we will impose,
\be
0 \leq Q_{\rm flux} \leq L_{\star} \, ,
\label{eq:tadpole}
\ee
where $L_{\star}$ denotes the model-dependent maximal contribution of the fluxes to the D3-tadpole, given a certain configuration of D7-branes and orientifold planes. 
The tadpole condition 
 does not ensure supersymmetry, and flux  satisfying \eqref{eq:tadpole} is generically non-supersymmetric.  
 
The fluxes induce a complex structure and axio-dilaton dependent energy density that in the four-dimensional theory is captured by  the flux induced superpotential \cite{Gukov:1999ya},
\be
W =
\int_{M_3}  G_3 \wedge \Omega =   \vec{N} \cdot \vec{\Pi} \, .
\label{eq:GVW}
\ee 
Clearly, $W$ is linear in $\tau$ and, classically,  at most cubic in the complex structure moduli in the large complex structure expansion.

We may now define the   `flux landscape'  
as the 
 ensemble of  four-dimensional ${\cal N}=1$ supergravities with a K\"ahler potential of the form \eqref{eq:K} and a superpotential of the form \eqref{eq:GVW} for the set of viable compactification manifolds and consistent choices of flux.

 Note that we do not restrict the flux to be supersymmetric (as we want to study the moduli space dependence of the spectra of \m~and \h), so that the tadpole condition \eqref{eq:tadpole} does not bound the flux choices to be finite for a given compactification manifold and flux tadpole. However, we will in addition require that the K\"ahler sector is stabilised in such a way that the four-dimensional  supergravity is consistently  the controlled low-energy limit of the corresponding string compactification. In other words, while we will not explicitly consider K\"ahler moduli stabilisation in this paper,  we will assume that the compactification volume is stabilised at a sufficiently large value to 
  justify the $\alpha'$-expansion, the neglect of higher KK-modes, and the validity of the supergravity action at energies below the KK-scale. For example, we implicitly  require that 
$
m_{3/2}|_{\rm full}/m_{KK} \ll1
$, 
where $m_{3/2}|_{\rm full}$ denotes the gravitino mass including the K\"ahler moduli. Phrased in terms of the gravitino mass of the truncated axio-dilaton and complex structure sector,
\be
m_{3/2} \equiv e^{(K_{(\tau)} +K_{\rm c.s.})/2} |W| \, ,
\label{eq:m32}
\ee
the corresponding constraint is given by \cite{Cicoli:2013swa}
\be
m_{3/2} \ll {\cal V}^{1/3} \, .
\ee 
This effectively bounds the ensemble of flux choices for any given manifold to be finite.

We close this section by 
making one additional comment on 
 our neglect of K\"ahler moduli, as we will in this paper predominantly  
consider the truncated system of the axio-dilaton and the complex structure moduli. This truncation is well-motivated as we aim to study the intrinsic structure and randomness of the flux superpotential, which does not depend on the K\"ahler moduli. While the inclusion of K\"ahler moduli may change the spectra of \h~and \m~through non-vanishing cross-terms, such changes are in many interesting cases small or easy to take into account, such as e.g.~in the case of complex structure moduli and the axio-dilaton being stabilised at a hierarchically higher scale than the K\"ahler moduli \cite{Kachru:2003aw}, or when the no-scale symmetry of the K\"ahler sector is only weakly broken \cite{LVS, Westphal:2006tn, Covi:2008ea, Kallosh:2014oja,decoupling}. Moreover, not very much is known about the distribution of  non-perturbative effects that may stabilise the K\"ahler moduli, and  statistical modelling based on e.g.~random matrix theory is less well-motivated (see however \cite{Easther:2005zr, Long:2014fba, Bachlechner:2014gfa} for some interesting developments in this direction). 
For similar reasons, 
%
several previous statistical studies of 
 flux vacua have  neglected K\"ahler moduli \cite{Ashok:2003gk, Denef:2004cf, Denef:2004ze}.





\section{Spectra of \m~and \h~in an explicit flux compactification}
\label{sec:num}
\label{sec:examples}
\label{sec:explicit}
We begin our study of the spectra of \m~and \h~with an instructive  example  
of 
flux compactifications on a particular  orientifold. We will find it useful to return to this example at several points throughout this paper, and we will refer to it as   `Model 1'. In this section, we numerically compute the spectra of \m~and \h~for canonically normalised fields in Model 1  as a function of the effectively four complex-dimensional moduli space \Cm.


As any given example manifold may have particularities that could bias the results, we will in Appendix \ref{app:B} compute the spectra in four additional flux compactifications as well as in a non-trivial modification of Model 1. All
 compactification manifolds 
that we consider in this paper can be constructed using toric geometry, and all have been previously studied in the literature: our Models 1--4 are taken directly from reference \cite{Krippendorf}, and our Model 5 is given by the degree 18 hypersurface in $\mathbb{CP}^4_{1,1,1,6,9}$,  which has previously been studied in~\cite{Candelas:1994hw, Denef:2004dm, Markus, LVS}. 
\subsection{An explicit Calabi-Yau orientifold compactification}
Model 1 is constructed  through compactification on one of a mirror-dual pair of Calabi-Yau hypersurfaces in certain four-dimensional toric varieties \cite{Krippendorf}.
The compactification admits a discrete $\Gamma =\mathbb{Z}^3_3$ action acting on the periods, and  by considering only the subsector that is invariant under this action, the corresponding low-energy effective theory includes four complex structure moduli (non-invariant moduli can be shown to be fixed supersymmetrically). The classical prepotential is obtained from the intersection numbers of the mirror pair, and is in the large complex structure limit given by,
\bea
&F_{\rm cl.} = +3u_1u_4+\frac{3}{2}u_2u_4+\frac{3 }{2}u_3u_4+\frac{15}{4}u_4^2+\frac{3}{2}u_1+u_2+u_3+\frac{33}{12}u_4-i\zeta(3)\frac{33}{4\pi^3}& \nonumber \\
&-\frac{3}{2}u_1^2u_4
-3u_1u_2u_4-3u_1u_3u_4-3u_2u_3u_4-\frac{9}{2}u_1u_4^2-3u_2u_4^2-3u_3 u_4^2-\frac{5}{2}u_4^3& \, ,~~ 
\eea
where $\zeta$ denotes the Riemann zeta function. 
World-sheet instantons correct the prepotential at the non-perturbative level, and the leading contributions are given by \cite{Krippendorf},
\bea
I &=& 3 e^{2 i \pi  u_1}+3 e^{2 i \pi  u_2}+3 e^{2 i \pi  u_3}+144 e^{2 i \pi  u_4}+
144 e^{4 i \pi  u_4}+
\nonumber  \\
&+&27 e^{2 i \pi 
   \left(u_1+u_4\right)}+27 e^{2 i \pi  \left(u_2+u_4\right)}+27 e^{2 i \pi  \left(u_3+u_4\right)} + \ldots 
   \label{eq:Fq}\, .
\eea
With the particular D7-brane and O-plane configuration considered in \cite{Krippendorf}, $u_2$ and $u_3$ are related by the orientifold involution and only fluxes that are symmetric under $u_2 \leftrightarrow u_3$ may consistently be turned on. Thus, this model has $u_2=u_3$ and effectively three complex structure moduli.  The flux tadpole of equation \eqref{eq:tadpole} is given by $L_{\star} = 22$.  


\subsection{The spectrum of \m}
We are now interested in characterising the spectra of \m~and \h~as a function of the moduli space \Cm. Throughout this paper, we consider the spectra of canonically normalised fields, and for our numerical study of Model 1, we include the world-sheet instantons up to  
the second order, cf.~equation \eqref{eq:Fq}.

\begin{figure}[t!]
\begin{center}
\subfigure[$\tau = u^i=10i$ ]{\label{fig:Mfirst}     
\begin{tikzpicture}
\begin{axis}[mz plot style main]
\addplot[mycolor,fill=mycolor] table[y index = 1] {Mz_histogramData_model1wrtSym_plot1.dat};
\end{axis}
\end{tikzpicture}}%
\subfigure[$u_{\textrm{Im}}^{\textrm{max}}=10,\tau_{\textrm{Im}}^{\textrm{max}}=10$] {\label{fig:Msecond}
\begin{tikzpicture}
\begin{axis}[mz plot style main]
\addplot[mycolor,fill=mycolor] table[y index = 1] {Mz_histogramData_model1wrtSym_plot2.dat};
\end{axis}
\end{tikzpicture}}%
\\%
\subfigure[$u_{\textrm{Im}}^{\textrm{max}}=5,\tau_{\textrm{Im}}^{\textrm{max}}=5$] {\label{fig:Mthird}
\begin{tikzpicture}
\begin{axis}[mz plot style main]
\addplot[mycolor,fill=mycolor] table[y index = 1] {Mz_histogramData_model1wrtSym_plot3.dat};
\end{axis}
\end{tikzpicture}}%
\subfigure[$u_{\textrm{Im}}^{\textrm{max}}=2,\tau_{\textrm{Im}}^{\textrm{max}}=5$] {\label{fig:Mfourth}  
\begin{tikzpicture}
\begin{axis}[mz plot style main]
\addplot[mycolor,fill=mycolor] table[y index = 1] {Mz_histogramData_model1wrtSym_plot4.dat};
\end{axis}
\end{tikzpicture}}%
\end{center}
\caption{Empirical eigenvalue densities  of $\mathcal{M}$ in units of $|W|$.}
\label{Mz spectra, model 1, wrt sym}
\label{fig:Mhist}
\end{figure}
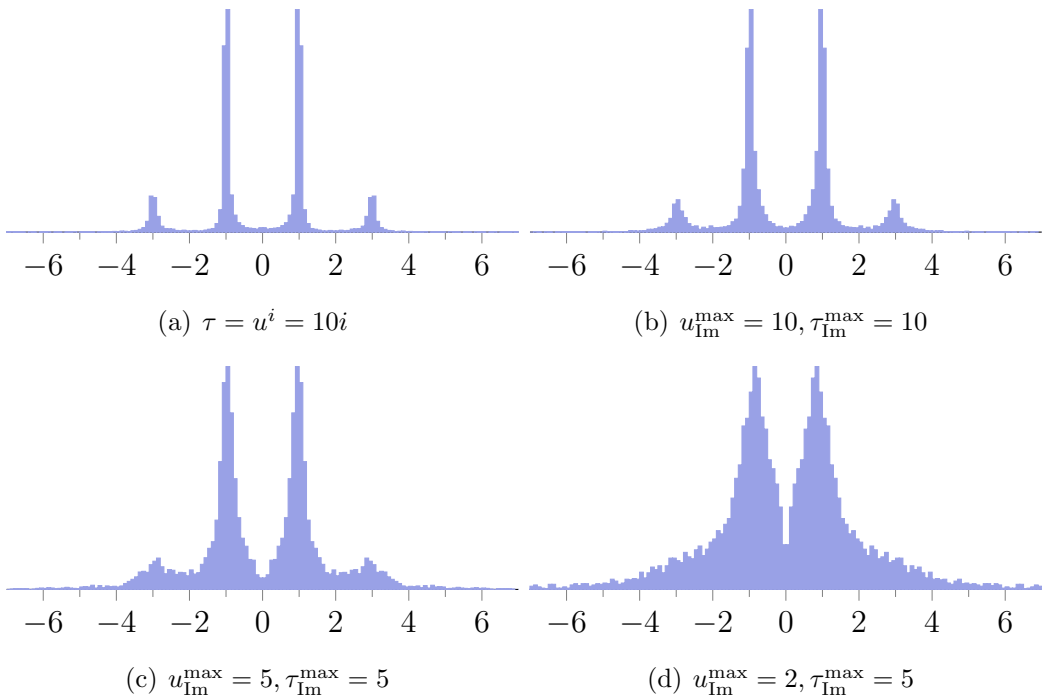%


We begin by computing the spectrum of \m~at a fixed point in the moduli space,   $\tau = u^i= 10i$ for $i=1, \ldots, 4$, while scanning over  flux integers randomly chosen in the range $[-5,5]$ with a uniform distribution, subject to the  tadpole condition \eqref{eq:tadpole}. For reasons that will become clear in section \ref{sec:analytic}, we exclude the case when both the RR and NS-NS  flux simultaneously vanish on the  cycle 
whose  period has  cubic terms in the complex structure moduli 
(cf. the cycle $A^0$ appearing in \eqref{eq:Period0}). The resulting eigenvalue histogram is shown in Figure \ref{fig:Mfirst}, when plotted in units of $|W|$ for each flux choice. The spectrum shows prominent  peaks at $\pm|W|$, and smaller peaks at $\pm 3|W|$, in stark contrast with the smooth `semi-circle-like' Altland-Zirnbauer CI (AZ-CI) spectrum of Figure \ref{fig:AZ-CI}.

To investigate the dependence of the spectrum on the values of the moduli fields, we have performed various joint scans of subspaces of the moduli space and of the flux numbers.   In  Figures \ref{fig:Msecond}--\ref{fig:Mfourth} 
we plot the resulting eigenvalue densities for  three cases in which complex structure moduli vevs are (for simplicity) sampled uniformly  in the range, 
\be
{\rm Re}(u^i) \in [-10,10] \, ,~~~{\rm Im}(u^i) \in [1,u_{\textrm{Im}}^{\textrm{max}}] \, ,
\label{eq:SampleDist}
\ee
with $u_{\textrm{Im}}^{\textrm{max}}\in \{10,5,2\}$. The axio-dilaton is simultaneously  
 sampled uniformly in the fundamental domain with $2<$Im($\tau$)$<\tau_{\textrm{Im}}^{\textrm{max}}$, where $\tau_{\textrm{Im}}^{\textrm{max}} \in \{10,5,5\}$. 
 
 The  shape of this spectrum and the implications that follow from it are the main themes of this paper. Here, we will merely make the following simple observations: \emph{i}) none of the densities resemble the AZ-CI spectrum, \emph{ii}) the peak at $\pm|W|$ is visible in all cases, while the peak at $\pm 3|W|$ is clearly visible in all but the last case. The peaks are less blurred for larger typical values of the  moduli. \emph{iii}) The peaks are only visible in the spectrum when plotted in units of $|W|$ (which varies from realisation to realisation) and for  canonically normalised fields. This could possibly explain why this effect has not been previously observed in the literature.


\subsection{The spectrum of \h}
\begin{figure}[t!]
\begin{center}
\subfigure[$\tau = u^i=10i$ ]{\label{fig:Hfirst}     
\begin{tikzpicture}
\begin{axis}[htot plot style main]
\addplot[mycolor,fill=mycolor] table[y index = 1] {Htot_histogramData_model1wrtSym_plot1.dat};
\end{axis}
\end{tikzpicture}}%
\subfigure[$u_{\textrm{Im}}^{\textrm{max}}=10,\tau_{\textrm{Im}}^{\textrm{max}}=10$] {\label{fig:Hsecond}
\begin{tikzpicture}
\begin{axis}[htot plot style main]
\addplot[mycolor,fill=mycolor] table[y index = 1] {Htot_histogramData_model1wrtSym_plot2.dat};
\end{axis}
\end{tikzpicture}}%
\\%
\subfigure[$u_{\textrm{Im}}^{\textrm{max}}=5,\tau_{\textrm{Im}}^{\textrm{max}}=5$] {\label{fig:Hthird}
\begin{tikzpicture}
\begin{axis}[htot plot style main]
\addplot[mycolor,fill=mycolor] table[y index = 1] {Htot_histogramData_model1wrtSym_plot3.dat};
\end{axis}
\end{tikzpicture}}%
\subfigure[$u_{\textrm{Im}}^{\textrm{max}}=2,\tau_{\textrm{Im}}^{\textrm{max}}=5$] {\label{fig:Hfourth}  
\begin{tikzpicture}
\begin{axis}[htot plot style main]
\addplot[mycolor,fill=mycolor] table[y index = 1] {Htot_histogramData_model1wrtSym_plot4.dat};
\end{axis}
\end{tikzpicture}}%
\end{center}
    \caption{Empirical eigenvalue densities of $\h$ in units of $m_{3/2}^2$.}
   \label{Htot spectra, model 1}
   \label{HtotModel1}
   \label{fig:Hhist}
\end{figure}
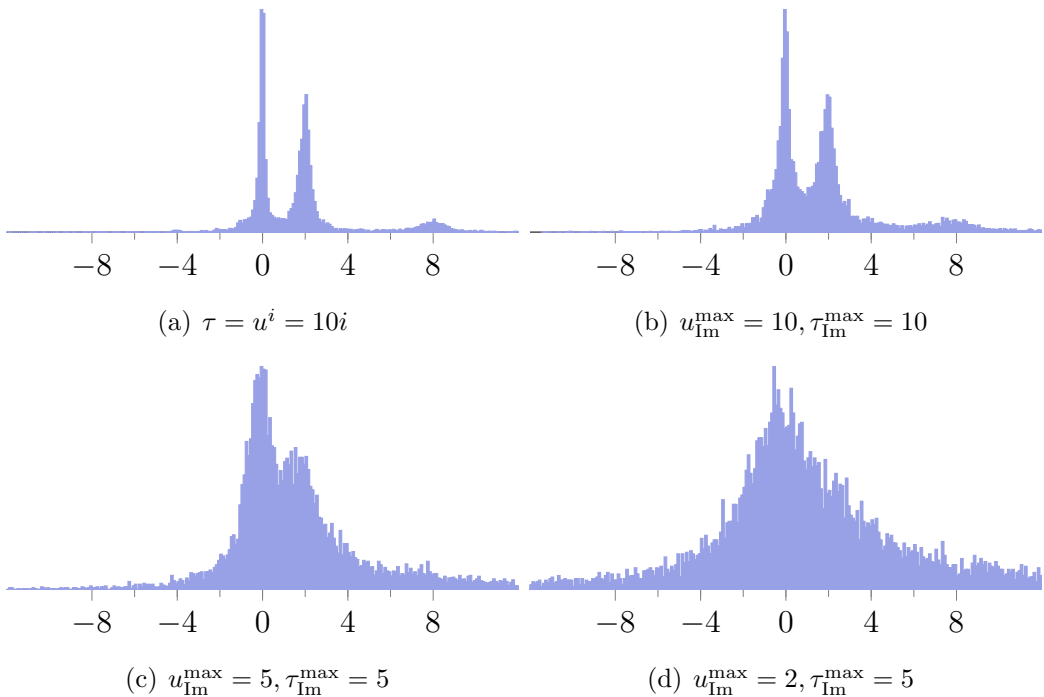

The eigenvalues of the Hessian matrix for canonically normalised fields give the squared physical masses of the scalar fields in the theory. We now compute the spectrum of \h~in Model 1 in the same cases considered above for the spectrum of \m.  Figure \ref{fig:Hfirst} shows the eigenvalue density of \h~when scanning over fluxes at  $\tau = u^i= 10i$. In stark contrast to the random matrix theory spectrum of the `WWW' model of Figure \ref{fig:WWW}, the spectrum exhibits sharp peaks at $0$ and $2m_{3/2}^2$, and a smaller bump at $8 m_{3/2}^2$. 

We furthermore consider joint scans of fluxes and moduli vevs within the subspaces defined as per the discussion around equation  \eqref{eq:SampleDist}.  The resulting spectra are shown in figures \ref{fig:Hsecond}--\ref{fig:Hfourth}. The peaks in the spectra remain distinctive and prominent in the first of these cases, while they blur as the sampling is restricted to regions with smaller moduli values.  In the final case plotted in Figure \ref{fig:Hfourth}, the peaks have been blurred into a broad feature that peaks around zero. 



We make the following observations: \emph{i}) In none of the cases is the spectrum well-described by the `WWW' random matrix model of Figure \ref{fig:WWW}. \emph{ii}) The peaks in the spectra are somewhat less sharp than those observed in the spectra of \m~for the same regions, but are similarly sharpened for large moduli values.  \emph{iii}) Again we note that the peaks would not appear very prominently had we not canonically normalised the fields and expressed the histogram in units of $m_{3/2}^2$.

In sum, we have in this section shown that  in a particular, explicit flux compactification, the spectra of \m~and \h~show curious peaks that become particularly prominent at large complex structure, but that appear to influence the spectrum over the entire large complex structure expansion. For both \m~and \h, we found the spectra to differ significantly from those of the  corresponding random matrix theory  models. We will now argue that this peaked structure is to be expected as a general feature of flux compactifications close to the large complex structure point, and that this structure has significant implications for moduli stabilisation and the statistics of flux compactifications in this region of moduli space.



\section{Analytic derivation of the spectra of \m~and \h}
\label{sec:analytic}
We are now interested in explaining the observed structure in the spectra analytically for  generic flux compactifications at large complex structure. Unfortunately, in general the superpotential \eqref{eq:GVW} is quite complicated and we know of no way of directly diagonalising \m~and \h. 
We will however find in sections \ref{sec:M} and \ref{sec:H} that it is possible to compute the spectra of these matrices algebraically for  general flux compactifications -- but only in a particular limit of the complex structure moduli space. We now define this subspace and briefly discuss its properties.

\subsection{The subspace \Sp}
\label{sec:Sp}

We will consider compactifications with non-zero R-R or NS-NS flux along at least one of the three-cycles of the compactification manifold.  Given such a compactification, we may choose the homogeneous coordinate $z^0$ to be the period of $\Omega$ corresponding  to a cycle,  $A^0$, with non-vanishing flux, and we may express the inhomogeneous coordinates as discussed in section \ref{sec:bkg} as $z^i = u^i/u^0$. For future convenience, we will denote the flux along this cycle by $N \neq 0$, without the vector arrow. 
In the large complex structure expansion, the superpotential \eqref{eq:GVW} is then a polynomial of degree three in the complex structure moduli, and  the cubic terms are all proportional to the single complex flux $N$.

 For sufficiently large complex structure moduli vevs, the superpotential is well-approximated as a homogeneous function of degree three in the complex structure moduli,
\be
W = \frac{N}{6} \ks_{ijk} u^i u^j u^k %
+ {\cal O}((u^i)^2)
\approx 
\frac{N}{6} \ks_{ijk} u^i u^j u^k 
\, .
\label{eq:cubic1}
\ee
We here assume that $W\neq 0$ so that, in particular, the `Yukawa couplings' $\ks_{ijk}$ do not all vanish. We note that the exact region for which \eqref{eq:cubic1} will be a good approximation will depend on the fluxes on all cycles of the compactification manifold. 
The complex structure  K\"ahler potential simplifies similarly at large values of the complex structure moduli to,
\be
K_{\rm c.s.} 
 \approx - \ln \left(
 \frac{i}{6} \ks_{ijk}(u - \bar u)^i(u - \bar u)^j(u - \bar u)^k 
 \right)
  \, .
 \label{eq:Kcd}
\ee

We will find it hard to make analytical progress for arbitrary phases of the complex structure moduli. However, the computation simplifies significantly in the subspace 
in which the complex structure moduli all have the same phase, i.e.
\be
u^i = u s^i  
\label{eq:2p} 
\, ,
\ee
for the $h^{1,2}_-$ real parameters $s^i$. We do not impose any restrictions on the value of the axio-dilaton, so the real dimensionality of the  subspace in satisfying \eqref{eq:2p}  is $h^{1,2}_-+3$.
We will denote the subspace of the form \eqref{eq:2p} in which \eqref{eq:cubic1} and \eqref{eq:Kcd} provide  good approximations by \Sp. For clarity, we will label equations that only apply in \Sp~by a $\mabb$ in the margin.


We now introduce a short-hand notation that we find useful in the explicit computations that will follow.  
For  some tensor $A_{ijklmn}$
we
denote tensor contractions with $u^i$ by subscript $u$, and contractions with $u^i - \bu^{\bi}$ by \w~so that for example,
\bea
A_{ijkl m n}\, u^i (u- \bu)^j u^k (u - \bu)^l = A_{u \w u \w m n} \, .
\eea
 In this notation the superpotential and K\"ahler potential at large complex structure are given by,
\be
W = \frac{N}{6} \ks_{uuu} \, ,~~~~ 
K = - \ln\left(\frac{i}{6} \ks_{\w\w\w}\right) - \ln(-i(\tau - \bar \tau)) \label{eq:cd} %
\, .
\ee

Using \eqref{eq:Kcd}, the components of the K\"ahler metric, $K_{a \bar b} = \partial^2_{a \bar b} K$, and its inverse $K^{a \bar b}$ are given by,
\bea
K_{\tau \bar \tau} &=& - \frac{1}{(\tau - \bar \tau)^2} \, ,~~~
K_{\tau \bi}= K_{\bar \tau i} = 0 \, ,\nonumber  \\
K_{i \bj} &=& 6 \frac{\ks_{i \bj \w}}{\ks_{\w\w\w}} - 9 \frac{\ks_{i \w \w} \ks_{\bj \w \w}}{\ks_{\w\w\w}^2}\nonumber\, , \\
K^{\tau \bar \tau} &=& - (\tau - \bar \tau)^2 \, , ~~~
K^{\tau \bi}= K^{\bar \tau i} = 0 \, , \nonumber \\
K^{i \bar  j} &=& \frac{1}{6}\ks_{\w\w\w} \ks^{-1}_{i \bj \w} - \frac{1}{2} (u - \bar u)^i (u -\bar u)^{\bj} 
\, .
\label{eq:metric}
\eea
We have here assumed that  $\ks_{i \bj \w}$ is invertible with the inverse $ \ks^{-1}_{i \bj \w}$ (in our numerical studies we simply ensure that the K\"ahler metric is positive definite).

In this notation, the F-terms of the complex structure moduli are given by,
\be
F_i =D_i W = \frac{N}{2} \left( \ks_{iuu} - \ks_{i \w\w} \frac{\ks_{uuu}}{\ks_{\w\w\w}}\right)
\, .
\label{eq:Fcdl}
\ee
%
In the subspace \Sp, the F-terms are given by,
\be
F_i = \frac{N}{2} \ks_{i ss} u^2 \left(1- \frac{u}{u - \bu} \right) = -3 W \frac{\ks_{iss}}{\ks_{sss}}  \left(\frac{\bu}{u} \right) \frac{1}{u - \bu}
 \marginnote{\mabb}\, ,
\label{eq:Fi}
\ee
where $\ks_{iss} = \ks_{ijk} s^j s^k$ and $\ks_{sss} = \ks_{ijk} s^j s^j s^k \neq 0$. Using the inverse metric of equation \eqref{eq:metric}, we find that 
\be
\bF^{ i} = - \bW s^{i} (u-\bu) \left(\frac{u}{\bu}\right) \marginnote{\mabb}\, ,
\label{eq:bFi}
\ee
so that $F_i \bF^i =3|W|^2$. Since furthermore
$F_{\tau} = K_{\tau} \left( \frac{u}{\bar u}\right)^3\bW$ so that
 $F_{\tau} \bF^{\tau} = |W|^2$,  the total contribution from the F-terms to the scalar potential is given by,
\be
F_a \bF^a = F_{\tau} \bF^{\tau} + F_i \bF^i = 4|W|^2 \marginnote{\mabb} \, .
\label{eq:Fsq}
\ee
Thus, the moduli F-terms source a positive vacuum energy, 
\be
V = e^K |W|^2= m_{3/2}^2 \marginnote{\mabb}  \, , 
\ee
where we (still) have ignored K\"ahler moduli. 






\subsection{The spectrum of ${\cal M}$}
\label{sec:M}
We are now interested in computing the spectrum of \m~for the canonically normalised complex structure and axio-dilaton fields. 
The symmetries of \m~dictate that its eigenvalues come in opposite sign pairs, and that the positive branch is given by the square root of the eigenvalues of $(\bZ Z)_{\bar a b} = \bZ_{\bar a \bar c} Z^{\bar c}_{~ b}$. The K\"ahler potential \eqref{eq:Kcd} leads to non-trivial kinetic terms for the scalar fields at a generic point, however, we can obtain the spectrum of $(\bZ Z)_{\bar a b}$ for canonically normalised fields by computing the spectrum of the non-Hermitian matrix $K^{a \bar a} (\bZ Z)_{\bar a b} = (\bZ Z)^{a}_{~ b}$.  
%
%
Thus, to find the spectrum of \m~for canonically normalised fields, we compute the square-root of the spectrum of $(\bZ Z)^{ a}_{~ b}$.

Before we go into the details of the computation of the spectrum of \m, we note that in flux compactifications,
 the tensor $Z_{ab}$ inherits much structure from the underlying Calabi-Yau geometry \cite{Denef:2004ze}. In particular, $Z_{ab}$ has only $h^{1,2}_-$   independent complex entries,  while  a generic complex symmetric tensor of the same dimensions would have $(h^{1,2}_- +2)(h^{1,2}_-+1)/2$ independent components.  The correlations that limit the number of independent degrees of freedom in $Z_{ab}$ in string theory can in part be traced to the simplicity of the axio-dilaton dependence of $K$ and $W$ from which it follows that, 
%
\be
Z_{\tau \tau} \equiv 0 \, .
\label{eq:Ztautau}
\ee
More significant however, are the correlations  in the complex structure sector that 
%
arise directly from the Hodge decomposition of covariant derivatives of the holomorphic three-form. 
While of course $\Omega$ is a (3,0)-form, $D_i \Omega$ is (2,1) and forms a symplectic basis of $H^{2,1}(M_3)$. One can furthermore show that  
$\D_i D_j \Omega$ is (1,2), and may then be expanded in terms of the basis vectors $\bar D_{\bi} \bar \Omega$  \cite{Candelas:1990pi}. The expansion coefficients are simply proportional to the `Yukawa couplings' as,
\be
\D_i D_j \Omega =  -i e^{K_{(\rm c.s.)}} \ks_{ij}^{~~\bar k} \bar D_{\bar k} \bar \Omega \, .
\label{eq:Omegarel}
\ee

We now note that the $ij$ components of $Z_{ab}$ can be simplified as,
\bea
Z_{ij} &=& \D_i D_j W =
\int G_3 \wedge \D_i D_j \Omega = 
 -i e^{K_{(\rm c.s.)}} \vec{N} \cdot \ks_{ij}^{~~\bar k} \bar D_{\bar k} \vec{\Pi}^*  \, .
 \eea
Furthermore, 
  the `mixed' components between the axio-dilaton and the complex structure moduli are given by,  
\be
Z_{\tau i} = \D_{\tau}(\vec{N} \cdot D_i \Pv) = K_{\tau} \N^* \cdot D_i \Pv \, .
\label{eq:Ztaui}
\ee
%
We thus have,
 \bea
Z_{ij}  &=&
 \frac{i }{K_{ \tau}} e^{K_{(\rm c.s.)}}  \ks_{ij}^{~~\bar k} \bZ_{\bar \tau \bar k} 
\, .
\label{eq:Zrel}
\eea 
and we may take $Z_{\tau i}$ to be the $h^{1,2}_-$ independent complex components of $Z_{ab}$ in flux compactifications. The relation \eqref{eq:Zrel} is quite useful in simplifying the computations in this section.

We are now ready to compute the spectrum of $(\bZ Z)^{ a}_{~ b}$.  Using equation \eqref{eq:cd}  we find that,  
\bea
Z_{\tau i} &=& \frac{K_{\tau } N^*}{2} \left( 
\ks_{i u u} - \ks_{i \w\w} \frac{\ks_{uuu}}{\ks_{\w\w\w}}
\right) \, , \nonumber \\
Z_{ij} &=&
-\frac{N}{2}
\Big[
  \ks_{ijm} \ks^{-1}_{m \bar k \w} \ks_{\bar k \bu \bu} -
\frac{ \ks_{ij \w}}{\ks_{\w\w\w}}  \Big(
 2  \ks_{\bu\bu\bu} 
  + 3   \ks_{\w \bu \bu} 
     \Big)
\Big]  \, .
\label{eq:Zij}
\eea 
We now specialise to the subspace \Sp~and raise  one index of $Z_{ab}$ to find  $Z^{\bar a}_{~b}$. The $\bar \tau \tau$ component is trivial, $Z^{\bar \tau}_{~\tau}\equiv 0$, and the other components are given by, 
\bea
 Z^{\bar m}_{~j} &=& W p\left(\delta^{\bar m}_{j} - 3  \frac{\ks_{jss}}{\ks_{sss}} s^{\bar m} \right) \, ,\nonumber \\ 
Z^{\bar \tau}_{j} &=&
-3 \frac{c }{p}\bW  \frac{\ks_{jss}}{\ks_{sss}} 
\, , \nonumber \\
Z^{\bar m}_{~\tau} 
&=&
- \bW \frac{1}{pc} s^{\bar m} \, , 
\marginnote{\mabb}
\label{eq:Zraised}
\eea
where we have introduced, 
\bea
c = \frac{\tau - \bar \tau}{u - \bar u} \, ,~~~~
p = \left( \frac{\bu}{u} \right)^2 \, ,~~~~
q= \frac{W}{\bW} \, .
\label{eq:constants}
\eea  
The matrix $(\bZ Z)^a_{~b}$ is then given by,
\bea
(\bZ Z)^a_{~b} &=&
\left(
\begin{matrix}
(\bZ Z)^{\tau}_{~\tau} & (\bZ Z)^{\tau}_{~j} \\
(\bZ Z)^{i}_{~\tau} & (\bZ Z)^i_j
\end{matrix}
\right)
=|W|^2\left(
\begin{matrix}
3 &  6 cp^2 q \frac{\ks_{jss}}{\ks_{sss}} \\
\frac{2 s^i}{cp^2 q} & \delta^i_j + 6 s^i \frac{\ks_{jss}}{\ks_{sss}}
\end{matrix}
\right)
\label{eq:matrix} \marginnote{\mabb} \, .
\eea
Using the expression for the F-term in \Sp, cf.~equations \eqref{eq:Fi} and \eqref{eq:bFi}, we find that this matrix can be written as,
\begin{empheq}[box=\fbox]{align}
\vspace*{.5 cm}
~~(\bZ Z)^a_{~b} = |W|^2 \delta^a_{b} + 2 \bF^a F_b  \marginnote{\mabb}~~  \, .
\label{eq:ZbZfinal}
\vspace*{.5 cm}
\end{empheq}
Deducing the spectrum is now trivial: 
any linearly independent set of $h^{1,2}_-$ vectors that are perpendicular to $F_b$ are eigenvectors of $(\bZ Z)^a_{~b}$ with the eigenvalue $|W|^2$. The final eigenvector is given by $\bF^b$, and since according to equation \eqref{eq:Fsq}, $F_a \bF^a = 4|W|^2$, the corresponding eigenvalue is equal to $9 |W|^2$. 

 
 The  spectrum of  \m~in \Sp~for canonically normalised fields is thus given by,
\begin{empheq}[box=\fbox]{align}
\vspace*{.5 cm}
{\rm Spectrum}(\m) = 
\left\{
\begin{array}{r l}
|W| &~~{\rm multiplicity}~h^{1,2}_- \, ,\\
3|W| &~~{\rm multiplicity}~1 \, , \\
-|W| &~~{\rm multiplicity}~h^{1,2}_- \, ,\\
-3|W| &~~{\rm multiplicity}~1 \marginnote{\mabb}\, . \\
\end{array}
\right.
\label{eq:hooray}
\vspace*{.5 cm}
\end{empheq}

Equation \eqref{eq:hooray} is one of our main results, and clearly exhibits `clustering' of the eigenvalues into delta-function peaks in the spectrum. The location of these peaks are exactly those where we observed the peaks in the spectrum in the explicit example of section \ref{sec:explicit}, cf.~Figure \ref{fig:Mfirst}. Thus, equation \eqref{eq:hooray} explains the presence of the peaks in the observed spectrum, and moreover proves that such peaks  are \emph{universal} in the large complex structure limit:  our analytical derivation applies to any non-vanishing choice of flux, any not all vanishing `Yukawa couplings' 
$\ks_{ijk}$, and any number of complex structure moduli, $h^{1,2}_-$.



\subsection{The spectrum of ${\cal H}$}
\label{sec:H}
We now show that the spectrum of the covariant Hessian, \h, similarly takes a very simple form in \Sp. The Hessian matrix is given by,
\bea
{\cal H} &=&
\left(
\begin{array}{c c}
\nabla^2_{a \bb} V & \nabla^2_{a b} V \\
\nabla^2_{\ba \bb} V & \nabla^2_{\ba b} V
\end{array}
\right)
\\
&=&
e^K
\left(
\begin{array}{c c}
 Z_{a}^{~\bar c}\ \bZ_{\bb \bc} -  F_a \bar{F}_{\bb} - R_{a \bb c \bar d} \bar{F}^c F^{\bar d}  & U_{a b c} \bar{F}^c - Z_{a b} \bW \\
\bU_{\ba \bb \bar c} F^{\bar c} - \bZ_{\ba \bb} W & \bZ_{\ba}^{~c}\ Z_{b c} -  F_b \bar{F}_{\ba} - R_{b \ba c \bar d} \bar{F}^c F^{\bar d}
\end{array}
\right) + \nonumber \\
&+& e^K
\left(
\begin{array}{c c}
K_{a \bb}  \Big(F^2 - 2 |W|^2 \Big) & 0 \\
0 & K_{\bar a b}  \Big(F^2 - 2 |W|^2 \Big) 
\end{array}
\right)
\,  ,
\eea
where $F^2= F_a \bF^a$,  $U_{a b c} = \D_a \D_b D_c W$ is complex and symmetric, and $R_{b \bar a c \bar d} = K_{b \bar f} \partial_c \Gamma^{\bar f}_{\bar a \bar d}$ denotes the non-trivial components of the Riemann curvature tensor on the field space. We will again find the spectrum of \h~for canonically normalised fields by computing the spectrum of \h~contracted with the inverse of the K\"ahler metric, schematically `$K^{-1} \h$'.

\subsubsection{The diagonal blocks of \h}
\label{sec:Hdiag}
Raising an index of the diagonal block matrices of \h~gives,
\be
e^{-K} K^{a \ba} \nabla^2_{\ba b} V= (\bZ Z)^a_{~b} -  F_b \bar{F}^{a} - K^{a \ba} R_{b \ba c \bar d} \bar{F}^c F^{\bar d} + \delta^a_b\Big(F^2 - 2 |W|^2 \Big) \, .
\ee 
As we have already computed $(\bZ Z)^a_{~b}$ in equation \eqref{eq:ZbZfinal}, the only non-trivial term remaining is the curvature term, $K^{a \ba} R_{b \ba c \bar d} \bar{F}^c F^{\bar d}$. We now show that also this contribution takes a very simple form in \Sp. 

As there are no cross-terms between the axio-dilaton and the complex structure moduli in the K\"ahler potential, the curvature tensor has vanishing mixed components between these sectors.  The axio-dilaton contribution is given by,
\bea
K^{\tau \bar \tau} R_{\tau \bar \tau \tau \bar \tau } \bF^{\tau} F^{\bar \tau} &=& (\tau- \bar \tau)^2
\left(
\frac{6}{(\tau- \bar \tau)^4} - \frac{4}{(\tau- \bar \tau)^4} \right)(\tau- \bar \tau)^2|W|^2
\\
&=& 2|W|^2   \, .
\eea
The complex structure components are not much harder to compute, and we find that,  
\bea
K^{\bar m i }K_{ik \bj}  K^{j \bj }K_{j \bar k \bar l} \bF^k F^{\bar k} &=& 4|W|^2 \delta^{\bar m}_{\bar l} \nonumber \, , \\
K^{j \bj} K_{i \bj k \bar l} \bF^k F^{\bar l} &=& 6 |W|^2 \delta^j_i  \marginnote{\mabb}\, ,
\eea
so that,
\be
K^{j \bj} R_{i \bj k \bar l} \bF^k F^{\bar l} 
=
K^{j \bj} \left( K_{i \bj k \bar l} - K_{i  k \bar m} K^{\bar m m} K_{\bj \bar l m} \right) \bF^k F^{\bar l} 
= 2|W|^2 \delta^j_i \marginnote{\mabb} \, .
\ee
The curvature  contribution to the Hessian matrix is evidently proportional to the unit matrix, 
\be
K^{b \bar b} R_{a \bb c \bar d} \bF^c F^{\bar d} = 2|W| \delta^a_{b} \marginnote{\mabb} \, .
\ee

The diagonal blocks of  $K^{-1} {\cal H}$ are then simply given by,
\bea
e^{-K} K^{a \ba} \nabla^2_{\ba b} V &=&
 |W|^2 \delta_b^{a} +  \bF^a F_b \, ,  \nonumber \\
e^{-K} K^{\ba a} \nabla^2_{a \bb} V &=& |W|^2 \delta^{\ba}_{\bb} +  F^{\ba} \bF_{\bb}  \marginnote{\mabb} \, . \label{eq:diag}
\eea

\subsubsection{The off-diagonal blocks of \h }
\label{sec:Hoff}
The off-diagonal blocks of $K^{-1} \h$ are given by
 $K^{a \bb} U_{abc} \bF^c - \bW Z^{\bb}_{b}$, and its complex conjugate. Equation \eqref{eq:Zraised} gives the components of  $Z^{\bb}_{b}$, and we here compute the components of $K^{a \bb} U_{abc} \bF^c$. The reader interested in the resulting simple expression --  but not the intermediate technical details of the computation -- may skip ahead to equation \eqref{eq:Uresult}.

The $U$-tensor contribution to `$K^{-1} \h$' is given by,
\bea
K^{a \bb} U_{abc} \bF^c  &=&K^{a \bb} \left(\partial_a Z_{bc} + K_a Z_{bc} - \Gamma_{ab}^d Z_{dc} - \Gamma^d_{ac} Z_{db}  \right) \bF^c \, .
\eea 
For the free indices $(\bar \tau, \tau)$, this expression vanishes identically since $\D_{\tau} Z_{\tau \tau} = \D_i Z_{\tau \tau} \equiv 0$ so that,
\bea
K^{\tau \bar \tau} U_{\tau \tau c} \bF^c  &=&
K^{\tau \bar \tau} U_{\tau \tau \tau} \bF^{\tau}+ K^{\tau \bar \tau} U_{\tau \tau i} \bF^i
=  0 \, .
\eea 
For free indices $(\bar \tau, i)$ we have,
\bea
K^{\tau \bar \tau} U_{\tau i c} \bF^c  &=&
K^{\tau \bar \tau} U_{\tau i \tau} \bF^{\tau}+ K^{\tau \bar \tau} U_{\tau j i} \bF^j = K^{\tau \bar \tau} U_{\tau j i} \bF^j \, ,
\eea 
while 
for $(\bj, \tau)$ we have,
\bea
K^{j \bj} U_{j \tau c} \bF^c  &=&
K^{j \bj} U_{j \tau \tau} \bF^{\tau}+ K^{j \bj} U_{j k \tau} \bF^k  = K^{j \bj} U_{j k \tau} \bF^k  \, .
\eea 
Clearly, the $(\bj, \tau)$ and $(\bar \tau, j)$ components are related by,  
$
 K^{j \bj} U_{j k \tau} \bF^k = K^{i \bj} K_{\tau \bar \tau} \left[
 K^{\tau \bar \tau} U_{\tau i l} \bF^l \right]
$. 
Finally, the $(\bj, i)$ components are given by,
\bea
K^{j \bj} U_{j i c} \bF^c  &=&
K^{j \bj} U_{j i \tau} \bF^{\tau}+ K^{j \bj} U_{j k i} \bF^k \, .
\eea 
Thus we need to compute $U_{\tau ij}$ and $U_{ijk}$ and contract these with the relevant F-terms and the inverse metric. 

Using equations \eqref{eq:Ztaui} and  \eqref{eq:Omegarel}, we  have that,
\bea
 U_{\tau ij} &=& \D_j Z_{i \tau} = K_{\tau} \N^* \cdot \D_j \D_i \vec{\Pi} \nonumber \\
 &=& - iK_{\tau}  e^{K_{c.s}} \ks_{ijk} \N^* \cdot \bar \D^k \vec{\Pi}^*= - i K_{\tau} e^{K_{c.s.}} \bF^k \ks_{ijk} \, .
\eea
In the subspace \Sp, this expression evaluates to,
\be
U_{\tau ij}=- 6 \left(\frac{u}{\bu} \right)\frac{1}{(u - \bu)^2 (\tau - \bar \tau)} \bW \frac{\ks_{ijs}}{\ks_{sss}} \marginnote{\mabb} \, .
\label{eq:Utauij}
\ee
Using \eqref{eq:Zrel} to simplify  $U_{kij}$ we have,
\bea
U_{kij} &=& \D_k(iK_{\tau}^{-1}e^{K_{c.s.}}\ks_{ijl}\bZ^l_{\bar{\tau}}) \nonumber \\ 
&=& iK_{\tau}^{-1}e^{K_{c.s.}}(2K_k\ks_{ijl}\bZ^l_{\bar{\tau}}+\ks_{ijl}\partial_k\bZ^l_{\bar{\tau}}-\Gamma^m_{ki}\ks_{mjl}\bZ^l_{\bar{\tau}}-\Gamma^m_{kj}\ks_{iml}\bZ^l_{\bar{\tau}}) \, .
\eea
We raise an index of equation \eqref{eq:Zij} to find, 
\bea
\bZ^k_{~\bar \tau} &=&  \frac{K_{\bar \tau } N}{2} \Big[
\frac{\ks_{\w\w\w}}{6} \ks^{-1}_{i k \w} \ks_{i \bu \bu} 
-(u - \bar u)^{k}
\left(
   \frac{1}{3}   \ks_{\bu\bu\bu}
 + \frac{1}{2}  \ks_{\w \bu \bu} 
  \right)
\Big] \, .
\eea
To find $\partial_k \bZ^l_{~\bar \tau} $ we need to differentiate $\ks^{-1}_{ij\w}$, which is perhaps most simply done by noting that $\partial_k(\ks^{-1}_{l m \w} \ks_{mn \w}) =\partial_k(\delta^l_{n}) = 0$, so that $\partial_k(\ks^{-1}_{n l \w}) = - \ks^{-1}_{lm\w} \ks_{mpk} \ks^{-1}_{pn\w}$. We then find that in \Sp,
\be
\partial_k \bZ^l_{~\bar \tau} = \delta^l_k  K_{ \tau } W \left( \frac{\bu}{u} \right)^2 \left(2  - \left( \frac{\bu}{u} \right)\right) \marginnote{\mabb} \, .
\ee

The relevant Christoffel symbols are given by, 
\bea
 \Gamma^l_{kj} &=& K^{l \bar k} K_{k j \bar k} = \ks^{-1}_{l \bar k \w} \ks_{\bar k kj} - \frac{3}{\ks_{\w\w\w}} \left(
 \delta^l_k \ks_{j \w\w}+ \delta^l_{j} \ks_{k\w\w} - (u - \bar u)^l \ks_{ij \w}
 \right)   \nonumber \\
 &=&
 \frac{1}{u - \bu} \left( 
 \ks_{kj \bar m} \ks_{\bar m l s}^{-1} + 3 s^l \frac{\ks_{kjs}}{\ks_{sss}} - 3 \delta^l_{k} \frac{\ks_{j ss}}{\ks_{sss}}
 - 3 \delta^l_{j} \frac{\ks_{kss}}{\ks_{sss}}
 \right) \marginnote{\mabb}
 \, .
\eea
It is then straightforward to show that,
\bea
\ks_{mjl}\bZ^l_{\bar{\tau}} &=& 
-W \left( \frac{\bu}{u} \right)^2 \frac{u - \bu}{\tau - \bar \tau} \ks_{mjs} \marginnote{\mabb} \, , \\ 
\Gamma^m_{ki}\ks_{mjl}\bZ^l_{\bar{\tau}} &=& 
-W\left( \frac{\bu}{u} \right)^2 \frac{1}{\tau - \bar \tau}\left(\ks_{kij}+3\frac{\ks_{jss}\ks_{kis}}{\ks_{sss}}-3\frac{\ks_{jks}\ks_{iss}}{\ks_{sss}}-3\frac{\ks_{ijs}\ks_{kss}}{\ks_{sss}}\right) \marginnote{\mabb} \, .
\eea
All but one term in $U_{jik}$ then cancels, and we have,
\be
U_{kij}=-6\left(\frac{\bu}{u}\right)^3\frac{1}{(u-\bu)^3}W\frac{\ks_{ijk}}{\ks_{sss}} \marginnote{\mabb} \, .
\label{eq:Uijk}
\ee

We are now ready to contract the components of the $U$-tensor found in equations \eqref{eq:Utauij} and \eqref{eq:Uijk} with the F-terms. The relevant expressions are given by,
\bea
U_{\tau ij}\bF^j &=& 6\left(\frac{u}{\bu}\right)^2\frac{1}{(\tau-\bar \tau)(u-\bu)}\bW^2\frac{\ks_{iss}}{\ks_{sss}} \, , \nonumber \\
U_{ijk}\bF^k &=& 6\left(\frac{\bu}{u}\right)^2\frac{1}{(u-\bu)^2}|W|^2\frac{\ks_{ijs}}{\ks_{sss}} \, , \nonumber \\
U_{ijc}\bF^c &=& =12\left(\frac{\bu}{u}\right)^2\frac{1}{(u-\bu)^2}|W|^2\frac{\ks_{ijs}}{\ks_{sss}} \marginnote{\mabb} \, .
\eea
Upon contraction with the inverse K\"ahler metric we find the components, 
\bea
(K^{-1} U \bar F)^{\bar \tau}_{\tau} &=& 0  \, , \nonumber \\
(K^{-1} U \bar F)^{\bar \tau}_{i} &=&-6 \frac{c}{pq} |W|^2 \frac{\ks_{iss}}{\ks_{sss}}  \, , \nonumber \\
(K^{-1} U \bar F)^{\bi}_{\tau} &=& -2 \frac{1}{cpq} |W|^2 s^{\bi} \, , \nonumber \\
(K^{-1} U \bar F)^{\bj}_{i} &=& 2p|W|^2\left( \delta^{\bj}_i - 3 s^{\bj} \frac{\ks_{iss}}{\ks_{sss}} \right) \marginnote{\mabb} \, .
\label{eq:Ucontrib}
\eea

By
comparing 
equation \eqref{eq:Ucontrib}  with equation \eqref{eq:Zraised}, we arrive at our final result, 
\be
(K^{-1} U \bar F)^{\bar a}_{b} = 2 \bW Z^{\bar a}_{b} \marginnote{\mabb} \, .
\label{eq:Uresult}
\ee
Thus, the off-diagonal blocks of $K^{-1} \h$ are simply given by,
\bea
e^{-K} K^{\bar a a} \nabla^2_{a b} V &=& (K^{-1} U \bar F)^{\bar a}_{b} - \bW Z^{\bar a}_{b} =  \bW Z^{\bar a}_{b} \nonumber \, , \\
e^{-K} K^{ a \bar a} \nabla^2_{\bar a \bb} V &=& (K^{-1} U \bar F)^{a}_{\bb} - W \bZ^{ a}_{\bb} =  W \bZ^{a}_{\bb}
 \label{eq:offdiag} \marginnote{\mabb}\, .
\eea

\subsubsection{The spectrum of \h}
Putting our results from section \ref{sec:Hdiag} and section \ref{sec:Hoff} together, we  have found that, 
\begin{empheq}[box=\fbox]{align}
~~K^{-1} \h &=
e^K
\left(
|W|^2 \mathbb{1} +
\left(
\begin{array}{c c}
0 & \bW Z^{\bar a}_{b} \\
W \bZ^{ a}_{\bar b} & 0  
\end{array}
\right)
+ \left(
\begin{array}{c c}
F^{\bar a} \bF_{\bar b} & 0 \\
0& \bF^a F_{b}  
\end{array}
\right)
\right)
\nonumber \\
&=
m_{3/2}^2 \mathbb{1} + m_{3/2}  e^{K/2} (K^{-1} {\cal M}) + e^K \left(
\begin{array}{c c}
F^{\bar a} \bF_{\bar b} & 0 \\
0& \bF^a F_{b}  
\end{array}
\right) \, .~~~~\marginnote{\mabb}
\label{eq:KinvH}
\end{empheq}
Just as for the spectrum of \m, the spectrum of \h~for canonically normalised fields can now be read off by inspection.
We first note that,
\be
Z^{\bar a}_{b} \bF^b = 3\bW F^{\bar a} \marginnote{\mabb}\,  \, ,
\label{eq:notCPeqn}
\ee
so that
the vectors $\hat F_{\pm}$ of equation \eqref{eq:v1}  are eigenvectors of $K^{-1} \m$  with the corresponding eigenvalues being equal to 
$\pm 3|W|$. It then immediately follows that $\hat F_{\pm}$ are also 
eigenvectors of $K^{-1} \h$ with the eigenvalues,
\be
m^2_{0\pm} = 
\left\{
\begin{array}{l c l}
8 m_{3/2}^2  \\
2 m_{3/2}^2 &&\marginnote{\mabb} \, .
\end{array}
\right.
\ee
The remaining eigenvectors of $K^{-1}\h$ are then formed from the eigenvectors of $K^{-1} \m$ with eigenvalues $\pm |W|$. These are all perpendicular to $\hat F_{\pm}$ and the corresponding canonically normalised squared masses are given by,
\be
m^2_{i\pm} 
=
\left\{
\begin{array}{l c l}
2 m_{3/2}^2 &~~~&  \\
0 && \marginnote{\mabb} \, ,
\end{array}
\right.
\ee
for $i=1, \ldots, h^{1,2}_-$.


In sum, the spectrum of \h~in \Sp~for canonically normalised fields is given by,
\begin{empheq}[box=\fbox]{align}
\vspace*{.5 cm}
{\rm Spectrum}(\h) = \left\{
\begin{array}{l c l}
0&~~~ & {\rm multiplicity}~h^{1,2}_-\, , \\
2m_{3/2}^2 &~~~ & {\rm multiplicity}~h^{1,2}_-+1 \, ,\\
8m_{3/2}^2 &~~~ & {\rm multiplicity}~1 ~\marginnote{\mabb}
\, 
\end{array}
\right.
\label{eq:Hspect} 
\end{empheq}
Equation \eqref{eq:Hspect} together with equation \eqref{eq:hooray} are the main analytical results in this paper.  Just as in the spectrum of \m, the spectrum of the Hessian matrix in \Sp~exhibits delta-function peaks, and we note that the location of these peaks are exactly as expected from the numerical analysis of  Model 1, cf.~Figure \ref{HtotModel1}. While the  peak at zero appears to be stronger in the numerical spectrum of Model 1, we note that this can be largely attributed  to the different widths of the peaks around zero and $2m_{3/2}^2$.  We thus conclude that equation \eqref{eq:Hspect} provides the analytical explanation of the observed peaked spectra of the Hessian matrix at large complex structure. 





\section{Discussion}
The main results presented in this paper have several important -- and perhaps unexpected -- implications for 
scenarios of moduli stabilisation as well as for the statistics of flux compactifications.  In this section, we discuss what we regard as the most important of these implications. 



\subsection{No critical points and large slow-roll parameters in \Sp}
\label{sec:nogo}
We begin by making a few simple observations regarding the subspace \Sp. In section \ref{sec:Sp} we found that $F_a \bF^a = 4|W|^2$ in \Sp. Since we have assumed that $|W|\neq 0$, there  are no supersymmetric vacua in this subspace. Furthermore, the spectrum of \m~in \Sp, cf.~equation \eqref{eq:hooray}, has no support at $2|W|$ so that the critical point equation \eqref{eq:cp} has no solutions. 
%
%
Thus,  there are in \Sp~no critical points  of the truncated axio-dilaton and complex structure moduli system.

Furthermore, the inflationary slow-roll parameters take universal, large values in \Sp. Since,
\be
\partial_a V =  e^K\bW F_a \, , ~\marginnote{\mabb} 
\ee
we find that, 
\be
\epsilon = \frac{1}{2} \left(
\frac{2 \partial_a V K^{a \bar b} \partial_{\bar b} V }{V^2}
\right) = 4 \, ,  ~\marginnote{\mabb} 
\label{eq:epsilon}
\ee
which is independent of the flux choice and details of the compactification manifold. We define the slow-roll $\eta_{\parallel}$ parameter to be given by,
\be
\eta_{\parallel} = \frac{e^A \nabla^2_{AB} V e^B }{ V} \, ,
\ee
where 
$e_A = - \partial_A V/||\partial V||$ for $||\partial V||= \sqrt{K^{A B} \partial_A V \partial_B V}$, and
$A$ runs over both holomorphic and anti-holomorphic indices.  Using our expressions from section \ref{sec:H}, we find that,
\bea
e^a \nabla^2_{a \bb} V e^{\bar b} &=&\frac{5}{2} m_{3/2}^2 \, , \nonumber \\
e^a \nabla^2_{a b} V e^{ b} &=& \frac{3}{2} m_{3/2}^2 \, .   ~\marginnote{\mabb} 
\eea 
It then follows that,
\be
\eta_{\parallel} = 8 \, ,   ~\marginnote{\mabb} 
\label{eq:eta}
\ee
in \Sp. 
We have thus shown that there are no vacua in \Sp~and that the slow-roll parameters are much too large to support slow-roll inflation.  

It is important to note that the 
 inclusion of additional moduli fields may  
alter these conclusions. We illustrate this with the example of the class of 
approximately no-scale supergravities considered in \cite{decoupling} where it was shown that de Sitter vacua can be constructed in string theory motivated ${\cal N}=1$ supergravities through only a very limited amount of tuning. In this scenario,  supersymmetry is predominantly broken by a `no-scale' field (cf.~K\"ahler modulus), with only a small amount of supersymmetry breaking arising from other fields (cf.~the axio-dilaton and the complex structure moduli). The critical point equation for these `other' fields
can be written as an eigenvalue equation of the truncated matrix \m, or perhaps simpler in terms of $Z\bZ$ as,
\be
(Z \bZ)_a^{~b}F_b = |W|^2 F_a \, ,
\ee
for the canonically normalised fields in this sector. In other words, while the full critical point equation requires \m~to have an eigenvalue equal to $2|W|$, when truncated to the fields perpendicular to the no-scale field, \m~should have an eigenvalue equal to $|W|$ to solve the critical point equation. This clearly illustrates that a study of the axio-dilaton and complex structure sector alone does not suffice to make general statements of the existence of non-supersymmetric critical points.\footnote{In \Sp,
 $(Z \bZ)_a^{~b}$ indeed has (several) eigenvalues equal to $|W|^2$, however, it is not hard to see that the 
 solutions of the type suggested in \cite{decoupling} cannot be constructed in this subspace:  the F-terms of the complex structure moduli and the axio-dilaton are not small with respect to $|W|$ in \Sp, and furthermore, the $F$-term is exactly the eigenvector of $Z\bZ$ with eigenvalue equal to $9|W|^2$, cf.~equation \eqref{eq:notCPeqn}. } Controlled moduli stabilisation is also necessary for the construction of viable models of inflation in string theory. For a discussion of some issues that arise in complex structure moduli inflation, see \cite{Hebecker:2014eua, Hebecker:2014kva, Hayashi:2014aua}.




\subsection{Highly peaked spectra at not-so-large complex structure}
We now return to our explicit example, Model 1, and ask the question: for a given choice of flux numbers and as a function of the moduli space, by how much does the spectrum deviate from the analytical spectrum that we found in section \ref{sec:analytic} for the subspace \Sp? How close does it come to the predictions of the random matrix theory models?

To quantify the discrepancy between eigenvalue spectra, we  introduce the \emph{`spectral deviation'}, $\mu$, as a measure of  the fractional difference between two given eigenvalue configurations. We define this measure as follows: take $\vec{\alpha} = (\alpha_1, \ldots, \alpha_N)$ and $\vec{\beta} = (\beta_1, \ldots, \beta_N)$ to be two sets of eigenvalues, listed in increasing order i.e.~$\alpha_1 \leq \alpha_2 \leq \ldots \leq \alpha_N$, and $\beta_1 \leq \beta_2 \leq \ldots \leq \beta_N$.  The spectral deviation of $\vec{\beta}$ from $\vec{\alpha}$ is then given by,
\be
\mu_{\vec{\alpha}}(\vec \beta) = \sqrt{\frac{(\vec{\alpha}- \vec \beta)^2 }{ \vec \alpha^2}} \, ,
\label{eq:SpectDev}
\ee 
where the product is taken to be the ordinary Cartesian inner product. 

We will consider two reference vectors $\vec{\alpha}$ when computing the spectrum of \m~as a function of moduli space. The first one is given by the mean positions of the eigenvalues in the  AZ-CI random matrix ensemble, which 
we find numerically to be 
given by,
\be
\vec{\alpha}_{\rm AZ-CI} = (0.35, 0.80, 1.27, 1.80, 2.45)|W| \, ,
\ee
for five fields, where we have 
included only the positive branch eigenvalues. The corresponding AZ-CI eigenvalue density (that we will not use but include here for clarity and reference)  is given by,
\be
\rho_{\rm CI}(\lambda) = \frac{1}{2\pi d \sigma^2 |\lambda|} \sqrt{(\eta_+- \lambda^2)(\lambda^2- \eta_-)} \, ,
\label{eq:AZCI}
\ee
where  $\eta_{\pm} = d \sigma^2 (1 \pm \sqrt{1+1/d})$ for \m~being $2d$ dimensional. The reference vector $\vec{\alpha}_{\rm AZ-CI}$ corresponds to a spectrum with  $\sigma^2 =2/5$.

The second reference vector  is the positive branch of the spectrum of \m~in \Sp, i.e.~$\vec \alpha_{\rm \Sp} = (1,1,1, 1, 3)|W|$ for five fields.  To illustrate its use, we note that the spectral deviation of $\vec{\alpha}_{\rm AZ-CI}$ from the `\Sp~spectrum' for  five fields is given by $\mu_{\rm \Sp}(\vec \alpha_{\rm AZ-CI}) = 0.34$, while  $\mu_{\rm AZ-CI}(\vec \alpha_{\rm \Sp}) = 0.36$.

 In Figure \ref{fig:SpectDev} we illustrate the moduli dependence of the spectral deviation for 
 `coincidental' complex structure moduli, $u^i = u$, and $\tau = 10i$ and the flux choice,
    \bea
\frac{1}{(2\pi)^2\alpha'} \int_{A^i} F_3 &=& \{2, 1, -5, -5, 4\} \, , ~~\frac{1}{(2\pi)^2\alpha'}\int_{B_i} F_3 = \{-4, -4, 3, 3, -4\} \, , \nonumber \\
\frac{1}{(2\pi)^2\alpha'}  \int_{A^i} H_3 &=& \{5, 2, 5, 5, 2\} \, , ~~ \frac{1}{(2\pi)^2\alpha'}\int_{B_i} H_3 = \{2, -5, -1, -1, -3\}
 \, ,
 \label{eq:fluxExpl}
 \eea
 that  contributes to the D3-tadpole by $Q_{\rm flux} = 3$. 

\newcommand{\xMin}{-2}%
\newcommand{\xMax}{2}%
\newcommand{\yMin}{.95}%
\newcommand{\yMax}{5}%
\begin{figure}[t!]
\begin{center}
\subfigure[$\mu_{\rm \Sp}$: lines at $\mu_{\rm \Sp}=\{0.05, 0.1,  0.3\}$.]
{\label{fig:SpectDev3}
\begin{tikzpicture}
\begin{axis}[height=\densityplotsize, width=\densityplotsize,xmin=\xMin,xmax=\xMax,ymin=\yMin,ymax=\yMax, xtick=\empty,ytick=\empty]
\addplot[thick,blue] graphics[xmin=\xMin,xmax=\xMax,ymin=\yMin,ymax=\yMax] {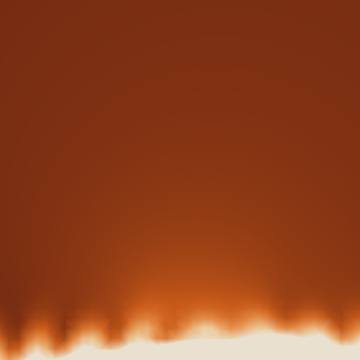};
\addplot[thick,blue] graphics[xmin=\xMin,xmax=\xMax,ymin=\yMin,ymax=\yMax] {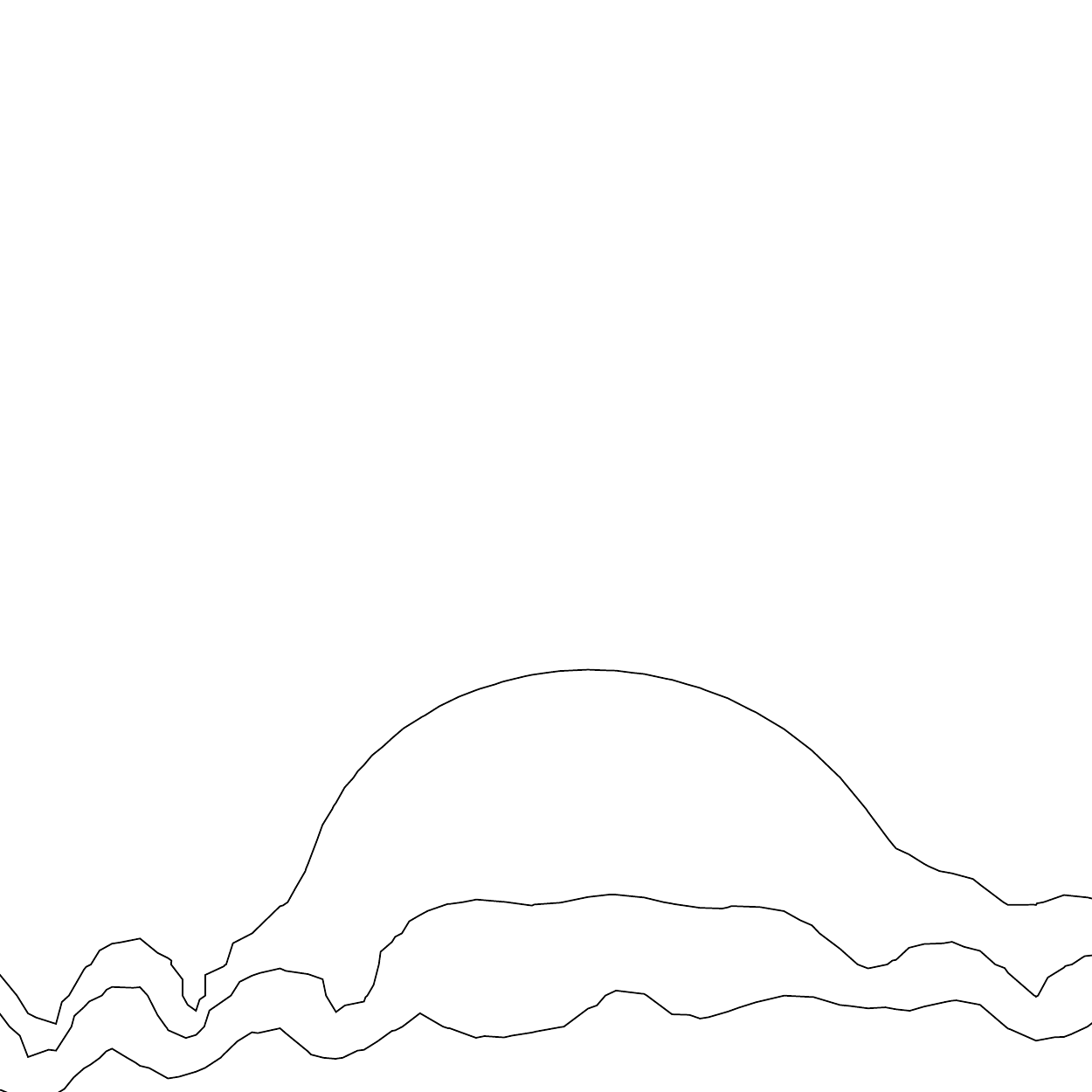};
\end{axis}
\begin{axis}[height=\densityplotsize, width=\densityplotsize, xlabel=Re$(u^i)$, ylabel=Im$(u^i)$, ylabel style={yshift=-.4cm},xmin=\xMin,xmax=\xMax,ymin=\yMin,ymax=\yMax,xtick={-2,-1,...,2},ytick={1,2,...,5},minor xtick={-2,-1.8,...,2},minor ytick={1,1.2,...,5}]
\end{axis}
\begin{scope}[shift={(.8\densityplotsize,0)}]
\begin{axis}[densityplot legend style,ymin=.015, ymax=.4, xtick=\empty,ytick={.1,.2,.3,.4},minor ytick={0,.02,...,.4}]
\addplot[thick,blue] graphics[xmin=-1, xmax=0, ymin=.015, ymax=.4] {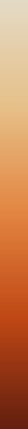};
\end{axis}
\end{scope}
\end{tikzpicture}
}
\subfigure[$\mu_{\rm AZ-CI}$: line at  $\mu_{\rm AZ-CI}=0.3$.] 
{\label{fig:SpectDevCI3}
\begin{tikzpicture}
\begin{axis}[height=\densityplotsize, width=\densityplotsize,xmin=\xMin,xmax=\xMax,ymin=\yMin,ymax=\yMax, xtick=\empty,ytick=\empty]
\addplot[thick,blue] graphics[xmin=\xMin,xmax=\xMax,ymin=\yMin,ymax=\yMax] {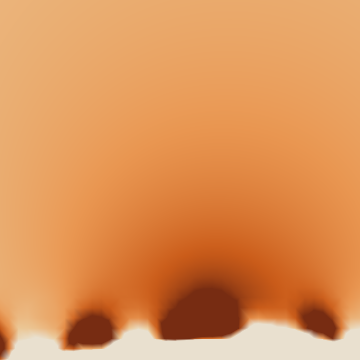};
\addplot[thick,blue] graphics[xmin=\xMin,xmax=\xMax,ymin=\yMin,ymax=\yMax] {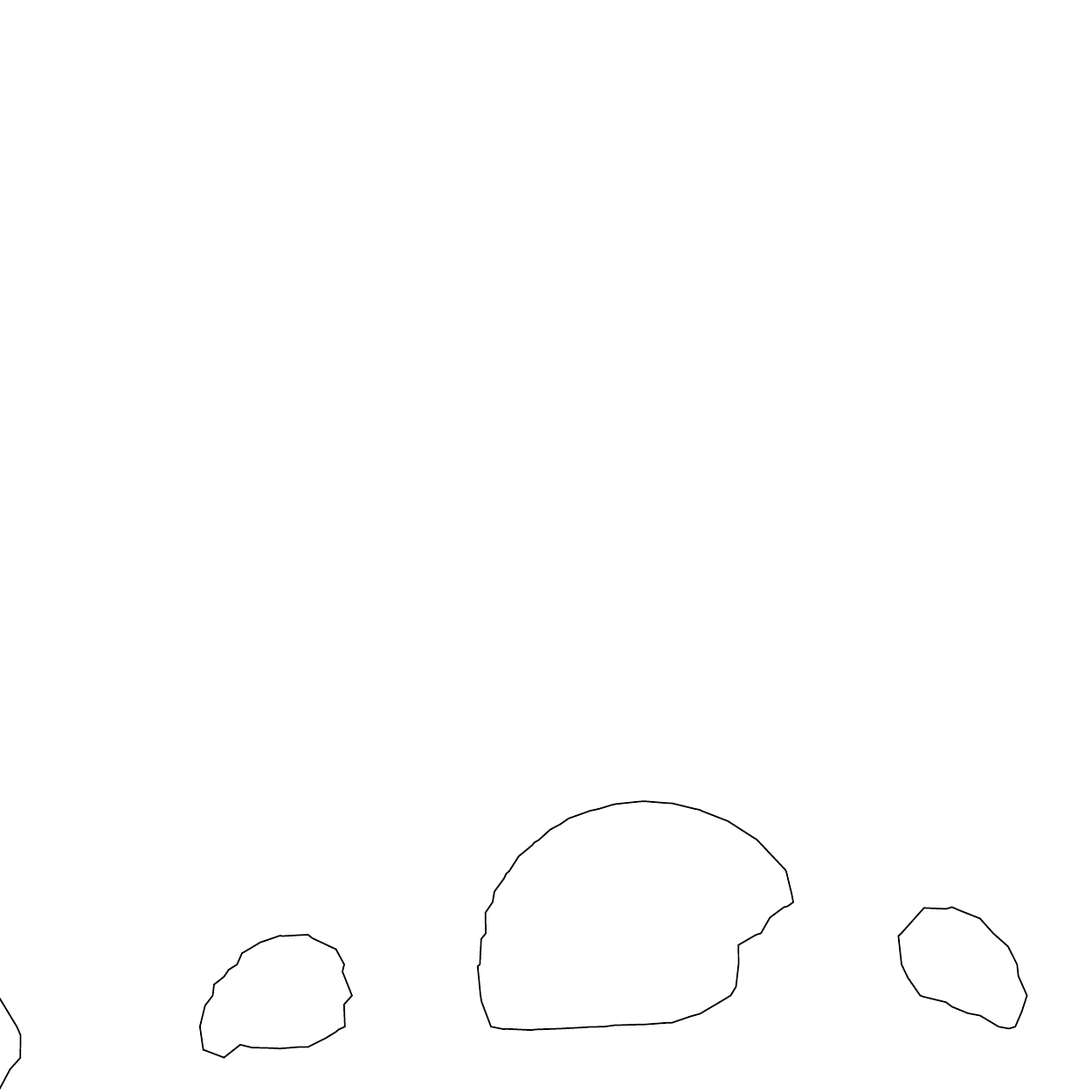};
\end{axis}
\begin{axis}[height=\densityplotsize, width=\densityplotsize, xlabel=Re$(u^i)$, ylabel=Im$(u^i)$, ylabel style={yshift=-.4cm},xmin=\xMin,xmax=\xMax,ymin=\yMin,ymax=\yMax,xtick={-2,-1,...,2},ytick={1,2,...,5},minor xtick={-2,-1.8,...,2},minor ytick={1,1.2,...,5}]
\end{axis}
\begin{scope}[shift={(.8\densityplotsize,0)}]
\begin{axis}[densityplot legend style,ymin=.2825, ymax=.374, xtick=\empty,ytick={.3,.32,.34,.36},minor ytick={.28,.285,...,.38}]
\addplot[thick,blue] graphics[xmin=-1, xmax=0, ymin=.2825, ymax=.374] {legend.png};
\end{axis}
\end{scope}
\end{tikzpicture}
}
\renewcommand{\xMin}{.2}
\renewcommand{\xMax}{.8}%
\renewcommand{\yMin}{.2}%
\renewcommand{\yMax}{.8}%
\subfigure[$\mu_{\rm \Sp}$: line at $\mu_{\rm \Sp}=0.02$]
{\label{fig:SpectDev14}
\begin{tikzpicture}
\begin{axis}[height=\densityplotsize, width=\densityplotsize,xmin=\xMin,xmax=\xMax,ymin=\yMin,ymax=\yMax, xtick=\empty,ytick=\empty]
\addplot[thick,blue] graphics[xmin=\xMin,xmax=\xMax,ymin=\yMin,ymax=\yMax] {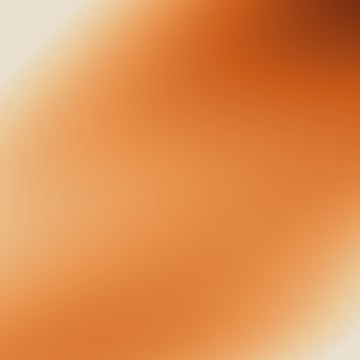};
\addplot[thick,blue] graphics[xmin=\xMin,xmax=\xMax,ymin=\yMin,ymax=\yMax] {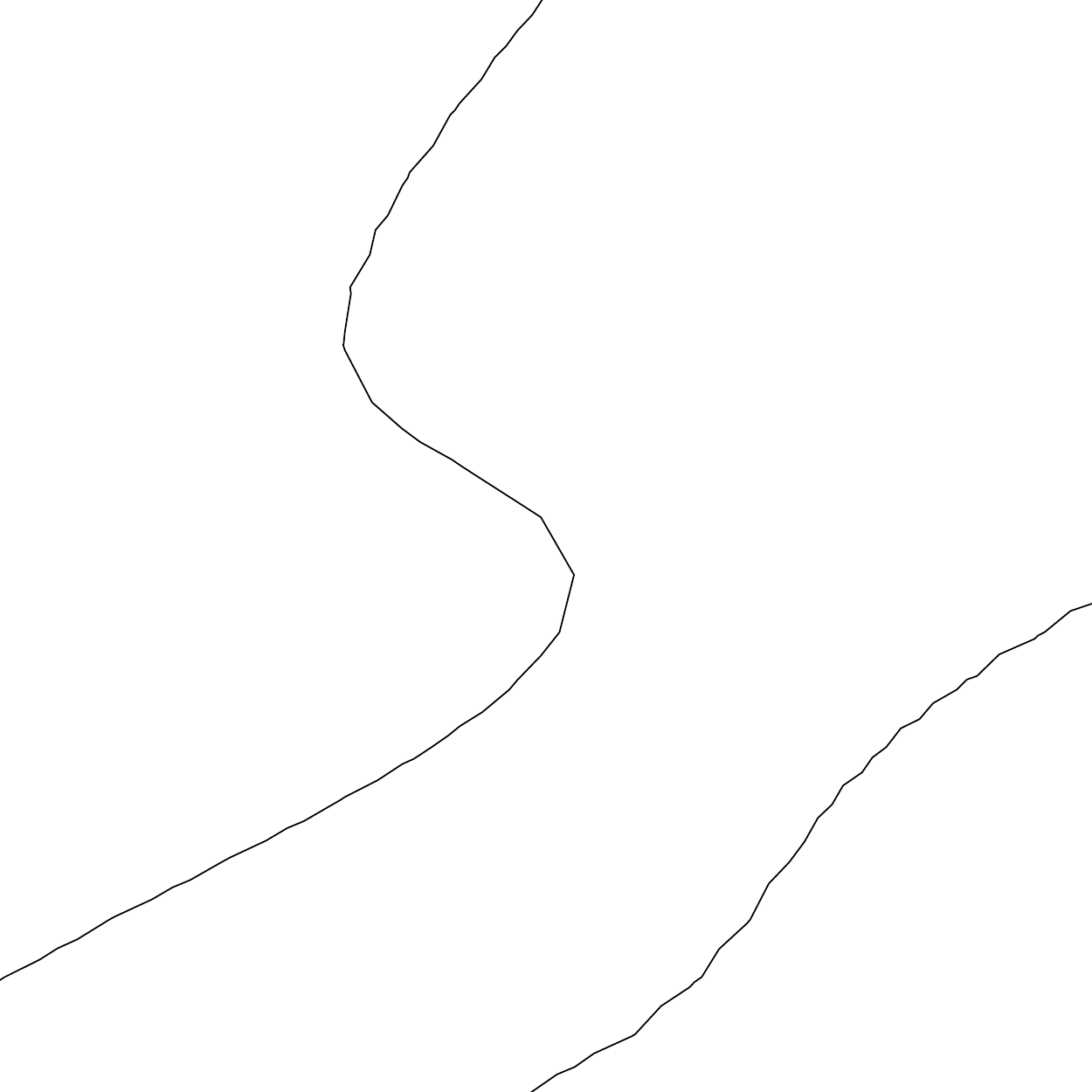};
\end{axis}
\begin{axis}[height=\densityplotsize, width=\densityplotsize, xlabel={$\theta_1$}, ylabel={$\theta_4$}, ylabel style={yshift=-.4cm},xmin=\xMin,xmax=\xMax,ymin=\yMin,ymax=\yMax,xtick={.25,.5,.75},ytick={.25,.5,.75},minor xtick={.2,.25,...,.8},minor ytick={.2,.25,...,.8},xticklabels={$\tfrac{\pi}{4}$,$\tfrac{\pi}{2}$,$\tfrac{3\pi}{4}$},yticklabels={$\tfrac{\pi}{4}$,$\tfrac{\pi}{2}$,$\tfrac{3\pi}{4}$},xlabel style={yshift=-.2cm}]
\end{axis}
\begin{scope}[shift={(.8\densityplotsize,0)}]
\begin{axis}[densityplot legend style,ymin=.0149, ymax=.0257, xtick=\empty,ytick={.016,.02,.024},minor ytick={0.015,0.0155,...,0.026},scaled y ticks=false,yticklabel style={/pgf/number format/fixed,/pgf/number format/precision=3}]]
\addplot[thick,blue] graphics[xmin=-1, xmax=0, ymin=.0149, ymax=.0257] {legend.png};
\end{axis}
\end{scope}
\end{tikzpicture}
}
\subfigure[$\mu_{\rm \Sp}$:  line at $\mu_{\rm \Sp}=0.02$]
{\label{fig:SpectDev123}
\begin{tikzpicture}
\begin{axis}[height=\densityplotsize, width=\densityplotsize,xmin=\xMin,xmax=\xMax,ymin=\yMin,ymax=\yMax, xtick=\empty,ytick=\empty]
\addplot[thick,blue] graphics[xmin=\xMin,xmax=\xMax,ymin=\yMin,ymax=\yMax] {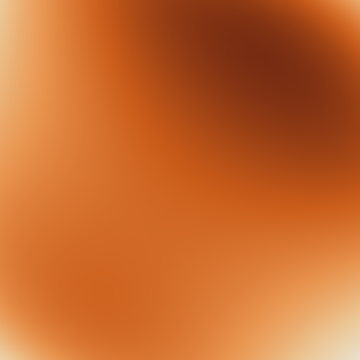};
\addplot[thick,blue] graphics[xmin=\xMin,xmax=\xMax,ymin=\yMin,ymax=\yMax] {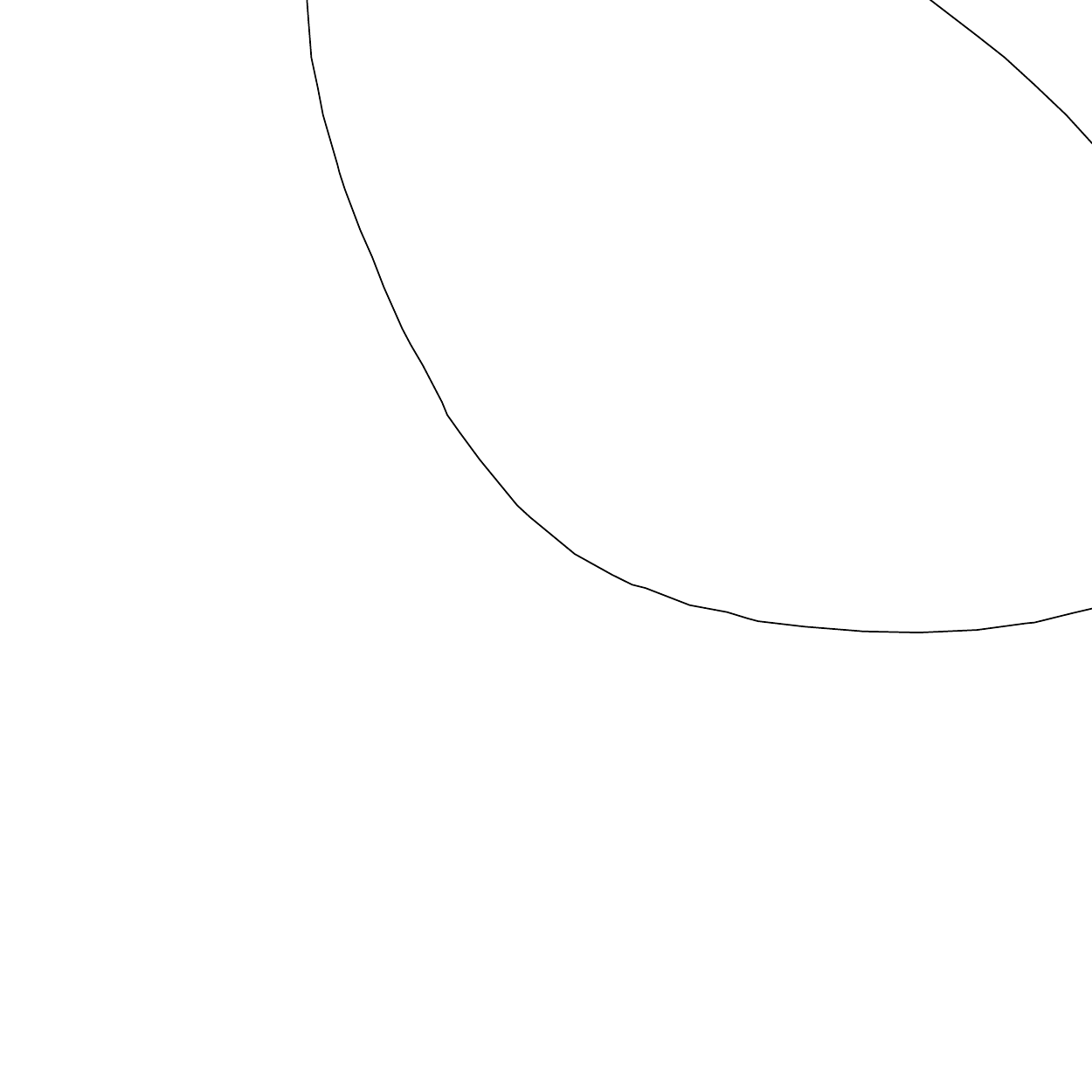};
\end{axis}
\begin{axis}[height=\densityplotsize, width=\densityplotsize, xlabel={$\theta_1$}, ylabel={$\theta_2=\theta_3$}, ylabel style={yshift=-.4cm},xmin=\xMin,xmax=\xMax,ymin=\yMin,ymax=\yMax,xtick={.25,.5,.75},ytick={.25,.5,.75},minor xtick={.2,.25,...,.8},minor ytick={.2,.25,...,.8},xticklabels={$\tfrac{\pi}{4}$,$\tfrac{\pi}{2}$,$\tfrac{3\pi}{4}$},yticklabels={$\tfrac{\pi}{4}$,$\tfrac{\pi}{2}$,$\tfrac{3\pi}{4}$},xlabel style={yshift=-.2cm}]
\end{axis}
\begin{scope}[shift={(.8\densityplotsize,0)}]
\begin{axis}[densityplot legend style,ymin=.0175, ymax=.0273, xtick=\empty,ytick={.018,.022,...,.026},minor ytick={.017,.0175,...,.028},scaled y ticks=false,yticklabel style={/pgf/number format/fixed,/pgf/number format/precision=3}]
\addplot[thick,blue] graphics[xmin=-1, xmax=0, ymin=.0175, ymax=.0273] {legend.png};
\end{axis}
\end{scope}
\end{tikzpicture}
}
\end{center}
\caption{Spectral deviations for the fluxes of equation \eqref{eq:fluxExpl}. In Figures (a) and (b)  $u^i = u$ and $\tau =10i$. In (c) and (d), ${\rm arg}(u_i) = \theta_i$ and $|u^i|=5$ and $\tau = 10i$. In (c) $\theta_2 = \theta_3 = \pi/2$, in (d) $\theta_4=\pi/2$. }
\label{fig:SpectDev}
\end{figure}
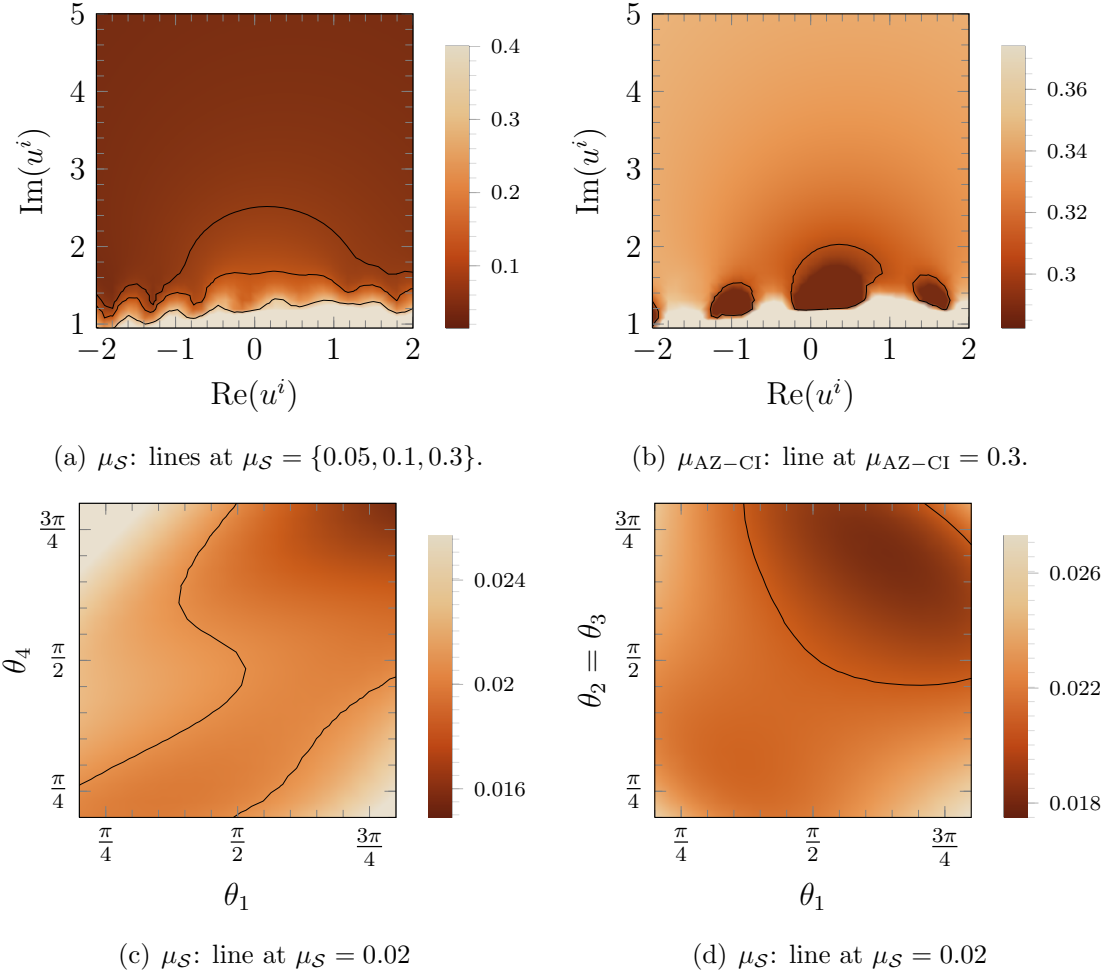


In this example, the AZ-CI spectrum does not provide a good approximation of the spectrum in any part of the sampled moduli space, even for small imaginary parts of the complex structure moduli, cf.~Figure \ref{fig:SpectDevCI3}.  

The \Sp-spectrum is quickly approached for imaginary parts of the moduli vevs of $~{\cal O}({\rm few})$, with $\mu_{\rm \Sp}< 0.05$ over a significant fraction of the moduli space, cf.~Figure \ref{fig:SpectDev3}. For large  complex structure moduli, the deviation is of the order of a few percent for  general values of the moduli phases.
In Figures \ref{fig:SpectDev14} and \ref{fig:SpectDev123} we illustrate this by scanning the phases 
of the complex structure moduli 
between $0.2 \pi$ and $0.8 \pi$ while keeping $|u_i|=5$. 


The moduli dependence of the spectral deviation certainly depends on the fluxes, but the example shown in Figure \ref{fig:SpectDev} is not atypical.  Consistent with the expectation from our analytical derivation, we find that large flux numbers on the $A^0$-cycle leads to the large regions with small $\mu_{\rm \Sp}$. In no case have we found that the AZ-CI spectrum provides a good approximation to the spectrum of \m~over a significant fraction of the moduli space.


\subsection{Strong linear correlation between $W$ and the supersymmetric masses}
\label{sec:corr}
 We now discuss one particularly interesting consequence of the strong peaks in the eigenvalue spectrum of \m. 
 From section \ref{sec:analytic} we have seen that in \Sp, the spectrum of \m~is given by integer multiples of $|W|$ so that, in particular, the spectrum of \m~is perfectly positively linearly correlated with $|W|$.  On the other hand, it is frequently assumed in the literature that the scale of the supersymmetric masses (set by \m), is statistically independent of the magnitude of the superpotential, i.e.~that the above correlation should instead vanish. We will discuss the motivation for this assumption in more detail in section \ref{sec:contFlux}. In this section we numerically  compute the correlation as a function of the moduli space for Model 1.

  The scale of the supersymmetric masses 
has important phenomenological consequences. If the supersymmetric 
 masses for a set of moduli fields can be arranged to be
  much larger than $W$ and the F-terms, then the Hessian matrix of this sector becomes positive definite, with the diagonal blocks given by $Z\bZ$. Such moduli are then supersymmetrically stabilised, and can consistently be integrated out. This strategy was famously employed in \cite{Kachru:2003aw} to stabilise the complex structure moduli and the axio-dilaton at the flux scale which, by tuning of flux numbers, was taken to be significantly larger than the value  of $|W|$.    
  
  The scale of \m~is clearly set by the scale of $Z_{ab}$, but   we have seen from equations 
  \eqref{eq:Ztautau} and \eqref{eq:Zrel} that the independent components of the $Z$-tensor can be taken to be $Z_{\tau i}$. The magnitudes of these components set the scale of the flux-induced supersymmetric masses, which we  define as,
 \be
 m_s = e^{K/2} \sqrt{ Z_{\tau i} \bZ^{i \tau}} \, .
 \label{eq:ms}
 \ee
We are then interested in the  correlation between $m_s$ and the gravitino mass,
$
m_{3/2} = e^{K/2} |W|
$, as defined in equation \eqref{eq:m32}. 

We emphasise that here, the prefactor ${\rm exp}(K/2)$ is taken to include  the K\"ahler potential truncated to the axio-dilaton and complex structure moduli sector, and an additional volume suppression of this scale will appear in the full compactification including K\"ahler moduli.\footnote{A particular consequence of not including K\"ahler moduli is that $m_{3/2}$ defined by \eqref{eq:m32} can (and typically do) take on values larger than the string scale. Such flux compactifications require  large volume suppressions to render the EFT to be controlled, as discussed in e.g.~\cite{Cicoli:2013swa}.}
 
\renewcommand{\xMin}{-2}%
\renewcommand{\xMax}{2}%
\renewcommand{\yMin}{.52}%
\renewcommand{\yMax}{4.9}%
\begin{figure}[t!]
\begin{center}
\subfigure[$\tau =i$]
{\label{fig:CorrCoeff1}
\begin{tikzpicture}
\begin{axis}[corrcoff plot back style]
\addplot[thick,blue] graphics[xmin=\xMin,xmax=\xMax,ymin=\yMin,ymax=\yMax] {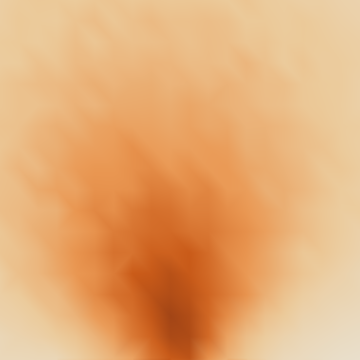};
\addplot[thick,blue] graphics[xmin=\xMin,xmax=\xMax,ymin=\yMin,ymax=\yMax] {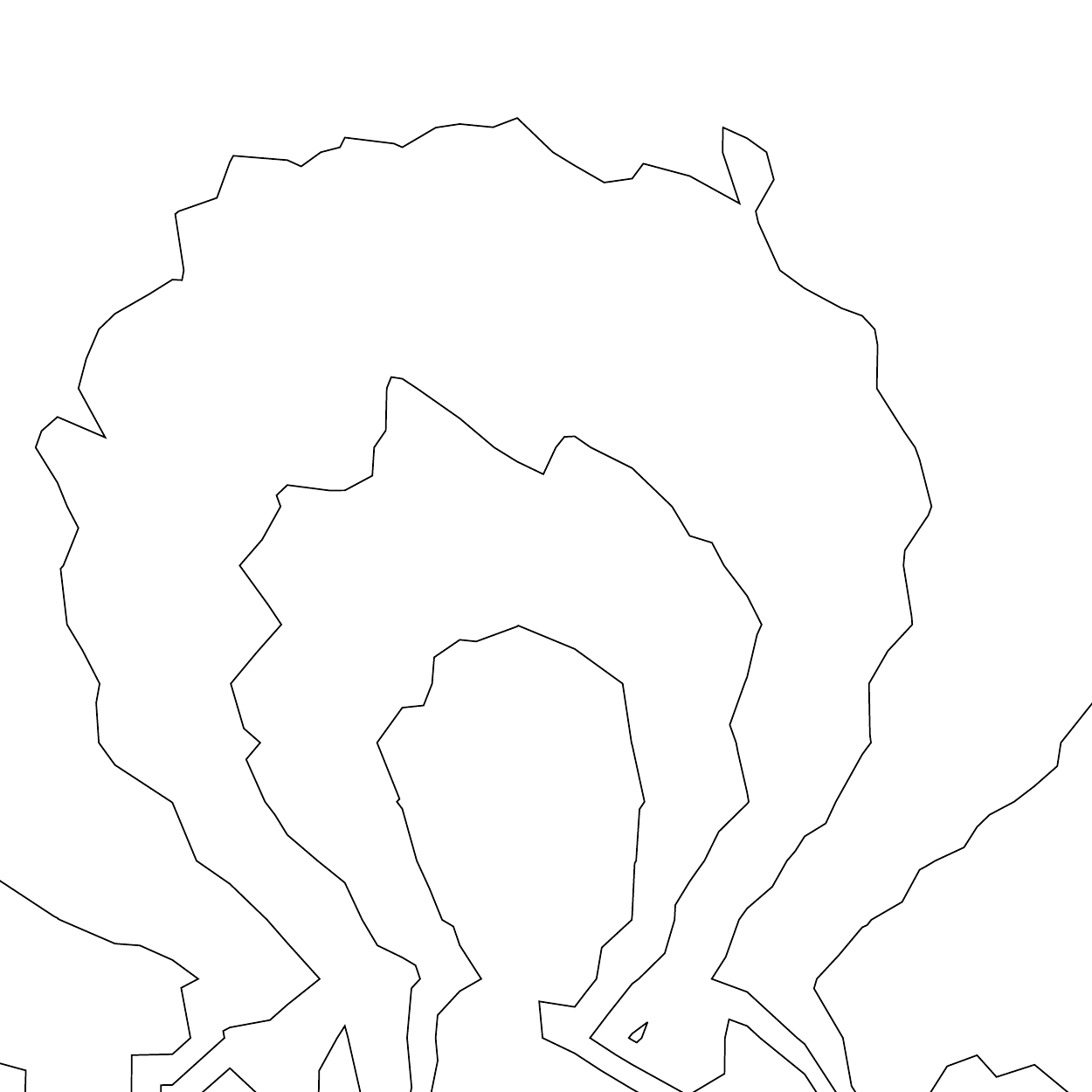};
\end{axis}%
\begin{axis}[corrcoff plot front style]
\end{axis}%
\begin{scope}[shift={(.8\densityplotsize,0)}]
\begin{axis}[densityplot legend style,ymin=-.43, ymax=.966, xtick=\empty,ytick={-.25,0,...,.75},minor ytick={-.4,-.35,...,.9}]
\addplot[thick,blue] graphics[xmin=-1, xmax=0, ymin=-.43, ymax=.966] {legend.png};
\end{axis}%
\end{scope}
\end{tikzpicture}%
}%
\subfigure[$\tau = 5i$] 
{
\begin{tikzpicture}
\begin{axis}[corrcoff plot back style]
\addplot[thick,blue] graphics[xmin=\xMin,xmax=\xMax,ymin=\yMin,ymax=\yMax] {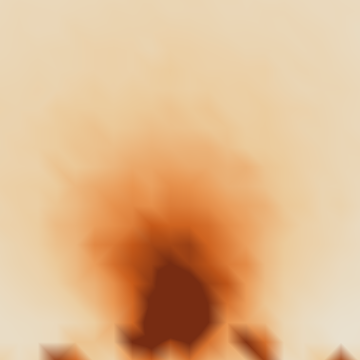};
\addplot[thick,blue] graphics[xmin=\xMin,xmax=\xMax,ymin=\yMin,ymax=\yMax] {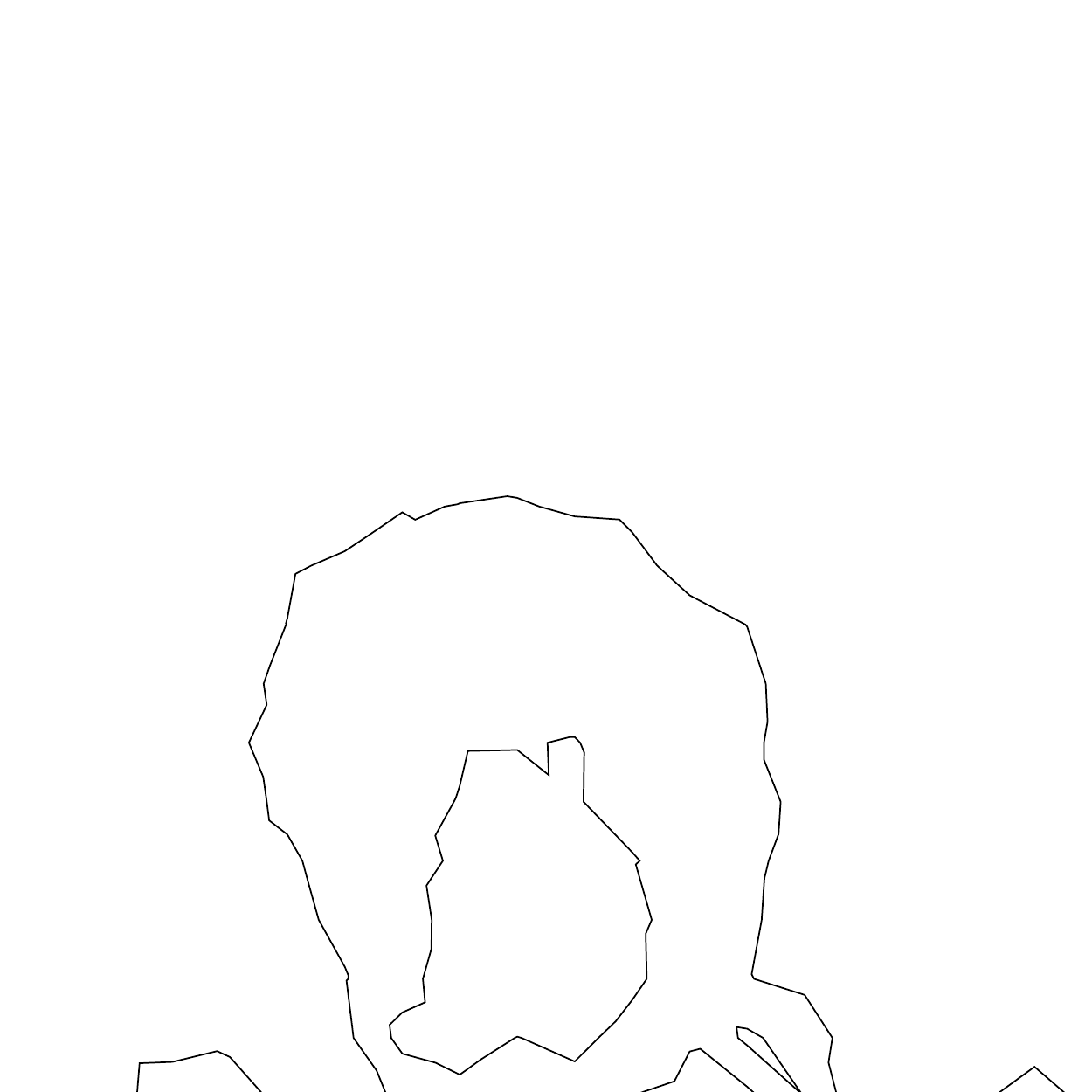};
\end{axis}%
\begin{axis}[corrcoff plot front style]
\end{axis}%
\begin{scope}[shift={(.8\densityplotsize,0)}]
\begin{axis}[densityplot legend style,ymin=.544, ymax=.988, xtick=\empty,ytick={.6,.7,.8,.9},minor ytick={.56,.58,...,1}]
\addplot[thick,blue] graphics[xmin=-1, xmax=0, ymin=.544, ymax=.988] {legend.png};
\end{axis}%
\end{scope}
\end{tikzpicture}%
}%
\\%
\subfigure[$\tau = 10i$]
{
\begin{tikzpicture}
\begin{axis}[corrcoff plot back style]
\addplot[thick,blue] graphics[xmin=\xMin,xmax=\xMax,ymin=\yMin,ymax=\yMax] {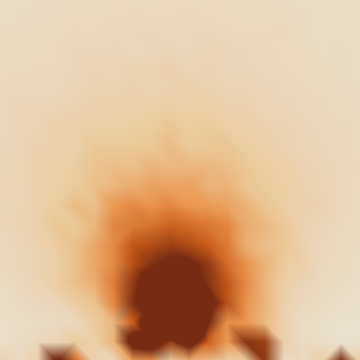};
\addplot[thick,blue] graphics[xmin=\xMin,xmax=\xMax,ymin=\yMin,ymax=\yMax] {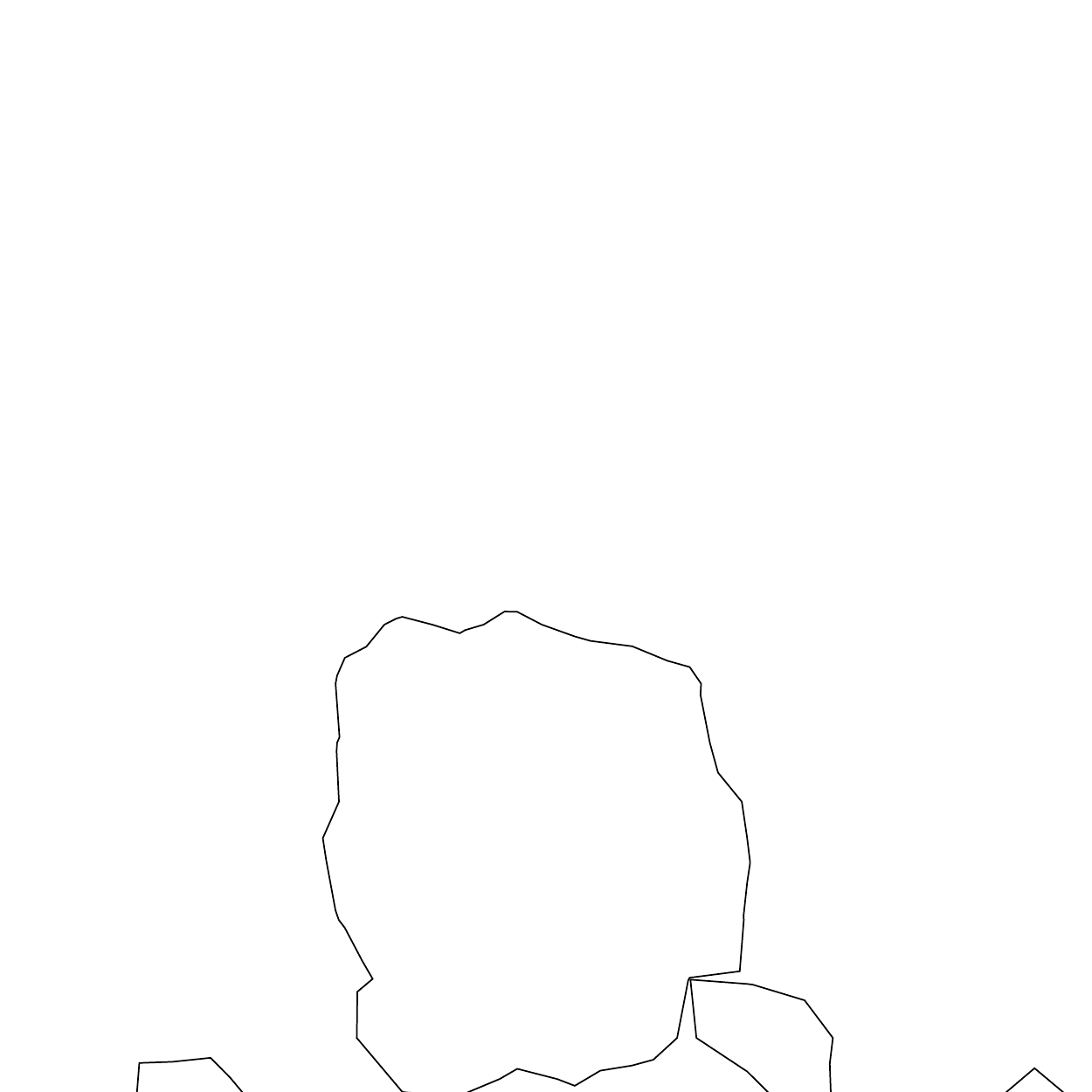};
\end{axis}%
\begin{axis}[corrcoff plot front style]
\end{axis}%
\begin{scope}[shift={(.8\densityplotsize,0)}]
\begin{axis}[densityplot legend style,ymin=.544, ymax=.988, xtick=\empty,ytick={.6,.7,.8,.9},minor ytick={.56,.58,...,1}]
\addplot[thick,blue] graphics[xmin=-1, xmax=0, ymin=.544, ymax=.988] {legend.png};
\end{axis}%
\end{scope}
\end{tikzpicture}%
}%
\subfigure[$\tau = 50i$]
{\label{fig:CorrCoeff50}
\begin{tikzpicture}
\begin{axis}[corrcoff plot back style]
\addplot[thick,blue] graphics[xmin=\xMin,xmax=\xMax,ymin=\yMin,ymax=\yMax] {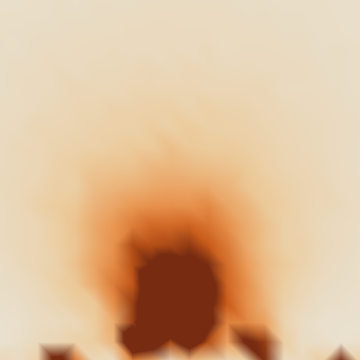};
\addplot[thick,blue] graphics[xmin=\xMin,xmax=\xMax,ymin=\yMin,ymax=\yMax] {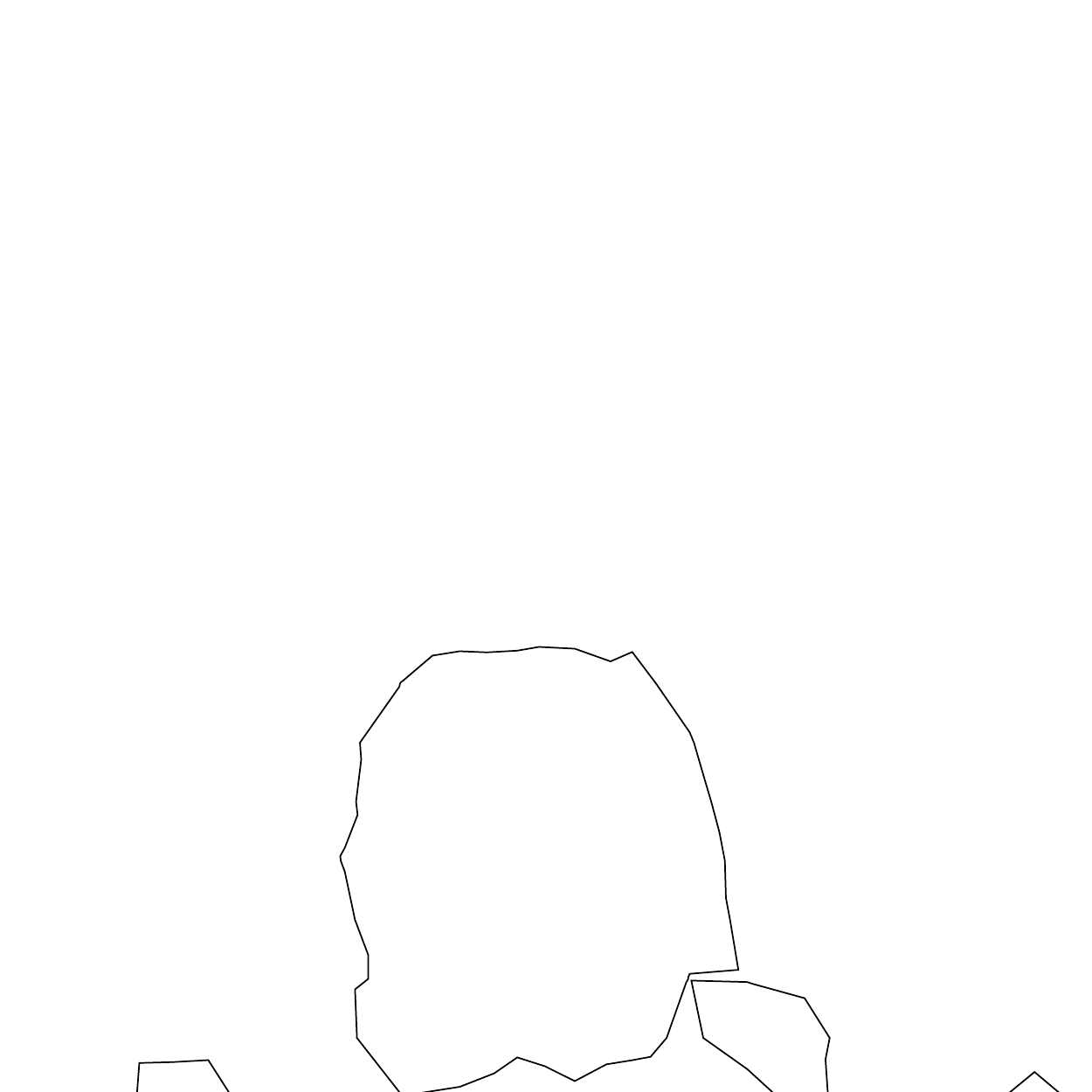};
\end{axis}%
\begin{axis}[corrcoff plot front style]
\end{axis}%
\begin{scope}[shift={(.8\densityplotsize,0)}]
\begin{axis}[densityplot legend style,ymin=.592, ymax=.99, xtick=\empty,ytick={.6,.7,.8,.9},minor ytick={.56,.58,...,1}]
\addplot[thick,blue] graphics[xmin=-1, xmax=0, ymin=.592, ymax=.99] {legend.png};
\end{axis}%
\end{scope}
\end{tikzpicture}%
}%
\end{center}
\caption{Pearson's correlation coefficient as a function of complex structure moduli with  $u^i = u$. Lines indicate $r=\{0, 0.25, 0.5, 0.75 \}$.}
\label{fig:CorrCoeff}
\end{figure}
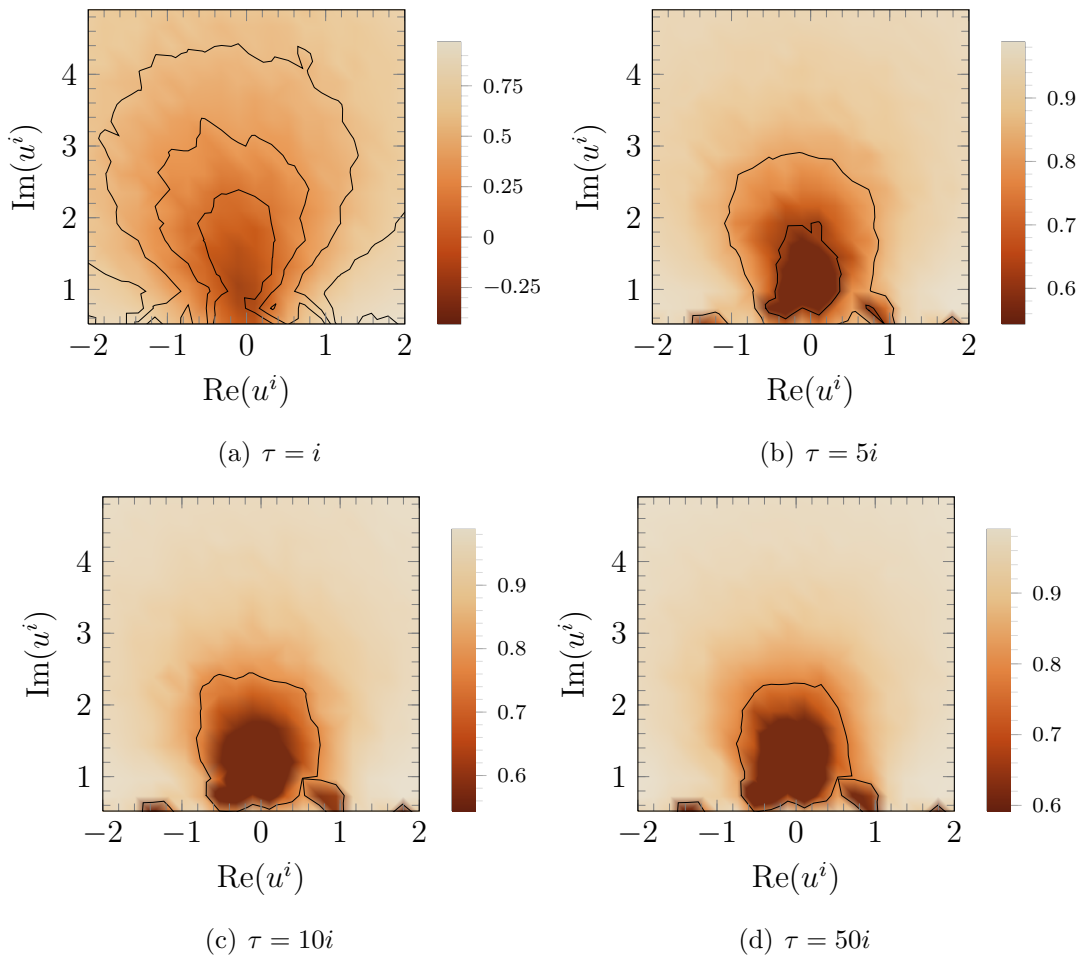
 

 To extract the strength of the  correlation between $m_s$ and $m_{3/2}$ as a function of the moduli space, we scan over  fluxes to create an ensemble of values of $(m_s, m_{3/2})$ at each given point in the moduli space. From this ensemble, we  compute Pearson's correlation coefficient,
 \be
 r = \frac{{\rm Covariance}(m_s, m_{3/2})}{\sigma_{m_s} \sigma_{m_{3/2}}} \, ,
 \ee
 where $\sigma$ denotes the relevant standard deviation. In practice, we consider ensembles generated from $1000$ random flux choices (subject to the tadpole condition $0 \leq Q_{\rm flux} \leq 22$) at each point of the densely sampled moduli space.  This enables us to study the dependence of the correlation coefficient $r$
on the moduli fields.  In \Sp, we expect perfect positive linear correlation with $r=1$, while in the regions in which 
$m_s$ and $m_{3/2}$ are independent we should find
$r = 0$. 

In Figure \ref{fig:CorrCoeff} we plot $r$ for the `coincident slice' of the complex structure moduli space for which $u^i = u$ for all $i$, and consider the string couplings  $g_s = 1, \frac{1}{5}, \frac{1}{10},~{\rm and}~\frac{1}{50}$. The strength of the correlation increases with decreasing string coupling, and for $g_s \leq 1/5$, $m_s$ and $m_{3/2}$ are strongly linearly correlated over a very large fraction of the sampled moduli space. We have again verified that this result does not significantly depend on the assumption of `coincidence' of the complex structure moduli vevs: qualitatively similar results arise even for random phases of the moduli.  



 \begin{figure}[t!]
    \begin{center}
    \subfigure[$\tau=5i$, $r=0.92$]{
\begin{tikzpicture}
   \begin{axis}[list plot style,xmin=-5,xmax=355,ymin=-5,ymax=210,xtick=\empty,ytick=\empty, xlabel=\empty,ylabel=\empty]
      	\addplot[thick,blue] graphics[xmin=-5,ymin=-5,xmax=355,ymax=210] {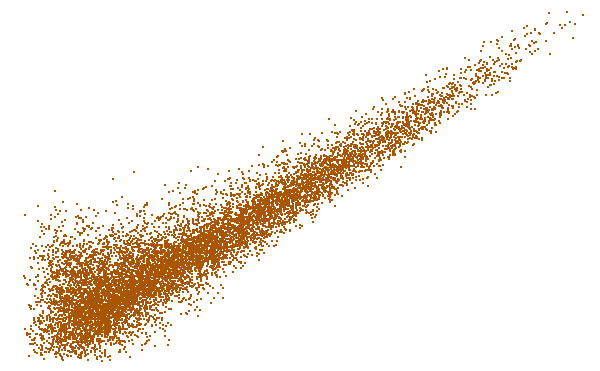};
      	\end{axis}
      \begin{axis}[list plot style, xmin=-5,xmax=355,ymin=-5,ymax=210,xtick={0,50,...,350},ytick={0,50,...,200}]
    \end{axis}
\end{tikzpicture}
      }
~~
    \subfigure[$\tau=10i$, $r=0.95$]{\label{fig:CorrData10}
\begin{tikzpicture}
   \begin{axis}[list plot style,xmin=-5,xmax=510,ymin=-5,ymax=310,xtick=\empty,ytick=\empty, xlabel=\empty,ylabel=\empty]
      	\addplot[thick,blue] graphics[xmin=-5,ymin=-5,xmax=510,ymax=310] {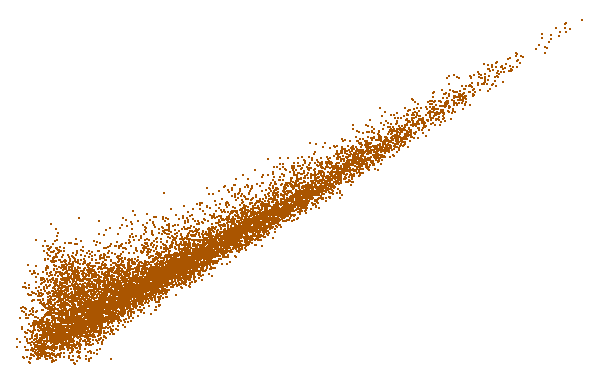};
      	\end{axis}
      \begin{axis}[list plot style, xmin=-5,xmax=510,ymin=-5,ymax=310,xtick={0,100,...,500},ytick={0,50,...,300}]
    \end{axis}
\end{tikzpicture}
      }
            \end{center}
    \caption{The distribution of $(m_s ,m_{3/2})$ at  $u^i =   -1.6 +  1.2 i$  for 10,000 random flux choices consistent with the tadpole condition.}
   \label{fig:CorrData}
\end{figure}


To illustrate the strength of the correlation between $m_s$ and $m_{3/2}$, in  Figure \ref{fig:CorrData} 
we plot the distribution of pairs $(m_s, m_{3/2})$ 
obtained from scanning over 10,000 flux choices at a particular point in the moduli space.  The strong linear correlation between the two quantities is plainly visible from the plot. Could this correlation be due to some particularity of the explicit models that we study? We will now argue that the answer to this question is no: a strong linear correlation should be expected for any compactification close to the large complex structure point.



\subsection{Why does the continuous flux approximation break down?}
\label{sec:contFlux}
A key tool in the derivation of 
many statistical results on the flux landscape -- including the famous derivation of the 
 index density of supersymmetric vacua in \cite{Ashok:2003gk} -- 
 is the approximation of integer quantised fluxes as  continuous variables. This 
approximation is often assumed to be valid for sufficiently large flux tadpoles
and allows for the replacement of sums over integers with continuous integrals, 
\be
\sum_{(\vec{N}_{\rm RR} \, ,~\vec{N}_{\rm NS})|_{0\leq Q_{\rm flux} \leq L_{\star}}  } \to \int \prod_{a=1}^{2(h^{1,2}_-+1)} {\rm d} N_{{\rm RR}\, a} {\rm d} N_{{\rm NS}\, a} \Theta(Q_{\rm flux}) \Theta(L_{\star} - Q_{\rm flux}) \, .
\label{eq:FluxApprox}
\ee
%
For our purposes, it is convenient to 
make the following  change of variables:
for any given value of the complex structure, $(\Omega, \bar \Omega, D_i \Omega, \bar D_{\bi} \bar \Omega)$ form a basis of $H^3(M)$, and we may correspondingly expand the flux vector with respect to the  basis of periods $(\Pv, \Pv^*, D_i \Pv, \bar D_{\bi} \Pv^*)$.\footnote{We are grateful to Kepa Sousa for discussion of this point.} This basis is not orthonormal in general, and the relevant symplectic inner products are given by,
\bea
\Pv\, \Sigma\, \Pv^* = -i e^{-K_{\rm c.s.}} \, ,~~D_i \Pv\, \Sigma\, \bar D_{\bj} \Pv^* = -i K_{i\bj} e^{-K_{\rm c.s.}} \, .
\eea
To simplify our expressions, we phrase our results in this section in terms of the transformed flux vector,
\be
\Nt
=
\left(
\begin{array}{c}
- \int_{A_i} G_3 \\
- \int_{B_i} G_3 
\end{array}
\right) = 
\Sigma\, \N \, .
\ee
In the new basis, this transformed flux vector has the expansion,
\be
\Nt = -i e^{K_{\rm c.s.}} \left[
W \Pv^* - \frac{\bar F_{\bar \tau}}{K_{\bar \tau}} \Pv + F^{\bj} \bar D_{\bj} \Pv^* - \frac{\bZ^i_{~\bar \tau}}{K_{\bar \tau}} D_i \Pv
\right] \, .
\label{eq:fluxExp}
\ee
This is a linear change of variables from the $4(h^{1,2}_-+1)$ real fluxes in the canonical basis to the $2+2(h^{1,2}_-+1) +2h^{1,2}_-$ real components of $(W, F_a, Z_{\tau i})$.\footnote{  
In this basis, the 
%
familiar `ISD condition' which stipulates  that  3-form flux, $G_3$, that preserves ${\cal N}=1$ supersymmetry has vanishing (3,0) and (1,2)-components becomes quite transparent: supersymmetric configurations have no support along $\Pv$ and $\bar D_{\bj} \Pv^*$.} 
The integral \eqref{eq:FluxApprox} is in this basis given by,
\be
 \int \prod_{a=1}^{2(h^{1,2}_-+1)} {\rm d} N_{RR\, a} {\rm d} N_{NS\, a} = {\cal C}\, \int {\rm d}^2 W\, {\rm d}^{2(h^{1,2}_-+1)} F_a\, {\rm d}^{2h^{1,2}_-} Z_{\tau i} \, ,
 \label{eq:measure}
\ee
where ${\cal C}$ is  a complex structure dependent -- but $W$, $F$, and $Z_{\tau i}$ independent -- constant. 

Using the expansion \eqref{eq:fluxExp}, we see that the flux contribution to the D3-tadpole, cf.~equation \eqref{eq:Qflux},  is  given by,
\be
[(2\pi)^4 \alpha'^2]\, Q_{\rm flux} = e^{K} \left( |W|^2 + Z_{\tau i} \bZ^{\tau i} - F_a \bF^a
\right) 
= m_{3/2}^2 + m_s^2 - e^K F_a \bF^a
\, .
\label{eq:tadpoleNewVars}
\ee
Thus,  in the continuous flux approximation the correlation between $m_s$ and $m_{3/2}$ arises solely from the tadpole condition \eqref{eq:tadpoleNewVars}  and not from the measure \eqref{eq:measure}.

There are two particularly interesting cases to consider: first, for a sufficiently large tadpole $L_{\star}$, there will be numerous flux choices giving $m_{3/2}^2 \ll [(2\pi)^4 \alpha'^2] L_{\star}$ for which the tadpole condition $Q_{\rm flux} \leq L_{\star}$ does not introduce a correlation between $m_s$ and $m_{3/2}$. In this case -- which is perhaps the most commonly considered in the literature -- we have $r=0$.  

Second, we may consider the general case in which $m_s$ and $m_{3/2}$ are random variables of the flux choices, and typically not much smaller than $[(2\pi)^2 \alpha'] \sqrt{L_{\star}}$. We will not discuss this general case in full detail, but rather consider the simpler, supersymmetric case.  We furthermore take $m_{3/2}$ and $m_s$ to be uniformly distributed in the quarter disc of radius $[(2\pi)^2 \alpha'] \sqrt{L_{\star}}$ in the upper-right quadrant of the $m_s$-$m_{3/2}$ plane. The covariance is then given by,
\be
{\rm Covar}(m_s, m_{3/2}) = \frac{L_{\star}}{2\pi} \left(
1- \frac{32}{9\pi}
\right) \, , 
\ee
and Pearson's correlation coefficient is given by,
\be
r =  \frac{2}{\pi} \left(
\frac{1- \frac{32}{9 \pi}}{1 - \frac{64}{9 \pi^2}}
\right) \approx -0.3 \, .
\ee

We note that the distributions of $m_s$ and $m_{3/2}$ can be found from equation \eqref{eq:measure} and are in fact not uniform, but peak at large values (e.g.~the distribution for $m_{3/2}$ is linear). Taking into account the full distribution would then lead to a stronger negative correlation, while including non-vanishing F-terms would effectively increase the radius of the permissible region in the $m_s$-$m_{3/2}$ plane, which would result in a smaller correlation.  However, as we are not  interested in the exact magnitude of this number but rather its sign, our simple discussion using uniform distributions suffices: 
%
%
we have shown that the continuous flux approximation generically gives  $r<0$, and  for the special subset of points with  
$m_{3/2} \ll  [(2\pi)^2 \alpha'] \sqrt{L_{\star}} $, it predicts $r\approx0$. Evidently, neither of these predictions explain the strong positive correlation we have observed in section \ref{sec:corr}, so we may now ask, why does the continuous flux approximation break down in the flux compactifications that we consider?

A
possible explanation of the observed correlation would be to note that while the flux tadpole of Model 1, $L_{\star} =22$, is larger than the number of flux cycles  ($2\times4+2=10$), the hierarchy between these numbers is not necessarily large enough to justify the continuous flux approximation. Similarly, to find flux choices that satisfy the tadpole condition with some frequency, we have restricted the largest flux number to $5$, which may well be much too small to justify the continuous flux approximation. For more general compactifications than those considered in reference \cite{Krippendorf}, significantly larger tadpoles can be found. 
We may then ask, does the relatively small flux tadpole drive the results of section \ref{sec:corr}? 

To address this question, we consider a hypothetical modification of the brane content of Model 1 that would give rise to a very large tadpole so that we may effectively take $L_{\star} \to \infty$.  Furthermore, we allow both positive and negative flux tadpoles and take the flux numbers to be uniformly distributed from $-50$ to $+50$. With these assumptions, the flux tadpole has no effect on the correlation coefficient, and the continuous flux approximation predicts $r=0$. 

\begin{figure}[t!]
    \begin{center}
    \subfigure[$\tau=5i$, $r=0.92$]{
\begin{tikzpicture}
   \begin{axis}[list plot style,xmin=-5,xmax=3400,ymin=-5,ymax=2200,xtick=\empty,ytick=\empty, xlabel=\empty,ylabel=\empty]
      	\addplot[thick,blue] graphics[xmin=-5,ymin=-5,xmax=3400,ymax=2200] {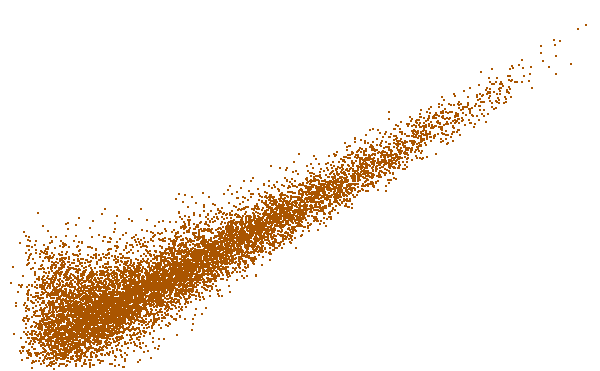};
      	\end{axis}
      \begin{axis}[list plot style, xmin=-5,xmax=3400,ymin=-5,ymax=2200,xtick={0,1000,...,3000},ytick={0,500,...,2000}]
    \end{axis}
\end{tikzpicture}
      }
~~
    \subfigure[$\tau=10i$, $r=0.95$]{
\begin{tikzpicture}
   \begin{axis}[list plot style,xmin=-5,xmax=4800,ymin=-5,ymax=2900,xtick=\empty,ytick=\empty, xlabel=\empty,ylabel=\empty]
      	\addplot[thick,blue] graphics[xmin=-5,ymin=-5,xmax=4800,ymax=2900] {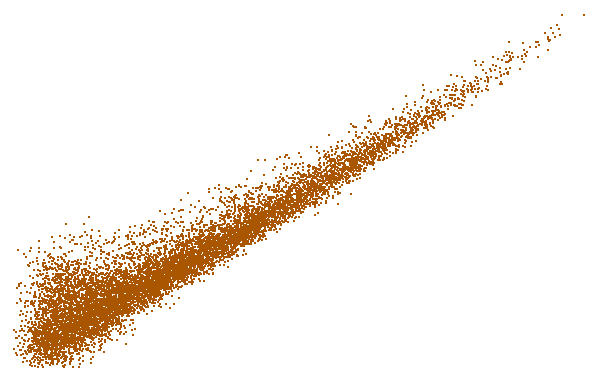};
      	\end{axis}
      \begin{axis}[list plot style, xmin=-5,xmax=4800,ymin=-5,ymax=2900,xtick={0,1000,...,4000},ytick={0,500,...,2500}]
    \end{axis}
\end{tikzpicture}
      }
            \end{center}
    \caption{The distribution of $(m_s ,m_{3/2})$ at  $u^i =   -1.6 +  1.2 i$  for 10,000 random flux choices drawn from a uniform distribution on $[-50,50]$ without imposing the tadpole condition. 
    }
   \label{fig:CorrDataNoTaddy}
\end{figure}

We show the resulting distribution of $m_{3/2}$ and $m_s$ in Figure \ref{fig:CorrDataNoTaddy} for the reference point $u^i =   -1.6 +  1.2 i$ and $\tau = 5i, 10i$. The resulting distribution is again strongly positively correlated with $r=0.92$ and $r=0.95$ -- just as in the case plotted in Figure \ref{fig:CorrData} for fluxes satisfying the tadpole condition. Thus, we conclude that the small tadpole of Model 1 does not affect or explain  the correlation between $m_s$ and $m_{3/2}$.

The reason for the breakdown of the continuous flux approximation can instead be understood as follows. The large complex structure limit ${\rm Im}(u^i) \to \infty$ is a so called `D-limit' in which the vectors  $(\Pv, \Pv^*, D_i \Pv, \bar D_{\bi} \Pv^*)$  no longer form a good basis, and $Q_{\rm flux}$ develops null directions \cite{Ashok:2003gk, Eguchi:2005eh, Magda2}.  Already in reference \cite{Ashok:2003gk} it was argued  that the continuous flux approximation should be expected to break down in the D-limit as the periods become comparable to the tadpole. 

In this paper we have seen that the continuous flux approximation breaks down quickly in the large complex structure expansion. This can be understood as a consequence of the fact that a single element (corresponding to the $A^0$-cycle) quickly comes to dominate the norm of $\Pv$, and that the corresponding contribution to $\vec{N} \cdot \Pv$ dominates the superpotential as a result.  All scales of the resulting theory are then set by the fluxes on the cycle $A^0$, with all other fluxes only contributing by subleading corrections. 

At weak string coupling, the contribution from the NS-NS fluxes to the superpotential are further enhanced by a factor of $1/g_s$, and the RR fluxes give a subdominant contribution. A single real flux number then sets the scale of 
 $m_s$ and $m_{3/2}$, thus explaining the  strong correlation in this case. 

We expect that the rapid dominance of the cubic terms in the superpotential is a generic feature of compactifications with a large number of complex structure moduli. The number of cubic terms in the superpotential scales like $\sim (h^{1,2}_-)^3$, as does the number of terms at quadratic order which come in sets of $\sim (h^{1,2}_-)^2$ terms, each proportional to one out of  $\sim h^{1,2}_-$ flux choices. Thus, we expect that the results found in this paper 
should be applicable to compactifications with a large number of complex structure moduli. 

\subsection{Universality of random matrix theory  and flux compactifications}
\label{sec:rmt}
We are now ready to make two final observations regarding the applicability of random matrix theory to the statistics of the flux landscape. 

First, we note that the spectra that we have obtained analytically for \m~and \h~in section \ref{sec:analytic} are not only simple and deterministic, but also have zero probability of appearing in the AZ-CI model, the `WWW' model, or any straight-forward generalisation thereof.  This can be understood as a direct consequence of the Vandermonde determinant that appears in the joint probability density of the eigenvalues. For example, the joint probability density of the $N$ positive eigenvalues $\nu_a$  of the AZ-CI ensemble
is given by,
\be
f(\nu_1,\ldots,\nu_N) \propto \ 
\prod_{a<b} |\nu^2_a-\nu^2_b|
{\rm{exp}}\left(-\frac{1}{2 \sigma^2} \sum_{a=1}^{N} \nu_{a}^2+   \sum^N_{a=1} \ln  \nu_a \right)\,, \label{eq:probdens}
\ee
where $\sigma^2$ denotes the variance of the normally distributed independent matrix entries. The measure \eqref{eq:probdens}
clearly gives zero weight to configurations with degenerate eigenvalues, such as the one we found in \eqref{eq:hooray}.

While the Vandermonde determinant
can be interpreted as giving rise to `eigenvalue repulsion', the observed spectra of flux compactifications considered in this paper are rather characterised by `eigenvalue attraction' with peaked spectra and  degenerate eigenvalues. We have shown that this spectacular difference arises due to
the 
fast growth of a single element of the period vector, and is therefore particular to compactifications close to the large complex structure point. Such 
compactifications are  interesting and much studied, but they only constitute 
 a small fraction of all possible flux compactifications.
 Is it then the case that random matrix theory models are broadly applicable to flux compactifications on manifolds that are sufficiently distant from any D-limit? We here show that  existing theorems in the Random Matrix Theory literature do not guarantee that the matrices \m~and \h~in the `flux landscape' reach universal limits, but,  we  argue,   there are good reasons to believe that the spectrum of \m~is well-described by the AZ-CI ensemble for generic compactifications with many moduli, as suggested in \cite{Denef:2004cf}.

  The 
  relevant RMT universality theorems can be understood as extensions of Wigner's 1958 derivation of the limiting `semi-circle law'  
for the eigenvalue density of an ensemble of  symmetric matrices with independent and identically distributed entries  \cite{Wigner2}.
While that derivation relied on  certain assumptions on the moments of the distribution of the independent matrix elements, it  was independent of  the details of the corresponding distribution and can therefore be regarded as `universal'. In the recent work \cite{2005SchenkerShulzBaldes}, Wigner's result was extended to matrices with correlated entries, and in \cite{2007HofmannCredner}, similar theorems were derived for the remaining symmetry classes in the Altland-Zirnbauer classification, including the AZ-CI ensemble. Thus, reference \cite{2007HofmannCredner} could provide the mathematical justification for asserting that the matrix \m~in flux compactifications -- despite string theory correlations -- should roughly be described by the CI matrix ensemble. 



References
\cite{2005SchenkerShulzBaldes, 2007HofmannCredner} 
consider $N\times N$ matrices $M_{AB}$ in which some of the entries are in some way correlated, $M_{AB} \sim M_{CD}$. The exact nature of the correlation is unimportant, as long as the correlations are not too many and that the number of null vectors does not grow too fast with increasing matrix size.  The three key assumptions are (in our notation) given by,
\bea
&{\rm C1:~~}&\max_{A} \left( \sum_{B,C,D}\delta\left(M_{AB}\sim M_{CD}\right) \right) = {\cal O}(N^{2- \epsilon}) \, ,
\label{eq:assump1} \\
&{\rm C2:~~}&\max_{A,B,C} \left( \sum_{D}\delta\left(M_{AB}\sim M_{CD}\right) \right) = {\cal O}(N^{\epsilon}) \, ,
\label{eq:assump2} \\
&{\rm C3:~~}&\sum_{\substack{A, B,C \\ A\neq C}} \delta\left(M_{AB}\sim M_{BC}\right)  = {\cal O}(N^{2}) \, ,
\label{eq:assump3} 
\eea
where $\epsilon\geq0$ and,
\be
\delta\left(M_{AB}\sim M_{CD}\right) =
\left\{
\begin{array}{l l l }
1 &~~&{\rm if}~M_{AB}\sim M_{CD} \, , \\
0 & & {\rm otherwise} \, .
\end{array} \right. 
\ee 
Assumption C1 ensures  that the total number of correlations between all elements of a row and all elements of the matrix does not grow faster than ${\cal O}(N^{2- \epsilon})$, while from assumption C2 it follows that no entry in the matrix has more than ${\cal O}(N^{\epsilon})$ correlations with any full row of the matrix. Assumption C3 finally states that the number of correlations between the $B$'th row and the $B$'th column (excluding elements related by the symmetries of $M_{AB}$), when summed over $B$ does not grow faster than ${\cal O}(N^{2})$.  

We now consider what these conditions mean for  the matrix \m~in flux compactifications that are described by the Gukov-Vafa-Witten superpotential \eqref{eq:GVW} and the leading order K\"ahler potential \eqref{eq:K}.\footnote{We are very grateful to Kepa Sousa for discussions on this point.} We will assume that the independent components of $Z_{ab}$ are $Z_{\tau i}$ so  that  $\delta(Z_{\tau i} \sim Z_{\tau j}) = \delta( Z_{\tau i} \sim \bZ^{j}_{\bar \tau}) = \delta^j_i$.

The non-trivial correlations of \m~arise from the Hodge decomposition of the holomorphic three-form $\Omega$ and are given by 
equation \eqref{eq:Zrel} which we here may write as $Z_{ij}  = {\cal C}  \ks_{ij}^{~~\bar k} \bZ_{\bar \tau \bar k}$, where ${\cal C}$ is an unimportant constant.  

It will suffice for our purposes to consider the implications of C1  for $A= \tau$, even though this row does not necessarily 
have the maximum number of correlations of any row in the matrix. We then have, 
\bea
 \sum_{i,j,k}\delta\left(Z_{\tau i}\sim Z_{jk}\right) &=& 
 \sum_{i,j,k}\delta\left(Z_{\tau i}\sim \ks_{jkl} \bZ^l_{\bar \tau } \right) 
 =
  \sum_{i,j,k}\delta\left( \ks_{jki} \right) \leq
  {\cal O}(N^{2- \epsilon})
\, ,
\label{eq:constr1}
 \eea
for $N= h^{1,2}_-+1$. Here $\delta\left( \ks_{jki} \right)=0$ if $\ks_{jki} = 0$ and $\delta\left( \ks_{jki} \right)=1$ if $\ks_{jki} \neq 0$.
 Condition C2 for $C=\tau$ implies that,
 \bea
\max_{i,j}\left( \sum_{k}\delta\left(Z_{ij }\sim Z_{\tau k}\right) \right)
&=&
\max_{i,j}\left( \sum_{k}\delta\left(\ks_{ijl} \bZ^l_{\bar \tau }\sim Z_{\tau k}\right) \right) \nonumber \\
&=&
\max_{i,j}\left( \sum_{k}\delta\left(\ks_{ijk} \right) \right) \leq  {\cal O}(N^{\epsilon}) 
 \, .
 \label{eq:constr2}
\eea
Finally, condition C3 gives that, 
\bea
\sum_{\substack{a,b,c \\ a\neq c}} \delta\left(Z_{ab}\sim Z_{bc}\right)  &=&
= 2\sum_{j, k} \delta( \ks_{jjk}) +
\sum_{\substack{j, l \\ j\neq l}} \sum_{p, k} \delta(\ks_{jkp}) \delta(\ks_{lkp})
\leq
{\cal O}(N^{2}) \, .
\label{eq:constr3}
\eea
Evidently, 
 when interpreted in the context of flux compactifications the conditions \eqref{eq:assump1}--\eqref{eq:assump3} give rise to non-trivial conditions on the number of non-vanishing Yukawa couplings. 
 
 For a set of compactification manifolds with an increasing number of complex structure moduli, 
 the   number of non-vanishing Yukawa couplings may in general scale with the total number of Yukawa couplings, i.e.~$\sim(h^{1,2}_-)^3$, in violation of \eqref{eq:constr1}.  Furthermore,  the sum $\sum_{k}\delta\left(\ks_{ijk} \right)$ naturally  scales  linearly with $h^{1,2}_-$, in violation of \eqref{eq:constr2} for $\epsilon <1$. In addition, the last term of equation \eqref{eq:constr3} may scale as fast as $\sim(h^{1,2}_-)^4$, making also condition C3 inapplicable to flux compactifications. Thus, we see that the `universality theorems' of \cite{2005SchenkerShulzBaldes, 2007HofmannCredner} do not automatically apply to flux compactifications. 
 
Obviously, this does not prove that random matrix theory is inapplicable to studies of the flux landscape. It may be possible to construct a sequence of flux compactifications that in some region of the moduli space satisfy the conditions \eqref{eq:constr1}--\eqref{eq:constr3}, thereby making  the existing universality theorems applicable. Moreover, it may be that for correlations of the form \eqref{eq:Zrel}, it is possible to derive stronger  results  than for the general correlations considered in \cite{2005SchenkerShulzBaldes, 2007HofmannCredner}.  Indeed, by numerically simulating random matrices \m~that satisfy the string theory conditions \eqref{eq:Ztautau} and \eqref{eq:Zrel} for a randomly chosen set of non-vanishing Yukawa couplings,  we find that the eigenvalue density quickly becomes quite similar to the AZ-CI spectrum for large matrices. 



We thus conclude that while existing universality arguments do not suffice to guarantee that the matrices \m~and \h~in flux compactifications approach the CI or `WWW' ensembles, respectively, it may well be possible to extend these arguments to the case that is particularly  interesting in string theory.  



\section{Conclusions}
\label{sec:concl}
We have considered four-dimensional effective supergravities arising in the low-energy limit of flux compactifications of type IIB string theory, and we have asked what are the spectra of the Hessian matrix \h~and the matrix \m~that governs the critical point equation. By direct computation, we have found these spectra analytically  
in a subspace of the large complex structure limit in which  the complex structure moduli  all have the same phase. 
The resulting eigenvalue distributions 
are remarkably given by highly degenerate eigenvalues at integer multiples of $|W|$ and $m_{3/2}^2$, independently of the details of the compactification manifold or the flux choice.  These results may thus be regarded as `universal' for type IIB flux compactifications at large complex structure. 

By computing the spectra of \h~and \m~numerically in explicit flux compactifications, we have found that the limiting spectra are quickly approached for ${\rm Im}(u^i)>1$, and that over most of the sampled moduli space, $m_{3/2}$ is strongly linearly correlated with the scale of the supersymmetric moduli masses, $m_s$.  Such a correlation makes  hierarchies of the form $m_{3/2} \ll m_s$ highly infrequent in this region of moduli space.  

Our results imply that proposed random matrix theory models 
for \h~and \m~
are inapplicable at large complex structure, and we have furthermore argued that 
in more general type IIB flux compactifications,  the number of correlations between matrix elements may grow too quickly
for existing `universality theorems' in the RMT literature to apply. It would be interesting to extend these theorems to include the particular relations that appear in flux compactifications. 
%
%

In this paper we have focussed on the moduli space dependence of  eigenvalue distributions, and we have not specialised to the  small subset of points that are supersymmetric vacua. It would be interesting to understand if the structure found in this paper can be found also among such vacua, or if on the contrary, supersymmetric vacua are preferably realised in the relatively small regions in which this structure is strongly broken. It would furthermore be interesting to extend the analysis of this paper to the full moduli sector of string compactifications. 

In sum, we have shown that in a particular region of the moduli space, the `flux landscape' has much non-random structure
that determines important aspects of the low-energy theory. 
Extending this study to broader regions of the moduli space   is an important question for the future.

\section*{Acknowledgments}
We would like to thank
Thomas Bachlechner, Joseph Conlon, Mafalda Diaz,  Jonathan Frazer,   Denis Klevers, Sven Krippendorf, Andre Lukas, Liam McAllister, Markus Rummel, Kepa Sousa and  Alexander Westphal for stimulating discussions. 
Part of this work was performed at the Aspen Center for Physics, which is supported by National Science Foundation grant PHY-1066293. This project was funded in part by the European Research Council grant 307605-SUSYBREAKING.

\appendix
\section{Additional explicit examples of flux compactifications}
\label{app:B}
In this appendix, we present  the results of numerical simulations of eigenvalue spectra in several additional examples of flux compactifications. We begin with investigating the influence of the orientifold involution of the eigenvalue density in Model 1, finding that 
breaking the $u_2 \leftrightarrow u_3$ symmetry by turning on general fluxes  
has no discernible effect on the 
eigenvalue distributions of \m~and \h.  We will refer to this model as Model 1*.  
%
%
 We then present the results for additional explicit flux compactifications, namely Models 2--4 of \cite{Krippendorf} (here referred to by the same name) and  the degree 18 hypersurface in $\mathbb{CP}^4_{1,1,1,6,9}$ (here referred to as Model 5).
 

The classical prepotentials are given in the large complex structure limit by \cite{Krippendorf, Candelas:1994hw},
\bea
F_2  &=&
 -u_1^2u_2-3u_1u_2^2-\tfrac{5}{3}u_2^3-2u_1u_2u_3-2u_2^2u_3-2u_1u_2u_4-2u_2^2u_4-2u_2u_3u_4+2u_1u_2
\nonumber \\
&&+u_2u_3+u_2u_4+\tfrac{5}{2}u_2^2+\tfrac{2}{3}u_1+\tfrac{8}{3}u_2+u_3+u_4-i\zeta(3)\tfrac{47}{4\pi^3}
\, , \\
F_3 &=&
 -\tfrac{4}{3}u_1^3-4u_1^2u_2-4u_1u_2^2-\tfrac{7}{6}u_2^3-4u_1^2u_3-8u_1u_2u_3-\tfrac{7}{2}u_2^2u_3-4u_1u_3^2 -\tfrac{7}{2}u_2u_3^2
 \nonumber \\
&& -u_3^3-\tfrac{3}{2}u_1^2u_4-3u_1u_2u_4-\tfrac{3}{2}u_2^2u_4-3u_1u_3u_4-3u_2u_3u_4-\tfrac{3}{2}u_3^2u_4-\tfrac{1}{2}u_1u_4^2 
\nonumber \\
&& -\tfrac{1}{2}u_2u_4^2-\tfrac{1}{2}u_3u_4^2-u_1^2u_5-2u_1u_2u_5-u_2^2u_5-2u_1u_3u_5-2u_2u_3u_5-u_3^2u_5 
\nonumber \\
&& -u_1u_4u_5-u_2u_4u_5-u_3u_4u_5+2u_1^2+4u_1u_2+4u_1u_3+u_1u_4+u_2u_4+u_3u_4
\nonumber \\
&& +\tfrac{7}{4}u_2^2+\tfrac{7}{2}u_2u_3+\tfrac{3}{2}u_3^2+u_1u_4+u_2u_4+u_3u_4+\tfrac{1}{2}u_1u_5+\tfrac{1}{2}u_2u_5+\tfrac{1}{2}u_3u_5
\nonumber \\
&&+\tfrac{23}{6}u_1+\tfrac{41}{12}u_2+3u_3+\tfrac{3}{2}u_4+u_5-i\zeta(3)\tfrac{45}{2\pi^3}
\, , \\
F_4 &=&
-u_1^3-\tfrac{3}{2}u_1^2u_2-\tfrac{1}{2}u_1u_2^2-u_1^2u_3-u_1u_2u_3-u_1^2u_4-u_1u_2u_4-u_1u_3u_4-\tfrac{7}{2}u_1^2u_5
\nonumber \\
&& -3u_1u_2u_5-\tfrac{1}{2}u_2^2u_5-2u_1u_3u_5-u_2u_3u_5-2u_1u_4u_5-u_2u_4u_5-u_3u_4u_5-\tfrac{7}{2}u_1u_5^2
\nonumber \\
&&-\tfrac{3}{2}u_2u_5^2-u_3u_5^2-u_4u_5^2-\tfrac{7}{6}u_5^3+\tfrac{3}{2}u_1^2+u_1u_2+\tfrac{1}{2}u_1u_3+\tfrac{1}{2}u_1u_4+\tfrac{7}{2}u_1u_5+u_2u_5
\nonumber \\
&&+\tfrac{1}{2}u_3u_5+\tfrac{7}{4}u_5^2+3u_1+\tfrac{3}{2}u_2+u_3+u_4+\tfrac{41}{12}u_5-i\zeta(3)\tfrac{45}{2\pi^3}
\, , \\
F_5 &=&
-\tfrac{3}{2}u_1^3-\tfrac{3}{2}u_1^2u_2-\tfrac{1}{2}u_1u_2^2+\tfrac{9}{4}u_1^2+\tfrac{3}{2}u_1u_2 + \tfrac{17}{4}u_1+\tfrac{3}{2}u_2-i\zeta(3)\tfrac{135}{4\pi^3} \, .
\eea
The leading-order instanton corrections to the prepotentials are given by \cite{Krippendorf, Candelas:1994hw},\footnote{We are very grateful to Denis Klevers and Sven Krippendorf for communicating to us corrected versions of these for Models 2 and 4.}  
\bea
I_2 &=& 
2 e^{2 i \pi u_1}+2 e^{2 i \pi  u_3}+232 e^{2 i \pi u_2}+2 e^{2 i \pi  u_4}+188 e^{4 i \pi u_2}+56 e^{2 i \pi  \left(u_1+u_2\right)} 
\nonumber \\
&&+56 e^{2 i \pi  \left(u_2+u_3\right)}+56 e^{2 i \pi  \left(u_2+u_4\right)} + \ldots
\, , \\
I_3 &=& 
e^{2 i \pi u_1}+252 e^{2 i \pi u_3}-2 e^{2 i \pi u_4}+e^{2 i \pi u_5}+ e^{2 i \pi  \left(u_1+u_2\right)}-9252e^{4 i \pi u_3}
\nonumber \\
&& +252 e^{2 i \pi  \left(u_2+u_3\right)}+ e^{2 i \pi  \left(u_1+u_4\right)}+3 e^{2 i \pi  \left(u_4+u_5\right)} + \ldots
\, , \\
I_4 &=& 
252 e^{2 i \pi u_1}+ e^{2 i \pi u_2}+ e^{2 i \pi u_3}+ e^{2 i \pi u_4}+ e^{2 i \pi u_5}-9252 e^{4 i \pi u_1}-2 e^{2 i \pi  \left(u_2+u_3\right)}
\nonumber \\
&& -2 e^{2 i \pi  \left(u_2+u_4\right)}+360 e^{2 i \pi  \left(u_1+u_5\right)} + \ldots
\, , \\
I_5 &=& 
-\tfrac{135}{2\pi^3}i e^{2 i \pi u_1}-\tfrac{3}{8\pi^3}i e^{2 i \pi u_2}-\tfrac{1215}{16\pi^3}i e^{4 i \pi u_1}+\tfrac{45}{16\pi^3}i e^{4 i \pi u_2}
\nonumber \\
&& +\tfrac{135}{\pi^3}i e^{2 i \pi \left(u_1+u_2\right)} + \ldots
\, ,
\eea

For Model 1* and Models 2--4 we consider a D3-tadpole contribution in the range $0 \leq Q_{\rm flux} \leq 22$ and uniformly sample the flux integers in the range $[-5,5]$, while for Model 5 we use a D3-tadpole range of $0 \leq Q_{\rm flux} \leq 182$ and a flux number range of $[-25,25]$. 
We here present the numerical results for the spectra of $\m$ and $\h$ computed in the same cases as was done for Model 1 in Figures \ref{fig:Mhist} and \ref{fig:Hhist}. 



\begin{figure}[ht!]
\begin{center}
\rowname{Model 1*}%
\subfigure {\label{fig:appM11}
\begin{tikzpicture}
\begin{axis}[mz plot style appendix]
\addplot[mycolor,fill=mycolor] table[y index = 1] {Mz_histogramData_model1_plot1.dat};
\end{axis}
\end{tikzpicture}}%
\subfigure {\label{fig:appM12}
\begin{tikzpicture}
\begin{axis}[mz plot style appendix]
\addplot[mycolor,fill=mycolor] table[y index = 1] {Mz_histogramData_model1_plot2.dat};
\end{axis}
\end{tikzpicture}}%
\subfigure {\label{fig:appM13}
\begin{tikzpicture}
\begin{axis}[mz plot style appendix]
\addplot[mycolor,fill=mycolor] table[y index = 1] {Mz_histogramData_model1_plot3.dat};
\end{axis}
\end{tikzpicture}}%
\subfigure {\label{fig:appM14}
\begin{tikzpicture}
\begin{axis}[mz plot style appendix]
\addplot[mycolor,fill=mycolor] table[y index = 1] {Mz_histogramData_model1_plot4.dat};
\end{axis}
\end{tikzpicture}}%
\\%
\rowname{Model 2}%
\subfigure {\label{fig:appM21}
\begin{tikzpicture}
\begin{axis}[mz plot style appendix]
\addplot[mycolor,fill=mycolor] table[y index = 1] {Mz_histogramData_model2_plot1.dat};
\end{axis}
\end{tikzpicture}}%
\subfigure {\label{fig:appM22}
\begin{tikzpicture}
\begin{axis}[mz plot style appendix]
\addplot[mycolor,fill=mycolor] table[y index = 1] {Mz_histogramData_model2_plot2.dat};
\end{axis}
\end{tikzpicture}}%
\subfigure {\label{fig:appM23}
\begin{tikzpicture}
\begin{axis}[mz plot style appendix]
\addplot[mycolor,fill=mycolor] table[y index = 1] {Mz_histogramData_model2_plot3.dat};
\end{axis}
\end{tikzpicture}}%
\subfigure {\label{fig:appM24}
\begin{tikzpicture}
\begin{axis}[mz plot style appendix]
\addplot[mycolor,fill=mycolor] table[y index = 1] {Mz_histogramData_model2_plot4.dat};
\end{axis}
\end{tikzpicture}}%
\\%
\rowname{Model 3}%
\subfigure {\label{fig:appM31}
\begin{tikzpicture}
\begin{axis}[mz plot style appendix]
\addplot[mycolor,fill=mycolor] table[y index = 1] {Mz_histogramData_model3_plot1.dat};
\end{axis}
\end{tikzpicture}}%
\subfigure {\label{fig:appM32}
\begin{tikzpicture}
\begin{axis}[mz plot style appendix]
\addplot[mycolor,fill=mycolor] table[y index = 1] {Mz_histogramData_model3_plot2.dat};
\end{axis}
\end{tikzpicture}}%
\subfigure {\label{fig:appM33}
\begin{tikzpicture}
\begin{axis}[mz plot style appendix]
\addplot[mycolor,fill=mycolor] table[y index = 1] {Mz_histogramData_model3_plot3.dat};
\end{axis}
\end{tikzpicture}}%
\subfigure {\label{fig:appM34}
\begin{tikzpicture}
\begin{axis}[mz plot style appendix]
\addplot[mycolor,fill=mycolor] table[y index = 1] {Mz_histogramData_model3_plot4.dat};
\end{axis}
\end{tikzpicture}}%
\\%
\rowname{Model 4}%
\subfigure {\label{fig:appM41}
\begin{tikzpicture}
\begin{axis}[mz plot style appendix]
\addplot[mycolor,fill=mycolor] table[y index = 1] {Mz_histogramData_model4_plot1.dat};
\end{axis}
\end{tikzpicture}}%
\subfigure {\label{fig:appM42}
\begin{tikzpicture}
\begin{axis}[mz plot style appendix]
\addplot[mycolor,fill=mycolor] table[y index = 1] {Mz_histogramData_model4_plot2.dat};
\end{axis}
\end{tikzpicture}}%
\subfigure {\label{fig:appM43}
\begin{tikzpicture}
\begin{axis}[mz plot style appendix]
\addplot[mycolor,fill=mycolor] table[y index = 1] {Mz_histogramData_model4_plot3.dat};
\end{axis}
\end{tikzpicture}}%
\subfigure {\label{fig:appM44}
\begin{tikzpicture}
\begin{axis}[mz plot style appendix]
\addplot[mycolor,fill=mycolor] table[y index = 1] {Mz_histogramData_model4_plot4.dat};
\end{axis}
\end{tikzpicture}}%
\\%
\rowname{Model 5}%
\addtocounter{subfigure}{-16}%
\subfigure{\label{fig:appM51}
\begin{tikzpicture}
\begin{axis}[mz plot style appendix]
\addplot[mycolor,fill=mycolor] table[y index = 1] {Mz_histogramData_model11169_plot1.dat};
\end{axis}
\end{tikzpicture}}%
\subfigure{\label{fig:appM52}
\begin{tikzpicture}
\begin{axis}[mz plot style appendix]
\addplot[mycolor,fill=mycolor] table[y index = 1] {Mz_histogramData_model11169_plot2.dat};
\end{axis}
\end{tikzpicture}}%
\subfigure{\label{fig:appM53}
\begin{tikzpicture}
\begin{axis}[mz plot style appendix]
\addplot[mycolor,fill=mycolor] table[y index = 1] {Mz_histogramData_model11169_plot3.dat};
\end{axis}
\end{tikzpicture}}%
\subfigure{\label{fig:appM54}
\begin{tikzpicture}
\begin{axis}[mz plot style appendix]
\addplot[mycolor,fill=mycolor] table[y index = 1] {Mz_histogramData_model11169_plot4.dat};
\end{axis}
\end{tikzpicture}}%
\\%
~~
\makebox[.23\textwidth][c]{\scriptsize{$\tau = u^i=10i$}}%
\makebox[.23\textwidth][c]{\scriptsize{$u_{\textrm{Im}}^{\textrm{max}}=10,\tau_{\textrm{Im}}^{\textrm{max}}=10$}}%
\makebox[.23\textwidth][c]{\scriptsize{$u_{\textrm{Im}}^{\textrm{max}}=5,~\tau_{\textrm{Im}}^{\textrm{max}}=5$}}%
\makebox[.23\textwidth][c]{\scriptsize{$u_{\textrm{Im}}^{\textrm{max}}=2,~\tau_{\textrm{Im}}^{\textrm{max}}=5$}}%
\end{center}%
\caption{Empirical eigenvalue densities of $\m$ in units of $|W|$.} 
\end{figure}%


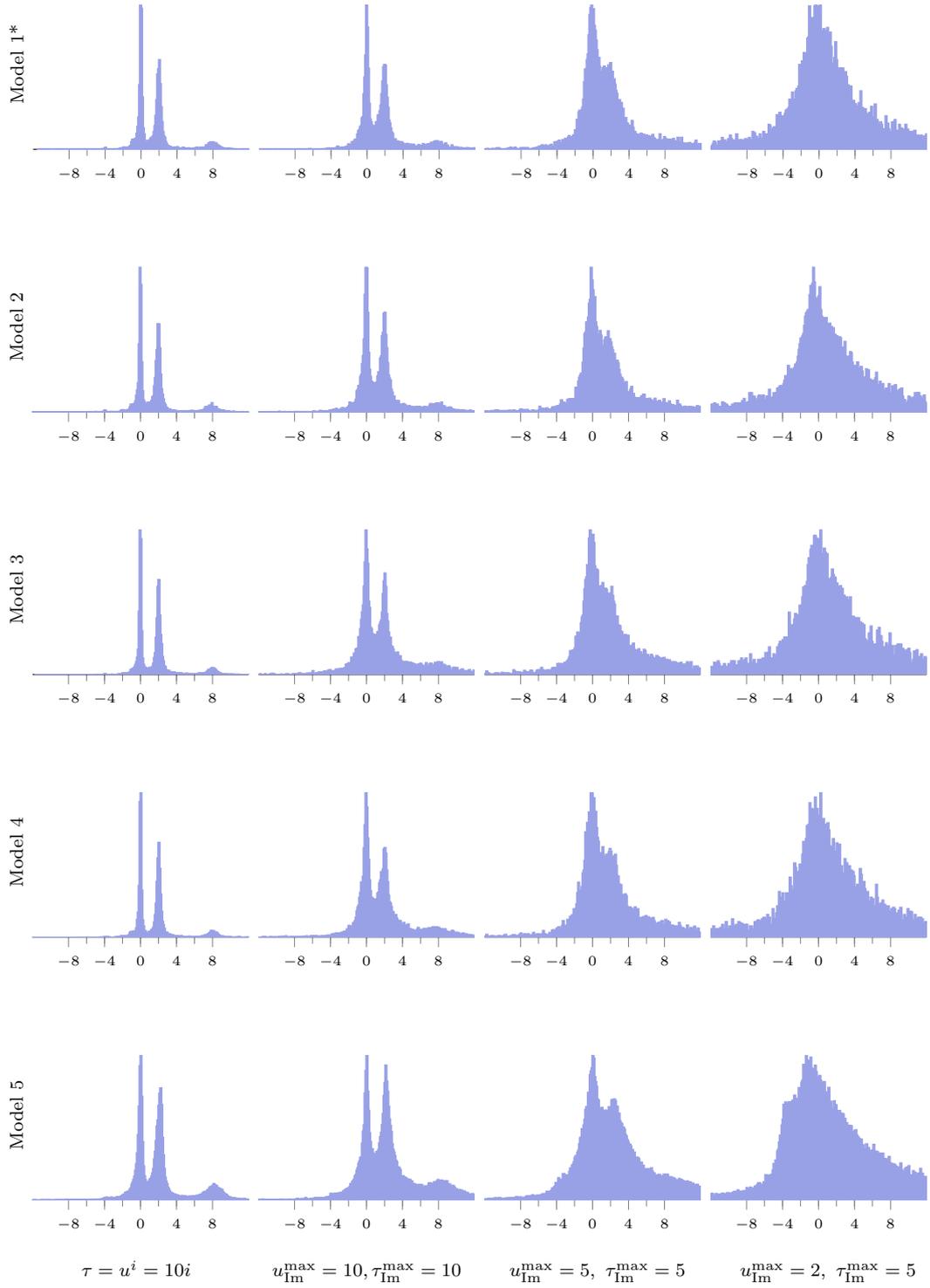
\begin{figure}[ht!]
\begin{center}
\rowname{Model 1*}%
\subfigure {\label{fig:appH11}
\begin{tikzpicture}
\begin{axis}[htot plot style appendix]
\addplot[mycolor,fill=mycolor] table[y index = 1] {Htot_histogramData_model1_plot1.dat};
\end{axis}
\end{tikzpicture}}%
\subfigure {\label{fig:appH12}
\begin{tikzpicture}
\begin{axis}[htot plot style appendix]
\addplot[mycolor,fill=mycolor] table[y index = 1] {Htot_histogramData_model1_plot2.dat};
\end{axis}
\end{tikzpicture}}%
\subfigure {\label{fig:appH13}
\begin{tikzpicture}
\begin{axis}[htot plot style appendix]
\addplot[mycolor,fill=mycolor] table[y index = 1] {Htot_histogramData_model1_plot3.dat};
\end{axis}
\end{tikzpicture}}%
\subfigure {\label{fig:appH14}
\begin{tikzpicture}
\begin{axis}[htot plot style appendix]
\addplot[mycolor,fill=mycolor] table[y index = 1] {Htot_histogramData_model1_plot4.dat};
\end{axis}
\end{tikzpicture}}%
\\%
\rowname{Model 2}%
\subfigure {\label{fig:appH21}
\begin{tikzpicture}
\begin{axis}[htot plot style appendix]
\addplot[mycolor,fill=mycolor] table[y index = 1] {Htot_histogramData_model2_plot1.dat};
\end{axis}
\end{tikzpicture}}%
\subfigure {\label{fig:appH22}
\begin{tikzpicture}
\begin{axis}[htot plot style appendix]
\addplot[mycolor,fill=mycolor] table[y index = 1] {Htot_histogramData_model2_plot2.dat};
\end{axis}
\end{tikzpicture}}%
\subfigure {\label{fig:appH23}
\begin{tikzpicture}
\begin{axis}[htot plot style appendix]
\addplot[mycolor,fill=mycolor] table[y index = 1] {Htot_histogramData_model2_plot3.dat};
\end{axis}
\end{tikzpicture}}%
\subfigure {\label{fig:appH24}
\begin{tikzpicture}
\begin{axis}[htot plot style appendix]
\addplot[mycolor,fill=mycolor] table[y index = 1] {Htot_histogramData_model2_plot4.dat};
\end{axis}
\end{tikzpicture}}%
\\%
\rowname{Model 3}%
\subfigure {\label{fig:appH31}
\begin{tikzpicture}
\begin{axis}[htot plot style appendix]
\addplot[mycolor,fill=mycolor] table[y index = 1] {Htot_histogramData_model3_plot1.dat};
\end{axis}
\end{tikzpicture}}%
\subfigure {\label{fig:appH32}
\begin{tikzpicture}
\begin{axis}[htot plot style appendix]
\addplot[mycolor,fill=mycolor] table[y index = 1] {Htot_histogramData_model3_plot2.dat};
\end{axis}
\end{tikzpicture}}%
\subfigure {\label{fig:appH33}
\begin{tikzpicture}
\begin{axis}[htot plot style appendix]
\addplot[mycolor,fill=mycolor] table[y index = 1] {Htot_histogramData_model3_plot3.dat};
\end{axis}
\end{tikzpicture}}%
\subfigure {\label{fig:appH34}
\begin{tikzpicture}
\begin{axis}[htot plot style appendix]
\addplot[mycolor,fill=mycolor] table[y index = 1] {Htot_histogramData_model3_plot4.dat};
\end{axis}
\end{tikzpicture}}%
\\%
\rowname{Model 4}%
\subfigure {\label{fig:appH41}
\begin{tikzpicture}
\begin{axis}[htot plot style appendix]
\addplot[mycolor,fill=mycolor] table[y index = 1] {Htot_histogramData_model4_plot1.dat};
\end{axis}
\end{tikzpicture}}%
\subfigure {\label{fig:appH42}
\begin{tikzpicture}
\begin{axis}[htot plot style appendix]
\addplot[mycolor,fill=mycolor] table[y index = 1] {Htot_histogramData_model4_plot2.dat};
\end{axis}
\end{tikzpicture}}%
\subfigure {\label{fig:appH43}
\begin{tikzpicture}
\begin{axis}[htot plot style appendix]
\addplot[mycolor,fill=mycolor] table[y index = 1] {Htot_histogramData_model4_plot3.dat};
\end{axis}
\end{tikzpicture}}%
\subfigure {\label{fig:appH44}
\begin{tikzpicture}
\begin{axis}[htot plot style appendix]
\addplot[mycolor,fill=mycolor] table[y index = 1] {Htot_histogramData_model4_plot4.dat};
\end{axis}
\end{tikzpicture}}%
\\%
\rowname{Model 5}%
\addtocounter{subfigure}{-16}%
\subfigure{\label{fig:appH51}
\begin{tikzpicture}
\begin{axis}[htot plot style appendix]
\addplot[mycolor,fill=mycolor] table[y index = 1] {Htot_histogramData_model11169_plot1.dat};
\end{axis}
\end{tikzpicture}}%
\subfigure{\label{fig:appH52}
\begin{tikzpicture}
\begin{axis}[htot plot style appendix]
\addplot[mycolor,fill=mycolor] table[y index = 1] {Htot_histogramData_model11169_plot2.dat};
\end{axis}
\end{tikzpicture}}%
\subfigure{\label{fig:appH53}
\begin{tikzpicture}
\begin{axis}[htot plot style appendix]
\addplot[mycolor,fill=mycolor] table[y index = 1] {Htot_histogramData_model11169_plot3.dat};
\end{axis}
\end{tikzpicture}}%
\subfigure{\label{fig:appH54}
\begin{tikzpicture}
\begin{axis}[htot plot style appendix]
\addplot[mycolor,fill=mycolor] table[y index = 1] {Htot_histogramData_model11169_plot4.dat};
\end{axis}
\end{tikzpicture}}%
\\%
~~
\makebox[.23\textwidth][c]{\scriptsize{$\tau = u^i=10i$}}%
\makebox[.23\textwidth][c]{\scriptsize{$u_{\textrm{Im}}^{\textrm{max}}=10,\tau_{\textrm{Im}}^{\textrm{max}}=10$}}%
\makebox[.23\textwidth][c]{\scriptsize{$u_{\textrm{Im}}^{\textrm{max}}=5,~\tau_{\textrm{Im}}^{\textrm{max}}=5$}}%
\makebox[.23\textwidth][c]{\scriptsize{$u_{\textrm{Im}}^{\textrm{max}}=2,~\tau_{\textrm{Im}}^{\textrm{max}}=5$}}%
\end{center}%
\caption{Empirical eigenvalue densities of $\h$ in units of $m_{3/2}^2$.} 
\end{figure}

\bibliographystyle{JHEP}
\bibliography{refs}

\providecommand{\href}[2]{#2}\begingroup\raggedright\begin{thebibliography}{10}

\bibitem{Grana:2000jj}
M.~Grana and J.~Polchinski, {\it {Supersymmetric three form flux perturbations
  on AdS(5)}},  {\em Phys. Rev.} {\bf D63} (2001) 026001,
  [\href{http://xxx.lanl.gov/abs/hep-th/0009211}{{\tt hep-th/0009211}}].

\bibitem{Gubser:2000vg}
S.~S. Gubser, {\it {Supersymmetry and F theory realization of the deformed
  conifold with three form flux}},
  \href{http://xxx.lanl.gov/abs/hep-th/0010010}{{\tt hep-th/0010010}}.

\bibitem{Giddings:2001yu}
S.~B. Giddings, S.~Kachru, and J.~Polchinski, {\it {Hierarchies from fluxes in
  string compactifications}},  {\em Phys.Rev.} {\bf D66} (2002) 106006,
  [\href{http://xxx.lanl.gov/abs/hep-th/0105097}{{\tt hep-th/0105097}}].

\bibitem{deWolfe:2002nn}
O.~DeWolfe and S.~B. Giddings, {\it {Scales and hierarchies in warped
  compactifications and brane worlds}},  {\em Phys. Rev.} {\bf D67} (2003)
  066008, [\href{http://xxx.lanl.gov/abs/hep-th/0208123}{{\tt
  hep-th/0208123}}].

\bibitem{Kachru:2003aw}
S.~Kachru, R.~Kallosh, A.~D. Linde, and S.~P. Trivedi, {\it {De Sitter vacua in
  string theory}},  {\em Phys. Rev.} {\bf D68} (2003) 046005,
  [\href{http://xxx.lanl.gov/abs/hep-th/0301240}{{\tt hep-th/0301240}}].

\bibitem{Kachru:2003sx}
S.~Kachru, R.~Kallosh, A.~D. Linde, J.~M. Maldacena, L.~P. McAllister, and
  S.~P. Trivedi, {\it {Towards inflation in string theory}},  {\em JCAP} {\bf
  0310} (2003) 013, [\href{http://xxx.lanl.gov/abs/hep-th/0308055}{{\tt
  hep-th/0308055}}].

\bibitem{Douglas:2006es}
M.~R. Douglas and S.~Kachru, {\it {Flux compactification}},  {\em Rev. Mod.
  Phys.} {\bf 79} (2007) 733--796,
  [\href{http://xxx.lanl.gov/abs/hep-th/0610102}{{\tt hep-th/0610102}}].

\bibitem{Denef:2007pq}
F.~Denef, M.~R. Douglas, and S.~Kachru, {\it {Physics of String Flux
  Compactifications}},  {\em Ann.Rev.Nucl.Part.Sci.} {\bf 57} (2007) 119--144,
  [\href{http://xxx.lanl.gov/abs/hep-th/0701050}{{\tt hep-th/0701050}}].

\bibitem{Denef:2008wq}
F.~Denef, {\it {Les Houches Lectures on Constructing String Vacua}},  in {\em
  {String theory and the real world: From particle physics to astrophysics.
  Proceedings, Summer School in Theoretical Physics, 87th Session, Les Houches,
  France, July 2-27, 2007}}, pp.~483--610, 2008.
\newblock \href{http://xxx.lanl.gov/abs/0803.1194}{{\tt arXiv:0803.1194}}.

\bibitem{Maharana:2012tu}
A.~Maharana and E.~Palti, {\it {Models of Particle Physics from Type IIB String
  Theory and F-theory: A Review}},  {\em Int. J. Mod. Phys.} {\bf A28} (2013)
  1330005, [\href{http://xxx.lanl.gov/abs/1212.0555}{{\tt arXiv:1212.0555}}].

\bibitem{Schellekens:2013bpa}
A.~Schellekens, {\it {Life at the Interface of Particle Physics and String
  Theory}},  {\em Rev.Mod.Phys.} {\bf 85} (2013), no.~4 1491--1540,
  [\href{http://xxx.lanl.gov/abs/1306.5083}{{\tt arXiv:1306.5083}}].

\bibitem{Tripathy:2002qw}
P.~K. Tripathy and S.~P. Trivedi, {\it {Compactification with flux on K3 and
  tori}},  {\em JHEP} {\bf 03} (2003) 028,
  [\href{http://xxx.lanl.gov/abs/hep-th/0301139}{{\tt hep-th/0301139}}].

\bibitem{Markus}
D.~Martinez-Pedrera, D.~Mehta, M.~Rummel, and A.~Westphal, {\it {Finding all
  flux vacua in an explicit example}},  {\em JHEP} {\bf 06} (2013) 110,
  [\href{http://xxx.lanl.gov/abs/1212.4530}{{\tt arXiv:1212.4530}}].

\bibitem{Ashok:2003gk}
S.~Ashok and M.~R. Douglas, {\it {Counting flux vacua}},  {\em JHEP} {\bf 0401}
  (2004) 060, [\href{http://xxx.lanl.gov/abs/hep-th/0307049}{{\tt
  hep-th/0307049}}].

\bibitem{Denef:2004ze}
F.~Denef and M.~R. Douglas, {\it {Distributions of flux vacua}},  {\em JHEP}
  {\bf 0405} (2004) 072, [\href{http://xxx.lanl.gov/abs/hep-th/0404116}{{\tt
  hep-th/0404116}}].

\bibitem{Denef:2004cf}
F.~Denef and M.~R. Douglas, {\it {Distributions of nonsupersymmetric flux
  vacua}},  {\em JHEP} {\bf 0503} (2005) 061,
  [\href{http://xxx.lanl.gov/abs/hep-th/0411183}{{\tt hep-th/0411183}}].

\bibitem{Eguchi:2005eh}
T.~Eguchi and Y.~Tachikawa, {\it {Distribution of flux vacua around singular
  points in Calabi-Yau moduli space}},  {\em JHEP} {\bf 01} (2006) 100,
  [\href{http://xxx.lanl.gov/abs/hep-th/0510061}{{\tt hep-th/0510061}}].

\bibitem{Wigner}
E.~P. Wigner, {\it {On the statistical distribution of the widths and spacings
  of nuclear resonance levels}},  {\em Math. Proc. Cambridge Philos. Soc.} {\bf
  47} (1951) 548--564.

\bibitem{Mehta}
M.~Mehta, {\it {Random Matrices}},  {\em Academic Press, Boston} (1991).

\bibitem{Deift}
P.~{Deift}, {\it {Universality for mathematical and physical systems}},  {\em
  ArXiv Mathematical Physics e-prints} (Mar., 2006)
  [\href{http://xxx.lanl.gov/abs/math-ph/0603038}{{\tt math-ph/0603038}}].

\bibitem{Kuijlaars}
A.~B.~J. {Kuijlaars}, {\it {Universality}},  {\em ArXiv e-prints} (Mar., 2011)
  [\href{http://xxx.lanl.gov/abs/1103.5922}{{\tt arXiv:1103.5922}}].

\bibitem{Wigner2}
E.~P. Wigner, {\it {On the distribution of the roots of certain symmetric
  matrices}},  {\em Ann.~of Math.} {\bf 67} (1958) 325.

\bibitem{2005SchenkerShulzBaldes}
J.~H. {Schenker} and H.~{Schulz-Baldes}, {\it {Semicircle law and freeness for
  random matrices with symmetries or correlations}},  {\em ArXiv Mathematical
  Physics e-prints} (May, 2005)
  [\href{http://xxx.lanl.gov/abs/math-ph/0505003}{{\tt math-ph/0505003}}].

\bibitem{2007HofmannCredner}
K.~{Hofmann-Credner} and M.~{Stolz}, {\it {Wigner theorems for random matrices
  with dependent entries: Ensembles associated to symmetric spaces and sample
  covariance matrices}},  {\em ArXiv e-prints} (July, 2007)
  [\href{http://xxx.lanl.gov/abs/0707.2333}{{\tt arXiv:0707.2333}}].

\bibitem{Marsh:2011aa}
D.~Marsh, L.~McAllister, and T.~Wrase, {\it {The Wasteland of Random
  Supergravities}},  {\em JHEP} {\bf 1203} (2012) 102,
  [\href{http://xxx.lanl.gov/abs/1112.3034}{{\tt arXiv:1112.3034}}].

\bibitem{Chen:2011ac}
X.~Chen, G.~Shiu, Y.~Sumitomo, and S.~H. Tye, {\it {A Global View on The Search
  for de-Sitter Vacua in (type IIA) String Theory}},  {\em JHEP} {\bf 1204}
  (2012) 026, [\href{http://xxx.lanl.gov/abs/1112.3338}{{\tt
  arXiv:1112.3338}}].

\bibitem{Bachlechner:2012at}
T.~C. Bachlechner, D.~Marsh, L.~McAllister, and T.~Wrase, {\it {Supersymmetric
  Vacua in Random Supergravity}},  {\em JHEP} {\bf 1301} (2013) 136,
  [\href{http://xxx.lanl.gov/abs/1207.2763}{{\tt arXiv:1207.2763}}].

\bibitem{Sousa:2014qza}
K.~Sousa and P.~Ortiz, {\it {Perturbative Stability along the Supersymmetric
  Directions of the Landscape}},  {\em JCAP} {\bf 1502} (2015) 017,
  [\href{http://xxx.lanl.gov/abs/1408.6521}{{\tt arXiv:1408.6521}}].

\bibitem{MH:2004gm}
A.~Kobakhidze and L.~Mersini-Houghton, {\it {Birth of the universe from the
  landscape of string theory}},  {\em Eur. Phys. J.} {\bf C49} (2007) 869--873,
  [\href{http://xxx.lanl.gov/abs/hep-th/0410213}{{\tt hep-th/0410213}}].

\bibitem{Aazami:2005jf}
A.~Aazami and R.~Easther, {\it {Cosmology from random multifield potentials}},
  {\em JCAP} {\bf 0603} (2006) 013,
  [\href{http://xxx.lanl.gov/abs/hep-th/0512050}{{\tt hep-th/0512050}}].

\bibitem{Easther:2005zr}
R.~Easther and L.~McAllister, {\it {Random matrices and the spectrum of
  N-flation}},  {\em JCAP} {\bf 0605} (2006) 018,
  [\href{http://xxx.lanl.gov/abs/hep-th/0512102}{{\tt hep-th/0512102}}].

\bibitem{Pedro:2013nda}
F.~G. Pedro and A.~Westphal, {\it {The Scale of Inflation in the Landscape}},
  {\em Phys. Lett.} {\bf B739} (2014) 439--444,
  [\href{http://xxx.lanl.gov/abs/1303.3224}{{\tt arXiv:1303.3224}}].

\bibitem{Long:2014fba}
C.~Long, L.~McAllister, and P.~McGuirk, {\it {Heavy Tails in Calabi-Yau Moduli
  Spaces}},  {\em JHEP} {\bf 1410} (2014) 187,
  [\href{http://xxx.lanl.gov/abs/1407.0709}{{\tt arXiv:1407.0709}}].

\bibitem{Marsh:2013qca}
M.~C.~D. Marsh, L.~McAllister, E.~Pajer, and T.~Wrase, {\it {Charting an
  Inflationary Landscape with Random Matrix Theory}},  {\em JCAP} {\bf 1311}
  (2013) 040, [\href{http://xxx.lanl.gov/abs/1307.3559}{{\tt
  arXiv:1307.3559}}].

\bibitem{Bachlechner:2014gfa}
T.~C. Bachlechner, C.~Long, and L.~McAllister, {\it {Planckian Axions in String
  Theory}},  \href{http://xxx.lanl.gov/abs/1412.1093}{{\tt arXiv:1412.1093}}.

\bibitem{Bachlechner:2014hsa}
T.~C. Bachlechner, M.~Dias, J.~Frazer, and L.~McAllister, {\it {Chaotic
  inflation with kinetic alignment of axion fields}},  {\em Phys. Rev.} {\bf
  D91} (2015), no.~2 023520, [\href{http://xxx.lanl.gov/abs/1404.7496}{{\tt
  arXiv:1404.7496}}].

\bibitem{Battefeld:2014qoa}
T.~Battefeld and C.~Modi, {\it {Local random potentials of high
  differentiability to model the Landscape}},  {\em JCAP} {\bf 1503} (2015),
  no.~03 010, [\href{http://xxx.lanl.gov/abs/1409.5135}{{\tt
  arXiv:1409.5135}}].

\bibitem{Altland:1997zz}
A.~Altland and M.~R. Zirnbauer, {\it {Nonstandard symmetry classes in
  mesoscopic normal-superconducting hybrid structures}},  {\em Phys. Rev.} {\bf
  B55} (1997) 1142--1161.

\bibitem{Grimm:2004uq}
T.~W. Grimm and J.~Louis, {\it {The Effective action of N = 1 Calabi-Yau
  orientifolds}},  {\em Nucl.Phys.} {\bf B699} (2004) 387--426,
  [\href{http://xxx.lanl.gov/abs/hep-th/0403067}{{\tt hep-th/0403067}}].

\bibitem{Gukov:1999ya}
S.~Gukov, C.~Vafa, and E.~Witten, {\it {CFT's from Calabi-Yau four folds}},
  {\em Nucl.Phys.} {\bf B584} (2000) 69--108,
  [\href{http://xxx.lanl.gov/abs/hep-th/9906070}{{\tt hep-th/9906070}}].

\bibitem{Cicoli:2013swa}
M.~Cicoli, J.~P. Conlon, A.~Maharana, and F.~Quevedo, {\it {A Note on the
  Magnitude of the Flux Superpotential}},  {\em JHEP} {\bf 01} (2014) 027,
  [\href{http://xxx.lanl.gov/abs/1310.6694}{{\tt arXiv:1310.6694}}].

\bibitem{LVS}
V.~{Balasubramanian}, P.~{Berglund}, J.~P. {Conlon}, and F.~{Quevedo}, {\it
  {Systematics of Moduli Stabilisation in Calabi-Yau Flux Compactifications}},
  {\em Journal of High Energy Physics} {\bf 3} (Mar., 2005) 7,
  [\href{http://xxx.lanl.gov/abs/hep-th/0502058}{{\tt hep-th/0502058}}].

\bibitem{Westphal:2006tn}
A.~Westphal, {\it {de Sitter string vacua from Kahler uplifting}},  {\em JHEP}
  {\bf 03} (2007) 102, [\href{http://xxx.lanl.gov/abs/hep-th/0611332}{{\tt
  hep-th/0611332}}].

\bibitem{Covi:2008ea}
L.~Covi, M.~Gomez-Reino, C.~Gross, J.~Louis, G.~A. Palma, et~al., {\it {de
  Sitter vacua in no-scale supergravities and Calabi-Yau string models}},  {\em
  JHEP} {\bf 0806} (2008) 057, [\href{http://xxx.lanl.gov/abs/0804.1073}{{\tt
  arXiv:0804.1073}}].

\bibitem{Kallosh:2014oja}
R.~Kallosh, A.~Linde, B.~Vercnocke, and T.~Wrase, {\it {Analytic Classes of
  Metastable de Sitter Vacua}},  {\em JHEP} {\bf 10} (2014) 11,
  [\href{http://xxx.lanl.gov/abs/1406.4866}{{\tt arXiv:1406.4866}}].

\bibitem{decoupling}
M.~C.~D. Marsh, B.~Vercnocke, and T.~Wrase, {\it {Decoupling and de Sitter
  Vacua in Approximate No-Scale Supergravities}},  {\em JHEP} {\bf 05} (2015)
  081, [\href{http://xxx.lanl.gov/abs/1411.6625}{{\tt arXiv:1411.6625}}].

\bibitem{Krippendorf}
M.~Cicoli, D.~Klevers, S.~Krippendorf, C.~Mayrhofer, F.~Quevedo, and
  R.~Valandro, {\it {Explicit de Sitter Flux Vacua for Global String Models
  with Chiral Matter}},  {\em JHEP} {\bf 05} (2014) 001,
  [\href{http://xxx.lanl.gov/abs/1312.0014}{{\tt arXiv:1312.0014}}].

\bibitem{Candelas:1994hw}
P.~Candelas, A.~Font, S.~H. Katz, and D.~R. Morrison, {\it {Mirror symmetry for
  two parameter models. 2.}},  {\em Nucl. Phys.} {\bf B429} (1994) 626--674,
  [\href{http://xxx.lanl.gov/abs/hep-th/9403187}{{\tt hep-th/9403187}}].

\bibitem{Denef:2004dm}
F.~Denef, M.~R. Douglas, and B.~Florea, {\it {Building a better racetrack}},
  {\em JHEP} {\bf 0406} (2004) 034,
  [\href{http://xxx.lanl.gov/abs/hep-th/0404257}{{\tt hep-th/0404257}}].

\bibitem{Candelas:1990pi}
P.~Candelas and X.~de~la Ossa, {\it {Moduli Space of {Calabi-Yau} Manifolds}},
  {\em Nucl.Phys.} {\bf B355} (1991) 455--481.

\bibitem{Hebecker:2014eua}
A.~Hebecker, S.~C. Kraus, and L.~T. Witkowski, {\it {D7-Brane Chaotic
  Inflation}},  {\em Phys. Lett.} {\bf B737} (2014) 16--22,
  [\href{http://xxx.lanl.gov/abs/1404.3711}{{\tt arXiv:1404.3711}}].

\bibitem{Hebecker:2014kva}
A.~Hebecker, P.~Mangat, F.~Rompineve, and L.~T. Witkowski, {\it {Tuning and
  Backreaction in F-term Axion Monodromy Inflation}},  {\em Nucl. Phys.} {\bf
  B894} (2015) 456--495, [\href{http://xxx.lanl.gov/abs/1411.2032}{{\tt
  arXiv:1411.2032}}].

\bibitem{Hayashi:2014aua}
H.~Hayashi, R.~Matsuda, and T.~Watari, {\it {Issues in Complex Structure Moduli
  Inflation}},  \href{http://xxx.lanl.gov/abs/1410.7522}{{\tt
  arXiv:1410.7522}}.

\bibitem{Magda2}
A.~P. Braun, N.~Johansson, M.~Larfors, and N.-O. Walliser, {\it {Restrictions
  on infinite sequences of type IIB vacua}},  {\em JHEP} {\bf 10} (2011) 091,
  [\href{http://xxx.lanl.gov/abs/1108.1394}{{\tt arXiv:1108.1394}}].

\end{thebibliography}\endgroup

\end{document}